\documentclass[aps, prl, amsmath, floats,floatfix, twocolumn, superscriptaddress]{revtex4-1}

\usepackage[normalem]{ulem}
\usepackage{amssymb}
\usepackage{amsmath}
\usepackage{verbatim}
\usepackage{mathrsfs}
\usepackage{amsfonts}
\usepackage{latexsym}
\usepackage{epsfig}
\usepackage{color}
\usepackage{graphicx,subfigure}
\usepackage{units}
\usepackage{slashbox}
\usepackage{envmath}
\usepackage{natbib}
\usepackage{hyperref}
\usepackage[utf8]{inputenc}

\begin{document}


\definecolor{orange}{rgb}{0.9,0.45,0}

\newcommand{\re}{\mbox{Re}}
\newcommand{\im}{\mbox{Im}}

\newcommand{\tf}[1]{\textcolor{red}{#1}}
\newcommand{\nsg}[1]{\textcolor{blue}{#1}}
\newcommand{\ch}[1]{\textcolor{green}{CH: #1}}
\newcommand{\fdg}[1]{\textcolor{orange}{FDG: #1}}
\newcommand{\pcd}[1]{\textcolor{magenta}{#1}}
\newcommand{\mz}[1]{\textcolor{cyan}{[\bf MZ: #1]}}

\def\CovDev{D}
\def\Res{{\mathcal R}}
\def\Gammaflat{\hat \Gamma}
\def\metricflat{\hat \gamma}
\def\Dflat{\hat {\mathcal D}}
\def\part_n{\partial_\perp}

\def\Lie{\mathcal{L}}
\def\A{\mathcal{X}}
\def\Aphi{\A_{\phi}}
\def\hAphi{\hat{\A}_{\phi}}
\def\E{\mathcal{E}}
\def\Ham{\mathcal{H}}
\def\M{\mathcal{M}}
\def\R{\mathcal{R}}
\def\p{\partial}

\def\hg{\hat{\gamma}}
\def\hA{\hat{A}}
\def\hD{\hat{D}}
\def\hE{\hat{E}}
\def\hR{\hat{R}}
\def\hcA{\hat{\mathcal{A}}}
\def\hDelt{\hat{\triangle}}

\def\be{\begin{equation}}
\def\ee{\end{equation}}

\renewcommand{\t}{\times}

\long\def\symbolfootnote[#1]#2{\begingroup%
\def\thefootnote{\fnsymbol{footnote}}\footnote[#1]{#2}\endgroup}


\title{Multi-field, multi-frequency bosonic stars and a stabilization mechanism} 	
 
\author{Nicolas Sanchis-Gual}
\affiliation{Departamento  de  Matem\'{a}tica  da  Universidade  de  Aveiro  and  Centre  for  Research  and  Development in  Mathematics  and  Applications  (CIDMA),  Campus  de  Santiago,  3810-183  Aveiro,  Portugal} 
\affiliation{Centro de Astrof\'\i sica e Gravita\c c\~ao - CENTRA, Departamento de F\'\i sica,
Instituto Superior T\'ecnico - IST, Universidade de Lisboa - UL, Avenida
Rovisco Pais 1, 1049-001, Portugal} 

 \author{Fabrizio Di Giovanni}
\affiliation{Departamento de
  Astronom\'{\i}a y Astrof\'{\i}sica, Universitat de Val\`encia,
  Dr. Moliner 50, 46100, Burjassot (Val\`encia), Spain}

    \author{Carlos Herdeiro}
\affiliation{Departamento  de  Matem\'{a}tica  da  Universidade  de  Aveiro  and  Centre  for  Research  and  Development in  Mathematics  and  Applications  (CIDMA),  Campus  de  Santiago,  3810-183  Aveiro,  Portugal}

  \author{Eugen Radu}
\affiliation{Departamento  de  Matem\'{a}tica  da  Universidade  de  Aveiro  and  Centre  for  Research  and  Development in  Mathematics  and  Applications  (CIDMA),  Campus  de  Santiago,  3810-183  Aveiro,  Portugal}

\author{Jos\'e~A. Font}
\affiliation{Departamento de
  Astronom\'{\i}a y Astrof\'{\i}sica, Universitat de Val\`encia,
  Dr. Moliner 50, 46100, Burjassot (Val\`encia), Spain}
\affiliation{Observatori Astron\`omic, Universitat de Val\`encia, C/ Catedr\'atico 
  Jos\'e Beltr\'an 2, 46980, Paterna (Val\`encia), Spain}


\date{March 2021}


\begin{abstract} 
Scalar bosonic stars (BSs) stand out as a multi-purpose model of exotic compact objects. We enlarge the  landscape of such (asymptotically  flat, stationary,  everywhere regular) objects by considering multiple fields (possibly) with different  frequencies.  This allows for new morphologies \textit{and} a stabilization mechanism for different sorts of unstable BSs. First, any odd number of complex fields, yields a continuous family of BSs departing from the spherical, equal frequency, $\ell-$BSs.  As the simplest illustration, we construct the \textit{$\ell$ = 1 BSs family}, 
that includes several  single frequency solutions, including even parity (such as spinning BSs and a toroidal, static BS) and  odd parity (a dipole BS) limits.
Second, these limiting solutions are dynamically unstable, but can be  stabilized by a \textit{hybrid}-$\ell$ construction: adding a  sufficiently large fundamental $\ell=0$ BS of another field, with a different frequency. Evidence for this dynamical robustness is obtained by non-linear numerical simulations of the corresponding Einstein-(complex, massive) Klein-Gordon system, both in formation and evolution scenarios, and a suggestive correlation between stability and energy distribution is observed. Similarities and differences with vector BSs are anticipated.

\end{abstract}


\maketitle

\vspace{0.8cm}

{\bf {\em Introduction.}} 
Recent observations of dark compact objects, via gravitational waves~\cite{Abbott2016,LIGOScientific:2018mvr,Abbott:2020niy}, very large baseline interferometry imaging of M87*~\cite{Akiyama:2019cqa} or orbital motions near SgrA*~\cite{2018A&A...618L..10G}  support the black hole hypothesis. 
Yet, the issue of degeneracy remains a central question. This has been sharpened by recent illustrations, in both the gravitational and electromagnetic channels~\cite{CalderonBustillo:2020srq,Herdeiro:2021lwl}, using \textit{dynamically robust} bosonic stars (BSs) to imitate the observed data.

In spite of many proposed black  hole mimicker models~\cite{Cardoso:2019rvt}, imposing an  established formation mechanism and dynamical stability, within a sound effective field theory, restricts considerably  the choices. The fundamental spherical (scalar~\cite{Kaup:1968zz,Ruffini:1969qy} or vector~\cite{Brito:2015pxa})  BSs, occuring in Einstein's gravity minimally coupled to a single complex, free bosonic field, fulfill these criteria~\cite{Liebling:2012fv}, having become prolific testing grounds for strong-gravity phenomenology.  The purpose of  this Letter is to enlarge the landscape of dynamically robust BSs, by considering  multi-field, multi-frequency solutions, which will open new avenues of research, both theoretical and phenomenological, for these remarkable gravitational solitons.

{\bf {\em Single and multi-field BSs.}} 
Single field BSs appear in  different varieties~\footnote{In this paper we focus on non-self-interacting fields.} besides the aformentioned fundamental spherical (\textit{monopole}) solutions~\cite{Herdeiro:2017fhv}, including spinning BSs~\cite{Schunck:1996he,Yoshida:1997qf,Brito:2015pxa,Herdeiro:2019mbz} and multipolar (static) BSs~\cite{Herdeiro:2020kvf}. Concerning  the former, only the vector case is dynamically robust~\cite{sanchis2019nonlinear}; concerning the latter, the simplest illustration is the \textit{dipole} BS, shown to be unstable below.

Single-field BSs provide building blocks for multi-field BSs, despite the non-linearity of the model. Appropriate superpositions, moreover,  change dynamical properties. An \textit{excited} monopole scalar BS, which is unstable against decaying to a fundamental BS~\cite{Balakrishna:1997ej},  is stabilized by adding a sufficiently large fundamental monopole BS of a second field~\cite{Bernal:2009zy} (see also~\cite{Matos:2007zza,UrenaLopez:2010ur,di2020dynamical}). In the same spirit,  for the non-relativistic BSs of the Schr\"odinger-Poisson system, a dipole configuration is stabilized by adding a sufficiently large fundamental monopole~\cite{Guzman:2019gqc} (see also~\cite{guzman2021stability}). These examples turn  out to be  illustrations of a stabilization \textit{mechanism},  as we shall discuss.

A particular type of multi-field BSs, composed by an odd number ($2\ell+1$, $\ell \in \mathbb{N}_0$) of (equal frequency) complex scalar fields was unveiled in~\cite{Alcubierre:2018ahf} and dubbed $\ell$-BSs. These are spherical and can be seen as a superposition of all $m$ multipoles, with the same amplitude, for a given $\ell$. $\ell$-BSs were shown to be stable in spherical symmetry~\cite{Alcubierre:2019qnh}; but non-spherical perturbations suggest new equilibrium configurations exist with different frequencies for different fields~\cite{Jaramillo:2020rsv}. This will be confirmed herein: $\ell$-BS are just the  symmetry-enhanced points of  larger continuous families of multi-field, multi-frequency BSs~\footnote{In spherical symmetry, multi-frequency BSs were discussed in~\cite{Choptuik:2019zji}; and BSs in multi-scalar theories were recently constructed in~\cite{Yazadjiev:2019oul,Collodel:2019uns}.}. 

{\bf {\em The model.}} 
%
 Einstein's gravity minimally coupled to a set of $N$ free, complex, massive scalar fields, $\Phi^{(j)}$, is  $S= (16\pi G)^{-1}\int  d^4x\sqrt{-g}\left[R
   -\mathcal{L}
 \right]$, where $G$  is Newton's constant, $R$ the Ricci scalar and the matter Lagrangian is
\begin{equation}
\label{action}
 \mathcal{L}=\sum _{j=1}^N
	\left(
	g^{\alpha\beta}\Phi_{, \, \alpha}^{(j)*} \Phi_{, \, \beta}^{(j)} + \mu^2 \Phi^{(j)*}\Phi^{(j)} 
	\right) \ ;
\end{equation}
 $\mu$ is the (common) mass of all fields $\Phi^{(j)}$ and `*' denotes complex conjugation.

All BSs studied herein are described by the metric ansatz
$
ds^2=-e^{2F_0(r,\theta)} dt^2+e^{2F_1(r,\theta)}(dr^2+r^2 d\theta^2)
+e^{2F_2(r,\theta)}r^2 \sin^2\theta (d\varphi-W(r,\theta) dt)^2,
$
in terms of four unknown metric functions of the coordinates
$(r,\theta)$; the two Killing coordinates $(t,\varphi)$ represent the time and azimuthal directions.
The $N$ scalar fields $\Phi^{(j)}$ are
\begin{equation}
\label{ansatz-general}
\Phi^{(j)}=\phi_j(r,\theta)e^{-i(w_j t-m_j\varphi)} \ ,
\end{equation}
where $w_j\in \mathbb{R}^+$ are the fields' frequencies and $m_j\in \mathbb{Z}$ the azimuthal harmonic indices.
The fields' amplitudes $\phi_j$  are real functions.  This ansatz illustrates \textit{symmetry non-inheritance}~\cite{Smolic:2015txa}: each $\Phi^{(j)}$ depends on the Killing coordinates but its energy-momentum tensor (EMT) does not~\footnote{For the $\ell$-BS,  moreover, the fields depend on both the Killing and $\theta$ coordinates but the total EMT does not, albeit the individual EMT of each scalar field may still depend on the $\theta$ coordinate.}. 

{\bf {\em Constructing  the enlarged $\ell$-BSs family.}} 
Taking an odd number of fields, $N=2\ell+1$, for a fixed $\ell\in \mathbb{N}_0$, a spherical ansatz ($W=0$, $F_1=F_2$, with no  angular dependence), equal frequencies ($w_j=w$) and equal radial amplitudes such that  $\phi_j(r,\theta)e^{im_j\varphi}=f(r)Y_\ell^{-\ell-1+j}(\theta,\varphi)$, where $Y_\ell^m$ are the standard spherical harmonics, one obtains $\ell$-BSs~\cite{Alcubierre:2018ahf}.

Taking still $N=2\ell+1$ but keeping the most general ansatz discussed above new possibilities emerge. We take $m_j=-\ell-1+j$, as for $\ell$-BSs. For concreteness we focus on the simplest non-trivial $\ell=1$ case. Then, the problem
reduces to solving a set of seven partial differential equations (PDEs),
for  $F_{0,1,2,},W$
and   $\phi_{1,2,3}$.
This number reduces for particular cases~\footnote{The non-vanishing Einstein equations are
  $E_t^t,E_r^r,E_\theta^\theta$, $E_\varphi^\varphi,E_\varphi^t$ and $E_r^\theta$. 
	Four of them
	are solved together with the three Klein-Gordon eqs., yielding 
	a coupled system 
	of seven PDEs on the unknown functions
	$F_{0,1,2},W,\phi_{1,2,3}$.
 The remaining two Einstein equations are treated as constraints 
and used to check the numerical accuracy.}.
These PDEs are solved with boundary conditions: $(i)$
at $r=0$, $\partial_r F_{0,1,2}=0,~~\partial_r W=0,~~\partial_r \phi_{2}=\phi_{1,3}=0$; $(ii)$
at infinity all functions vanish, $F_{0,1,2}= W=\phi_i=0$; $(iii)$ at $\theta=0,\pi$,
$\partial_\theta F_{0,1,2}=0,~~\partial_\theta \phi_2=\phi_{1,3}=0$; $(iv)$
 the geometry is invariant
under a reflection along the equatorial plane $\theta=\pi/2$, 
and, as for $\ell-$BSs, $\phi_2$ and $\phi_{1,3}$ 
are parity odd and even functions, respectively. Thus, at $\theta=\pi/2$, $\partial_\theta F_{0,1,2} =\partial_\theta W= \phi_{2}=\partial_\theta \phi_{1,3}=0$. All configurations reported here are \textit{fundamental}, with $n=0$, 
where $n$ is the number of nodes along the equatorial plane of $\phi_{1,3}(r,\pi/2)$~\footnote{Excited solutions with $n>0$
 exist as well.}. The solutions are constructed numerically by employing the same approach as for the case of single-field BSs - 
see $e.g.$ the description in~\cite{Herdeiro:2015gia}.

{\bf {\em The single-frequency, multi-field limits.}} 
There are special limits where all fields have the same frequency ($w_j=w$). First, there are two types of single-field configurations: 
$(i)$ \textit{dipole BSs} (DBS$_0$), which are odd parity, obtained by taking only the $m=0$ mode, $\phi_2\neq 0$~\footnote{It is $\Phi_{(2,1,0)}$ in the notation of~\cite{Herdeiro:2020kvf}.} (see also~\cite{Herdeiro:2021mol}). Their angular momentum density vanishes ($W=0$) and so does their total angular momentum, $J=0$; 
$(ii)$ \textit{spinning BSs}, (SBS$_{\pm 1}$)~\cite{Schunck:1996he,Yoshida:1997qf,Brito:2015pxa,Herdeiro:2019mbz}, which are even parity and have $J\neq 0$, obtained by taking only either $\phi_1\neq 0$ (SBS$_{-1}$) or $\phi_3\neq 0$ (SBS$_{+1}$).

Second, combinations of single-field configurations lead to two types of two-field configurations: 
$(iii)$ \textit{spinning dipolar BSs} (DBS$_0$+SBS$_{\pm1}$), in which case only either $\phi_1\neq \phi_2\neq 0$ (SBS$_{-1}$+DBS$_0$) or $\phi_2\neq \phi_3\neq 0$ (DBS$_0$+SBS$_{+1}$). These are novel solutions with $J\neq 0$, carried by the even-parity scalar field; 
$(iv)$ \textit{toroidal static BSs} (SBS$_{-1}$+SBS$_{+1}$), for which $\phi_1= \phi_3\equiv \phi\neq 0$.
Each field $\Phi^{(1)}, \Phi^{(3)}$
 carries a $local$ angular momentum density,
with the corresponding EMT component
$T_\varphi^{t(1)}=-T_\varphi^{t(3)}=
2e^{-2F_0} w \phi^2$,
such that their sum is zero, $T_\varphi^{t}=0$, and
the spacetime is locally and globally static, with $J=0$.

\begin{figure}[h!]
\centering
\includegraphics[width=1\linewidth]{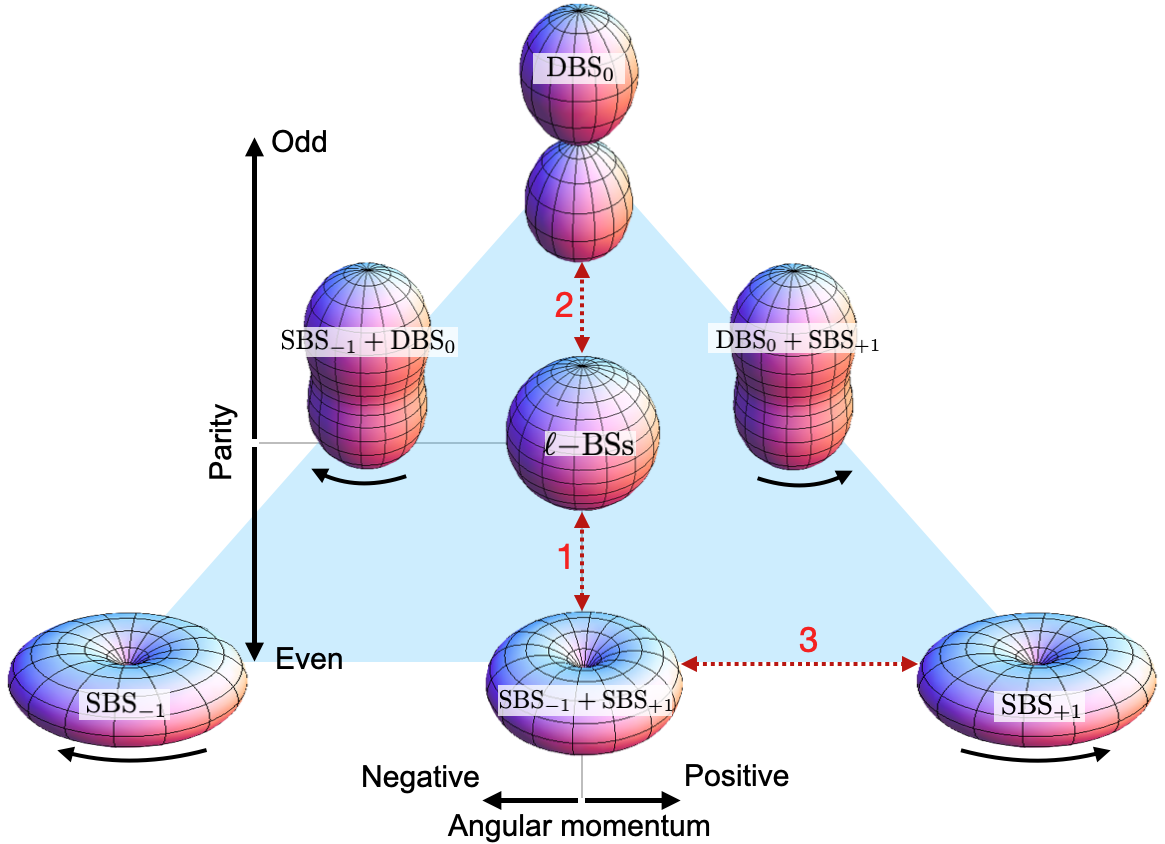}
\caption{$\ell=1$ BSs family.}
\label{fig1}
\end{figure}

Finally, $(v)$ \textit{$\ell$-BSs} (SBS$_{-1}$+DBS$_{0}$+SBS$_{+1}$), which are static, spherical and have $\phi_i=\phi(r)(\frac{\sin \theta}{\sqrt{2}}, \cos \theta,\frac{\sin \theta}{\sqrt{2}})$. Fig.~\ref{fig1} illustrates the single-frequency limits of the enlarged $\ell=1$ BSs family as 3D plots.

\begin{figure}
\centering
\includegraphics[width=1\linewidth]{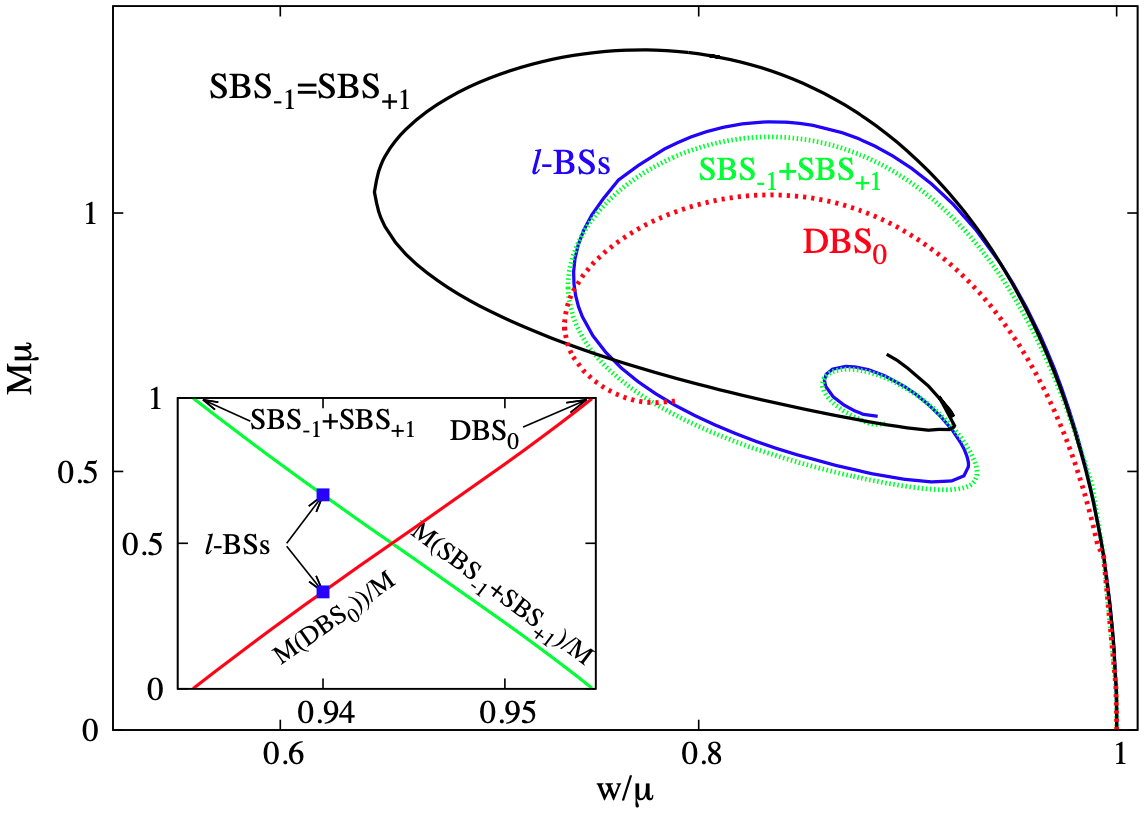}
\caption{ADM mass $vs.$ frequency for some of the single-frequency limits of the $\ell=1$ BSs family. (Inset) Fraction of the total mass  in SBS$_{-1}$+SBS$_{+1}$ and in DBS$_{0}$ along the static sequence {\bf 1}+{\bf 2} in Fig.~\ref{fig1} for $w_2=0.94\mu$ (and $w_1/\mu=w_3/\mu$ in the $x$-axis).}
\label{fig2}
\end{figure}

For each of the five types of solutions described above, there is a 1-dimensional family of BSs with $w_{\rm min}<w<\mu$,
where $w_{\rm min}$ is family dependent. In an ADM mass, $M$ $vs.$ frequency $w$ diagram, they describe a spiral-type curve (costumary for BSs) - Fig.\ref{fig2}~\footnote{
The same holds for the Noether charge of the fields.}.
As  the frequency is decreased from the maximal value, $\mu$,
 the ADM mass increases up to a maximum value 
$M^{\rm (max)}$ which is family dependent.
The energy density distribution of the solutions is illustrated by the morphologies in Fig.~\ref{fig1}. 

{\bf {\em The multi-frequency, multi-field interpolations.}} 
Relaxing the equal frequency requirement a larger solution space emerges (blue triangle in Fig.~\ref{fig1}). There are multi-frequency BSs interpolating between the  single-frequency ones, which are particular points in a manifold of solutions~\footnote{Bare in mind each of these points is in fact a continuous family, spanning a range of frequencies.}. 
As an illustration consider the interpolation SBS$_{-1}$+SBS$_{+1}$ $\longleftrightarrow$ DBS$_{0}$, which  goes through an $\ell$-BS. Fix,   $e.g.$, $w_2=0.94\mu$  for the $\ell$-BS along the sequence - Fig. \ref{fig2} (inset).
On the one hand, decreasing  $w_1=w_3$  ($w_2$ fixed), the toroidal static BS (SBS$_{-1}$+SBS$_{+1}$)
is approached for $w_1=w_3\simeq 0.933 \mu$ - sequence {\bf 1} in Fig.~\ref{fig1}. On the other hand, increasing  $w_1=w_3$ ($w_2$ fixed), 
the dipole DBS$_0$ is obtained for $w_1=w_3 \simeq 0.955 \mu$ - sequence {\bf 2}. These are static BSs sequences; thus $W=0$.

Similar interpolations occur between configurations with and without angular momentum, as in the  
transition SBS$_{-1}$+SBS$_{+1}$ $\longleftrightarrow$ SBS$_{+1}$ - sequence {\bf 3} in Fig.~\ref{fig1}. Starting from a static, toroidal BS  with $w_1=w_3=0.8 \mu$,  varying $w_3$ ($w_1$ fixed),
the amplitude of $\Phi^{(1)}$
vanishes for a critical value of
$w_3 \simeq 0.829 \mu$,
yielding the single-field SBS$_{+1}$.
All intermediate solutions with $w_1 \neq w_3$
 possess a nonvanishing angular momentum.
In all sequences, a  similar picture holds considering other frequencies.

The manifold of solutions  of the $\ell=1$ BSs family is as follows. Starting from an $\ell$-BS with a fixed frequency $w_1=w_2=w_3$, the line of static $(J=0)$ BSs is obtained keeping $w_1=w_3$ and  varying the ratio $y\equiv w_1/w_2$ ($=w_3/w_2$). Then $y\in [y_{\rm min},y_{\rm max}]$. $y$ varies the parity of the BSs; the boundary values are the parity even and odd solutions, respectively. Then, for each fixed $y$ one can vary $x\equiv w_3/w_1$, with $x\in [x_{\rm min},x_{\rm max}]$,  where the limits are $y$-dependent. $x$ varies $J$; for $x>1$ ($x<1$), $J$ is positive (negative)~\footnote{Since $J$ is carried by even parity fields, the $x$  range is maximal (minimal) for $y_{\rm min}$ ($y_{\rm max}$).  This explains the triangular shape in Fig.~\ref{fig1}.}. Finally,  varying the frequency of the starting $\ell$-BS yields a 3D manifold of solutions. Thus, we expect a $(2\ell+1)$D manifold of multi-frequency, multi-field BSs for a model with $(2\ell+1)$ complex scalar fields, including $\ell$-BSs as symmetry-enhanced solutions. 

{\bf {\em Dynamical (in)stability.}} 
We assess the dynamical stability of representative solutions in the $\ell=1$ BS family by resorting to fully non-linear dynamical evolutions of the corresponding Einstein--(multi-)Klein-Gordon system.  The infrastructure used in the numerical evolutions is the same as in~\cite{sanchis2019nonlinear}.

\begin{figure}
\centering
\includegraphics[width=0.195\linewidth]{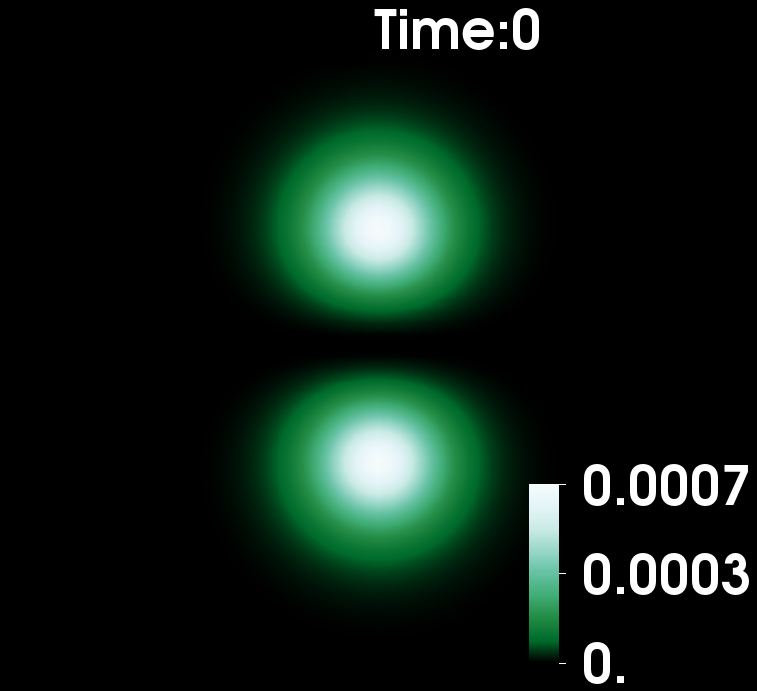}\hspace{-0.01\linewidth}
\includegraphics[width=0.195\linewidth]{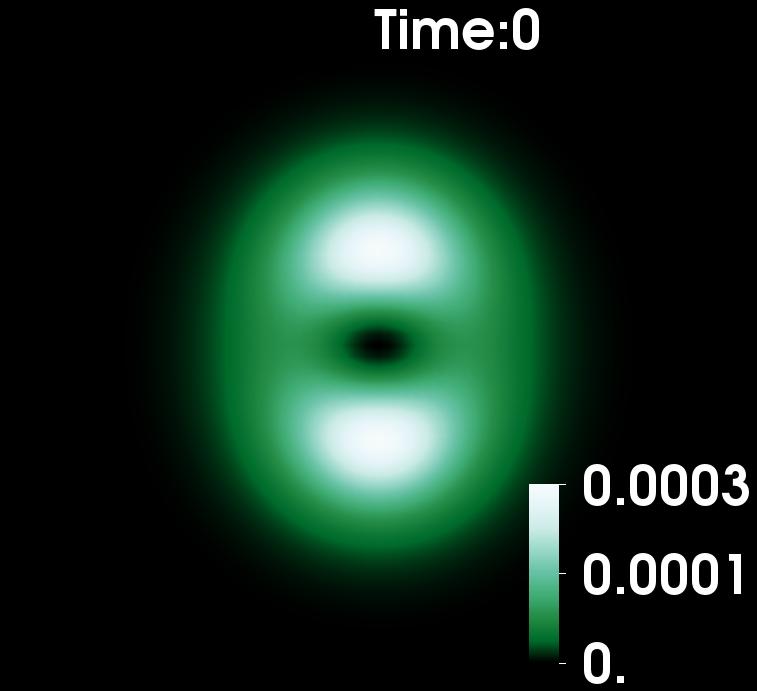}\hspace{-0.01\linewidth}
\includegraphics[width=0.195\linewidth]{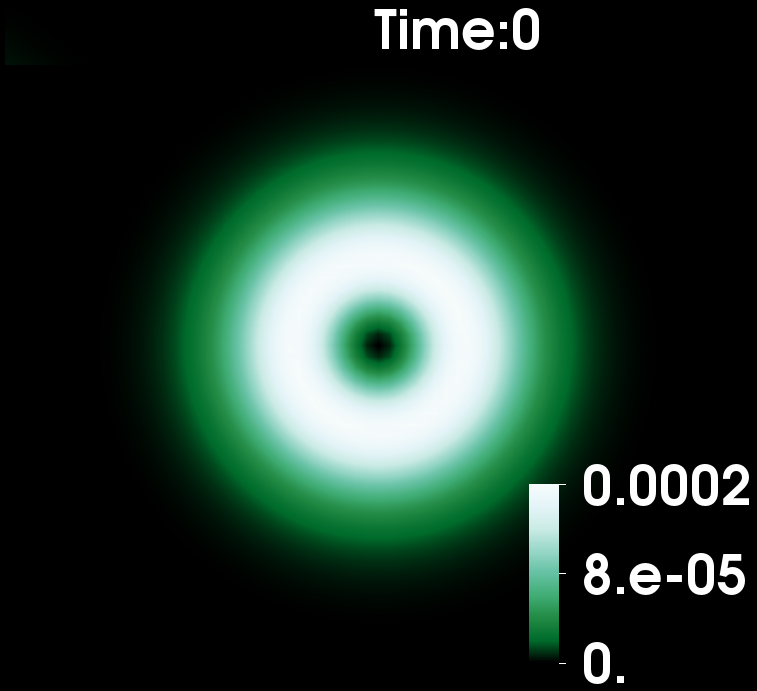}\hspace{-0.01\linewidth}
\includegraphics[width=0.195\linewidth]{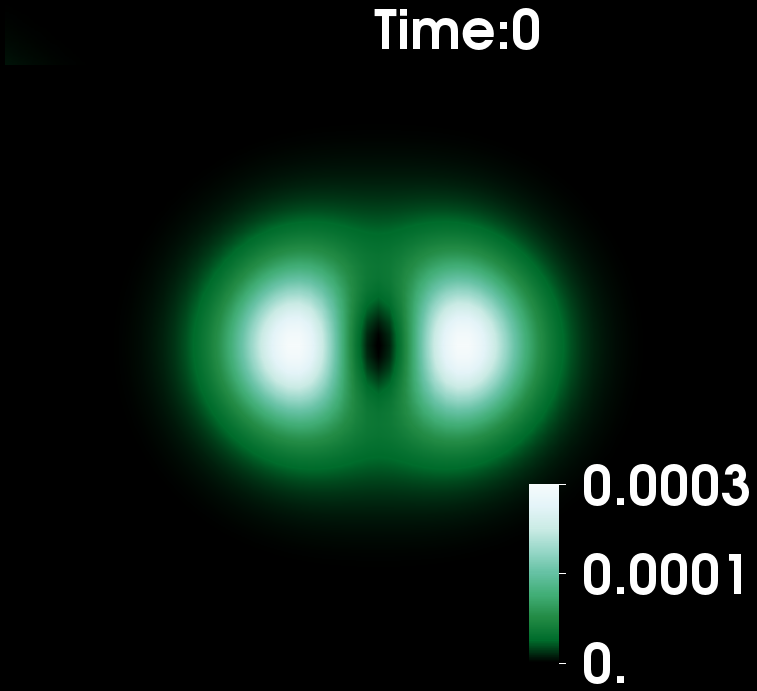}\hspace{-0.01\linewidth}
\includegraphics[width=0.195\linewidth]{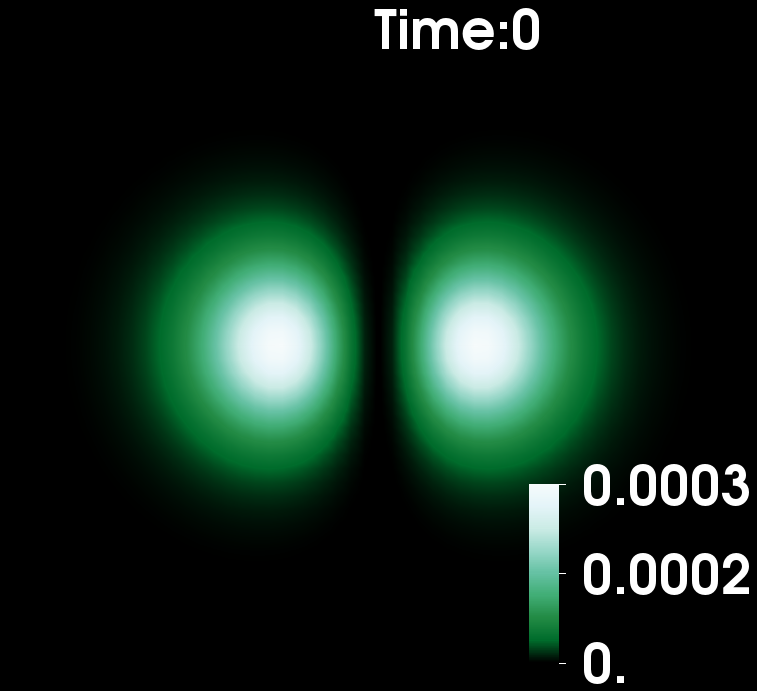}\hspace{-0.01\linewidth}\\
\includegraphics[width=0.195\linewidth]{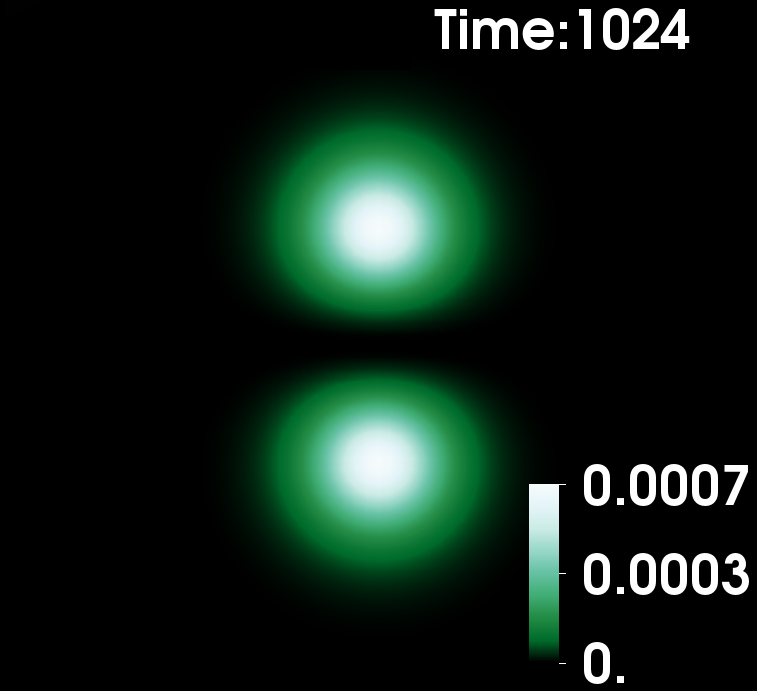}\hspace{-0.01\linewidth}
\includegraphics[width=0.195\linewidth]{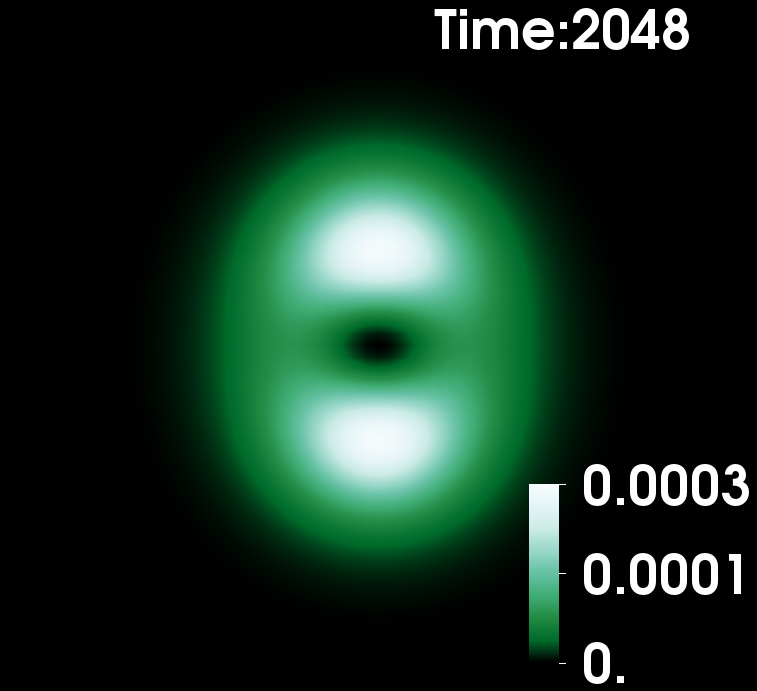}\hspace{-0.01\linewidth}
\includegraphics[width=0.195\linewidth]{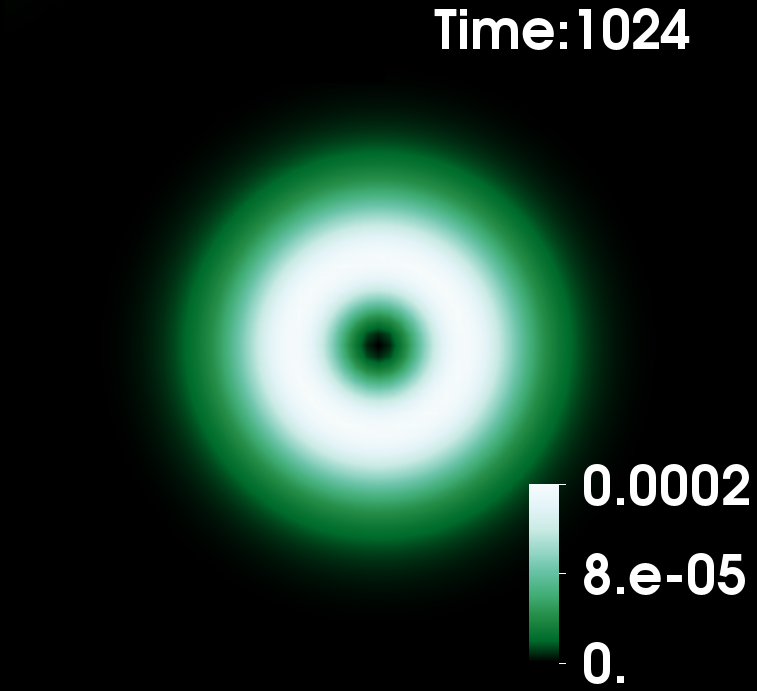}\hspace{-0.01\linewidth}
\includegraphics[width=0.195\linewidth]{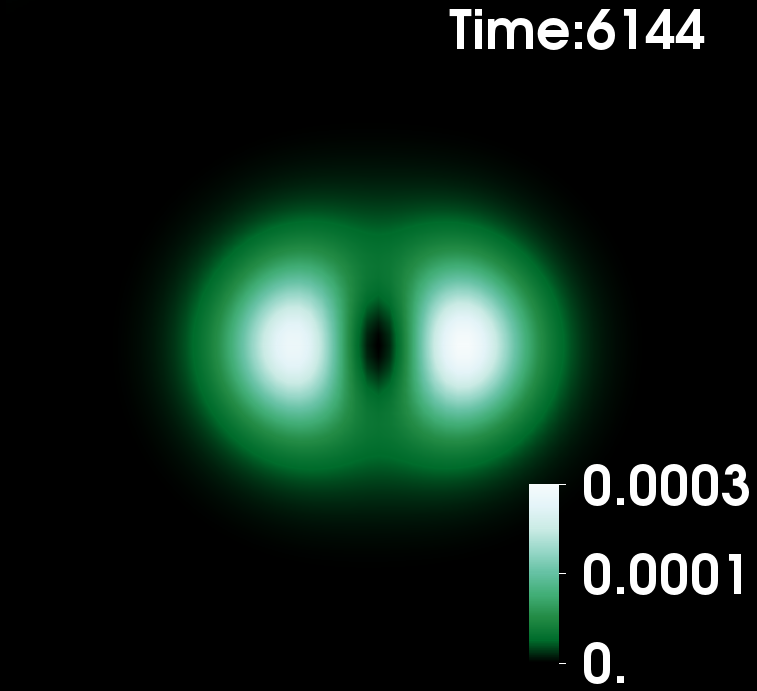}\hspace{-0.01\linewidth}
\includegraphics[width=0.195\linewidth]{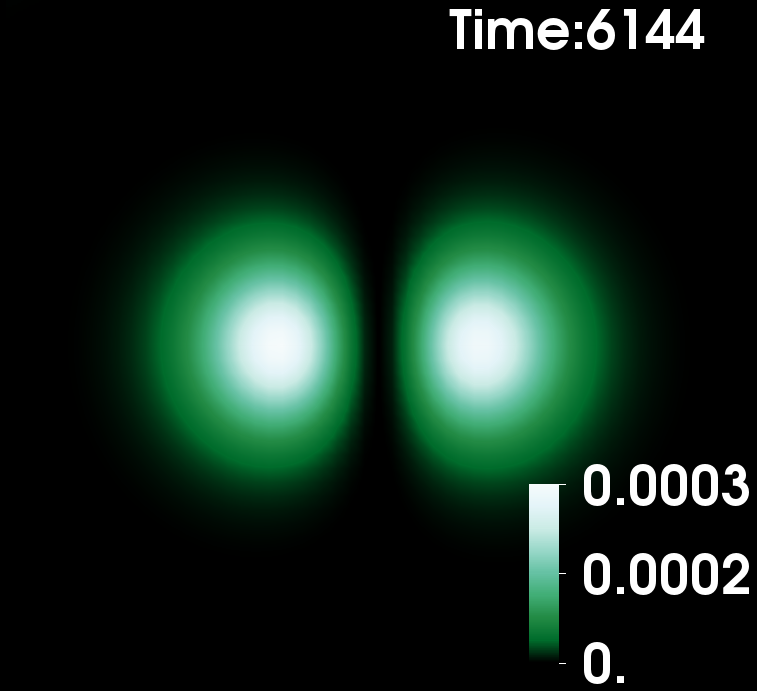}\hspace{-0.01\linewidth}\\
\includegraphics[width=0.195\linewidth]{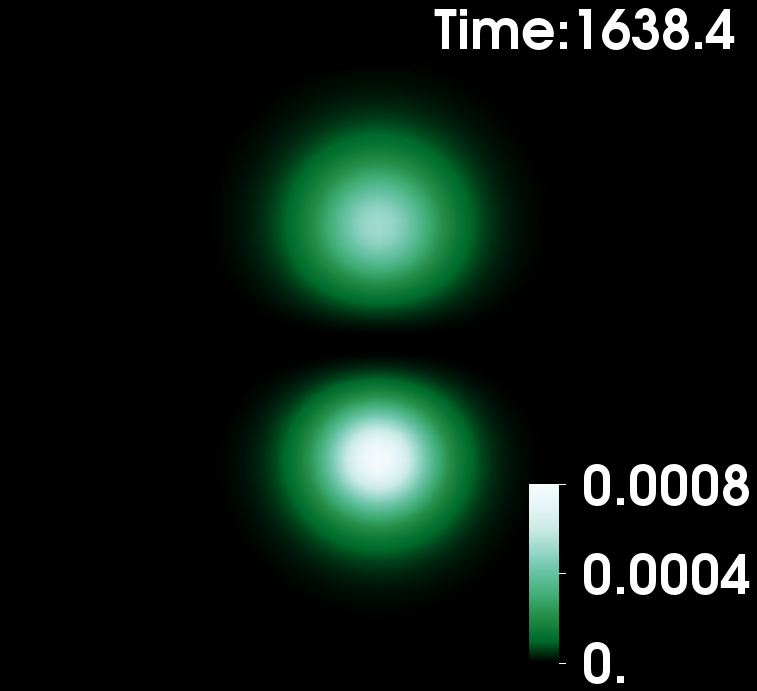}\hspace{-0.01\linewidth}
\includegraphics[width=0.195\linewidth]{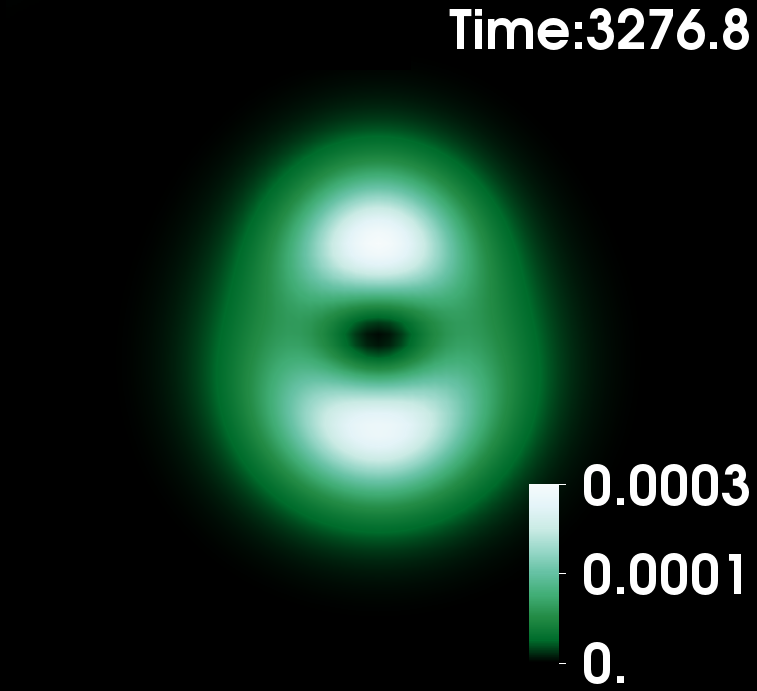}\hspace{-0.01\linewidth}
\includegraphics[width=0.195\linewidth]{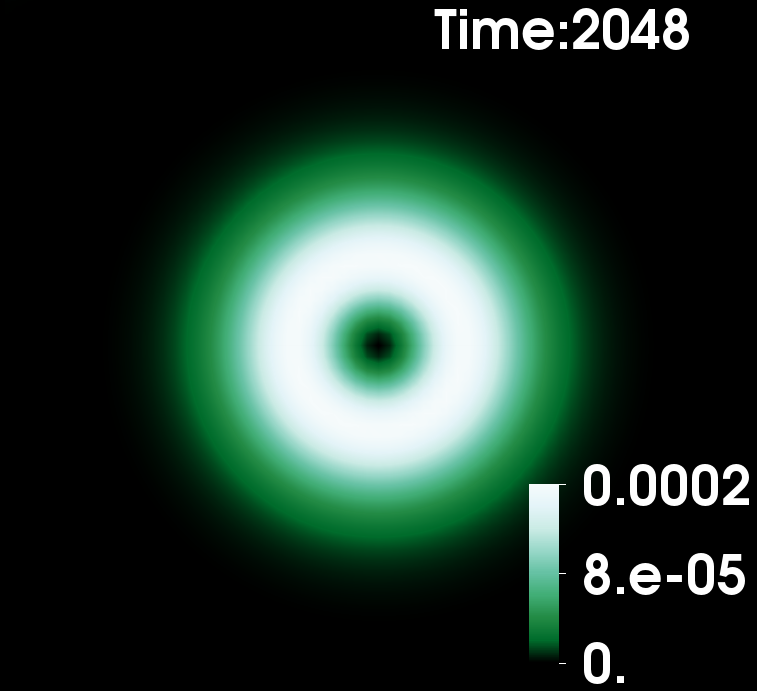}\hspace{-0.01\linewidth}
\includegraphics[width=0.195\linewidth]{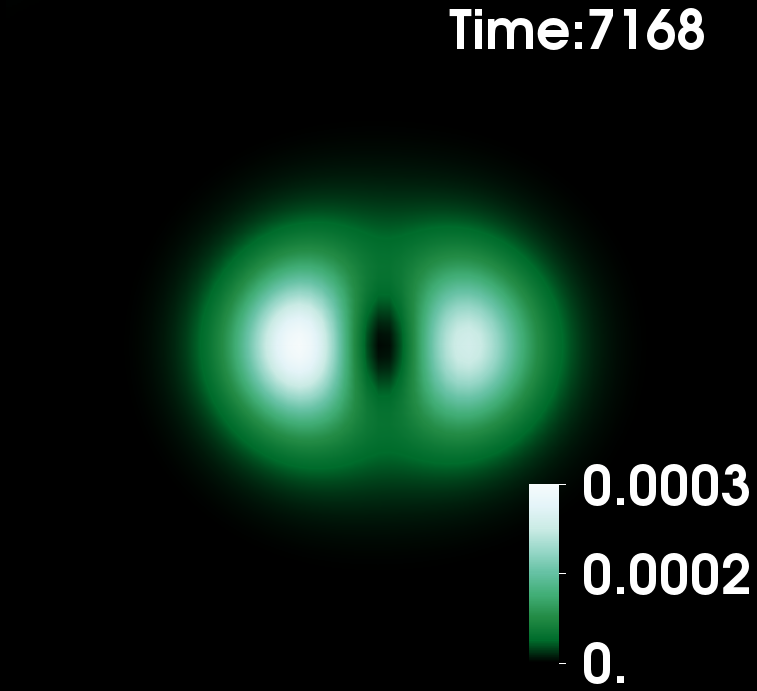}\hspace{-0.01\linewidth}
\includegraphics[width=0.195\linewidth]{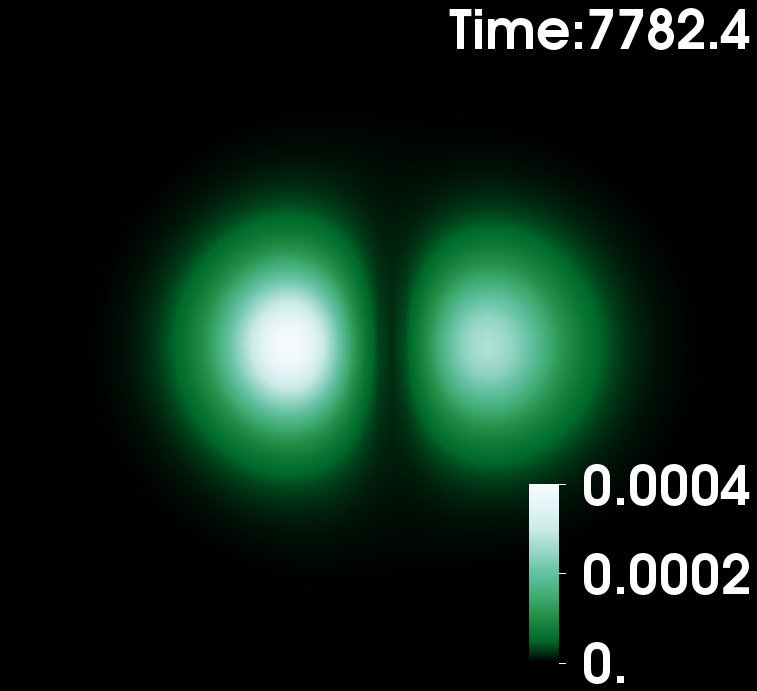}\hspace{-0.01\linewidth}\\
\includegraphics[width=0.195\linewidth]{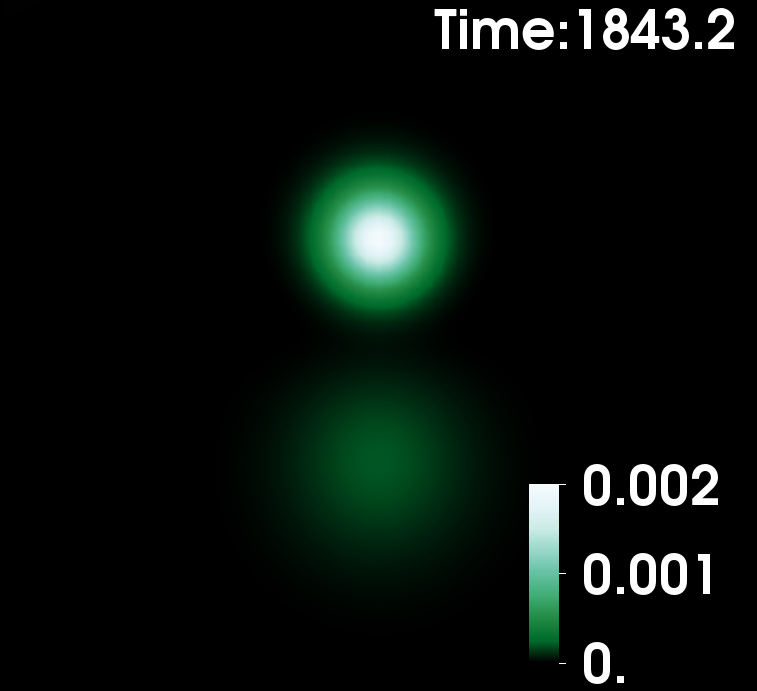}\hspace{-0.01\linewidth}
\includegraphics[width=0.195\linewidth]{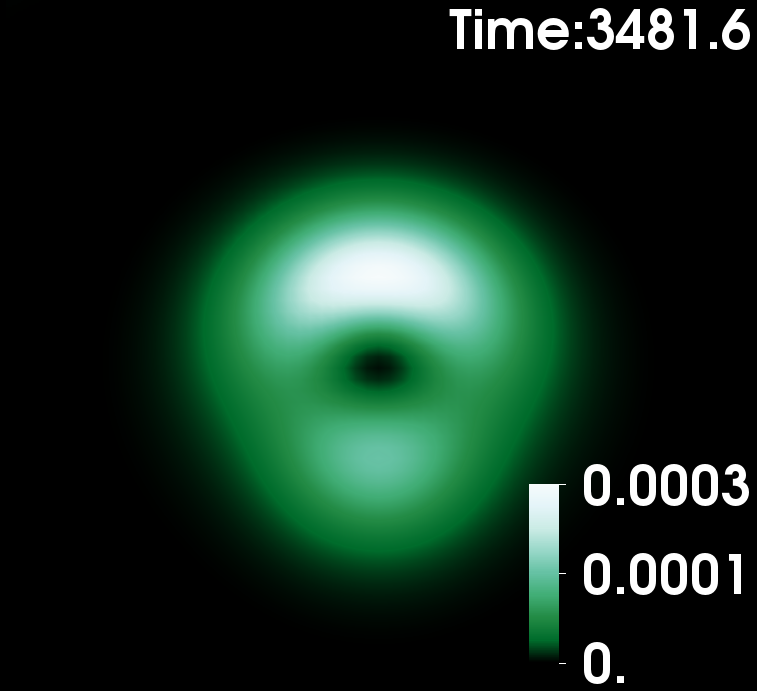}\hspace{-0.01\linewidth}
\includegraphics[width=0.195\linewidth]{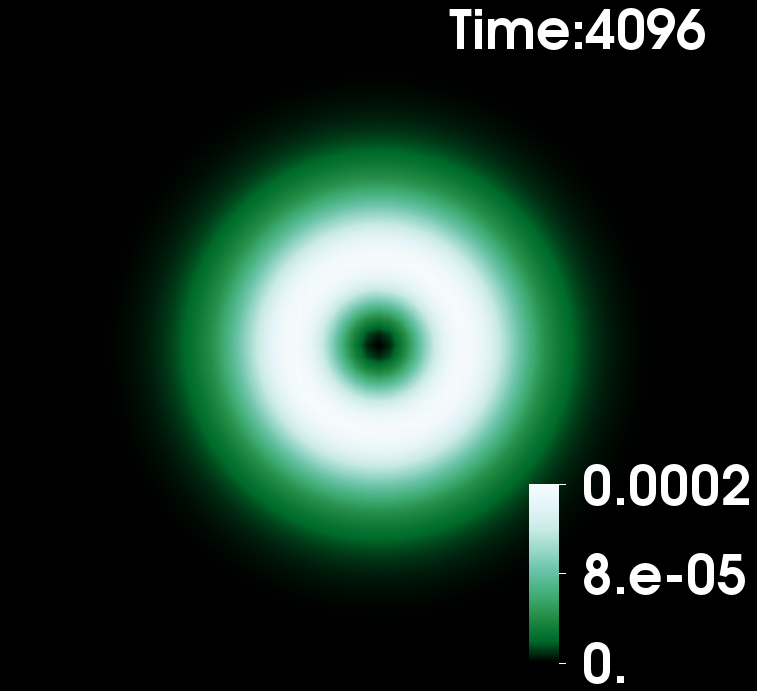}\hspace{-0.01\linewidth}
\includegraphics[width=0.195\linewidth]{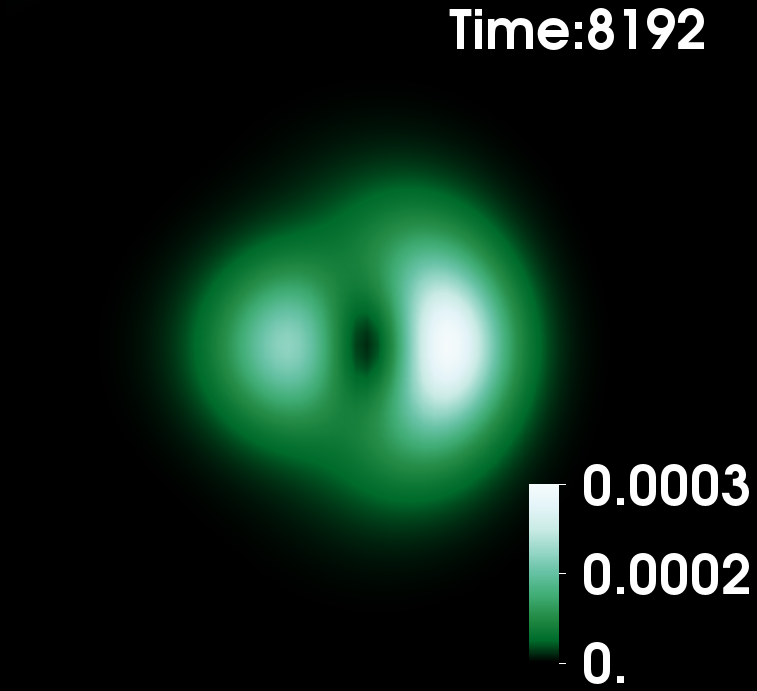}\hspace{-0.01\linewidth}
\includegraphics[width=0.195\linewidth]{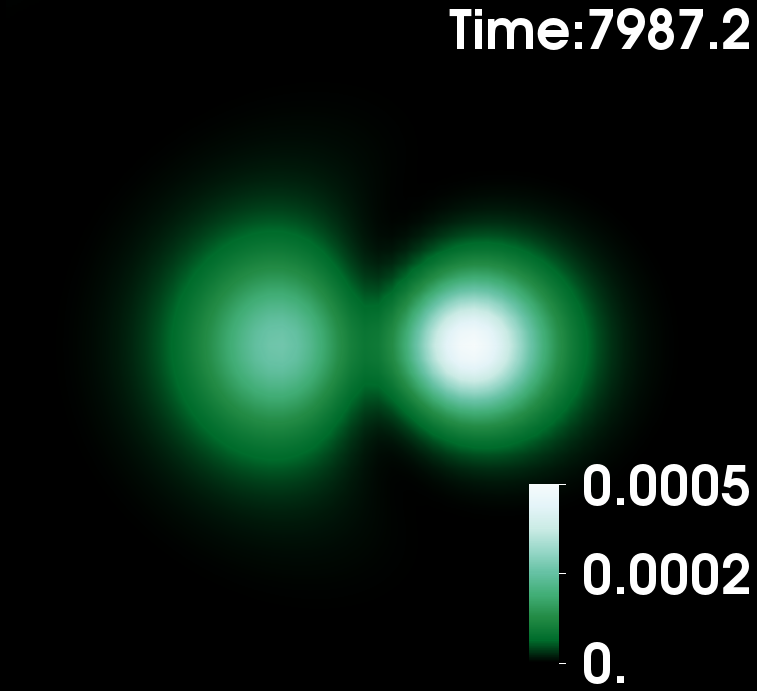}\hspace{-0.01\linewidth}\\
\includegraphics[width=0.195\linewidth]{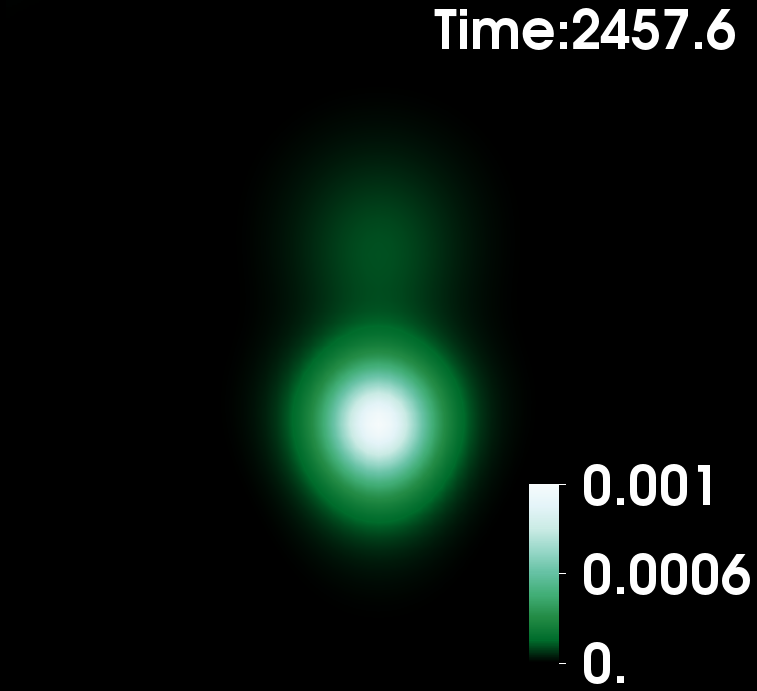}\hspace{-0.01\linewidth}
\includegraphics[width=0.195\linewidth]{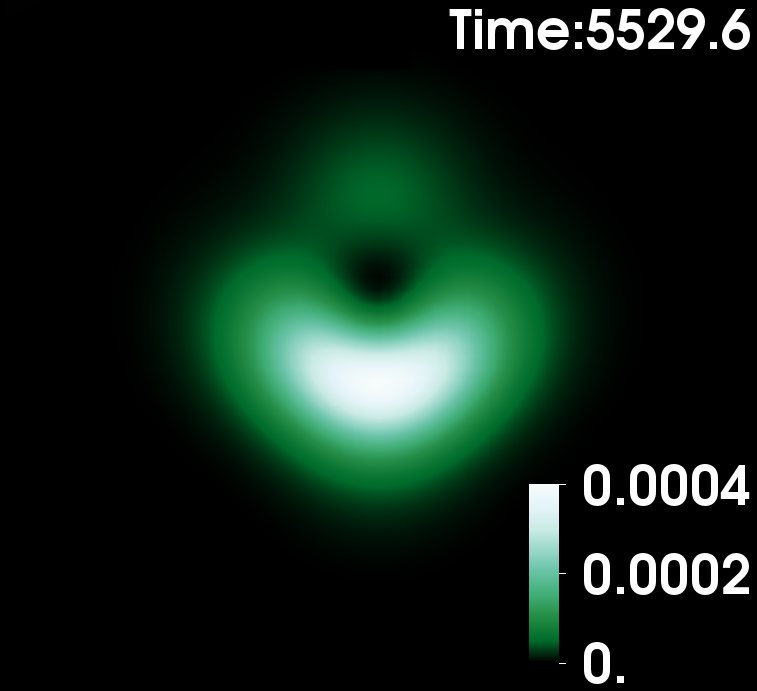}\hspace{-0.01\linewidth}
\includegraphics[width=0.195\linewidth]{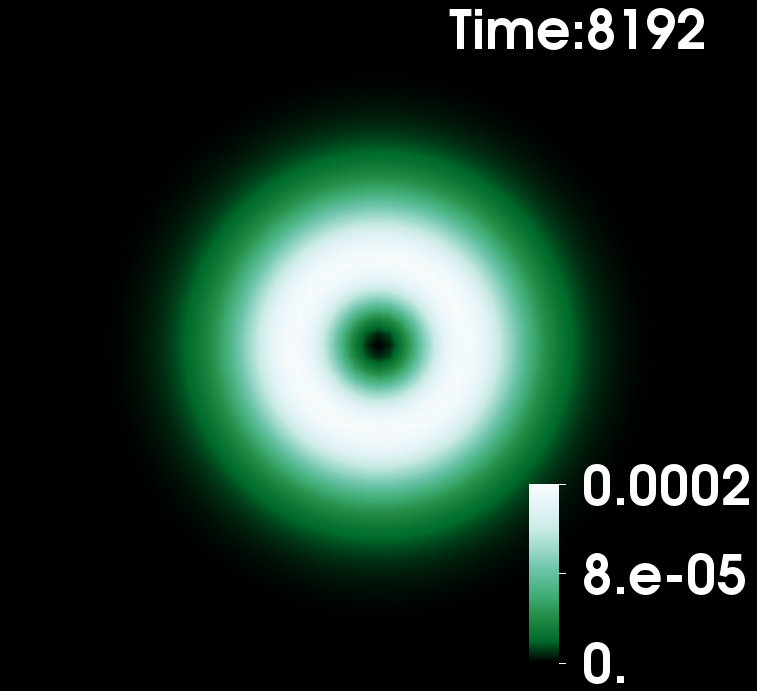}\hspace{-0.01\linewidth}
\includegraphics[width=0.195\linewidth]{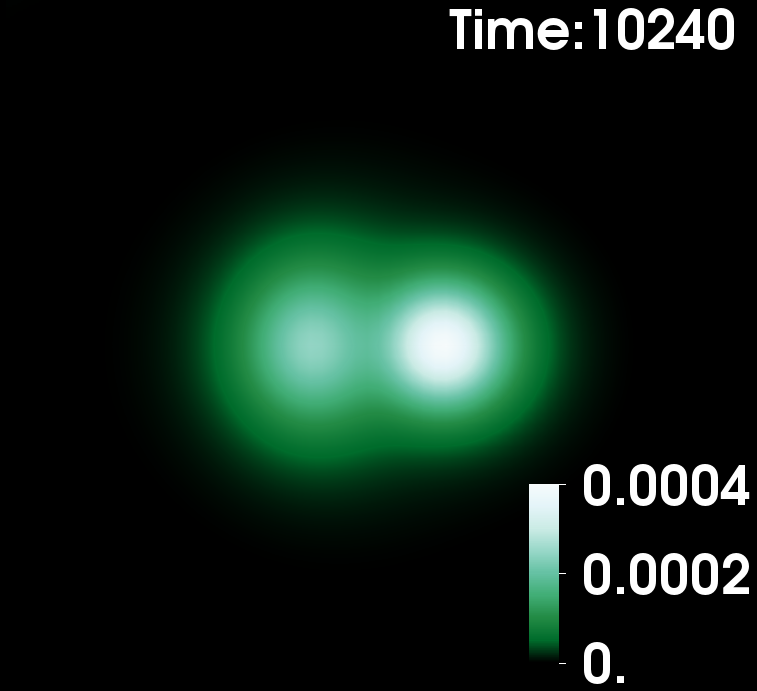}\hspace{-0.01\linewidth}
\includegraphics[width=0.195\linewidth]{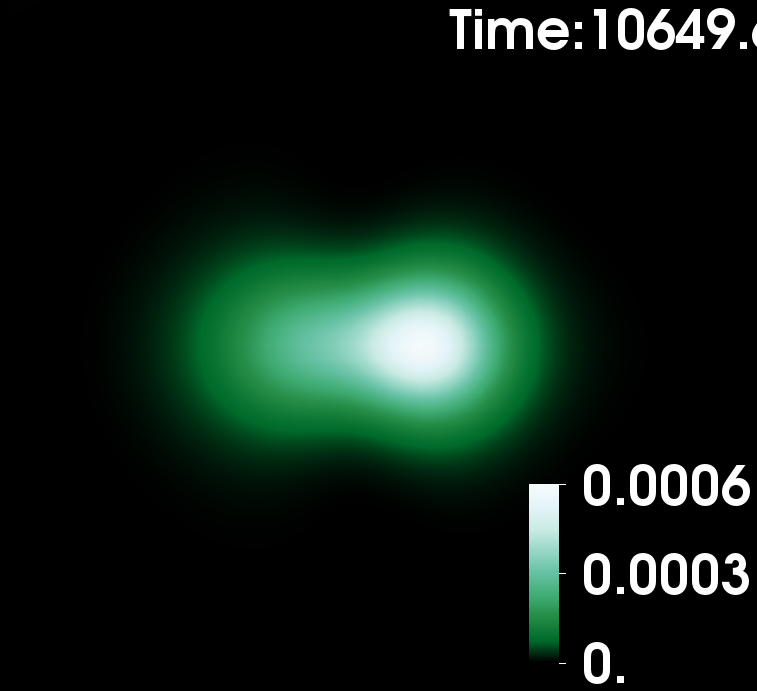}\hspace{-0.01\linewidth}\\
\includegraphics[width=0.195\linewidth]{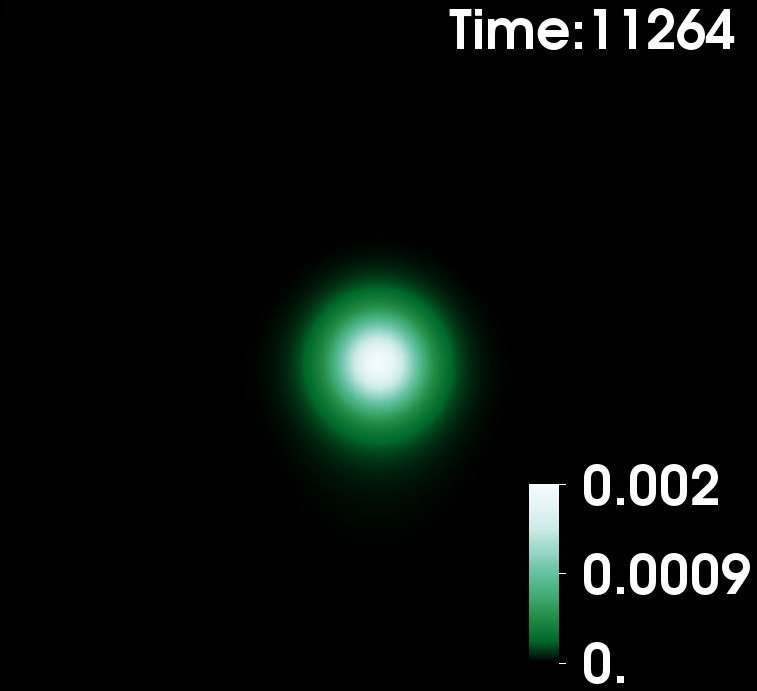}\hspace{-0.01\linewidth}
\includegraphics[width=0.195\linewidth]{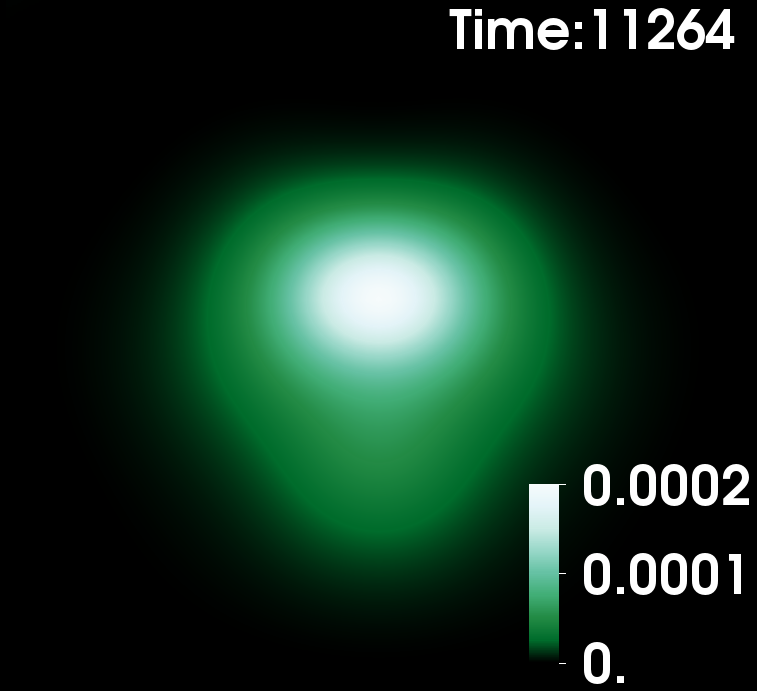}\hspace{-0.01\linewidth}
\includegraphics[width=0.195\linewidth]{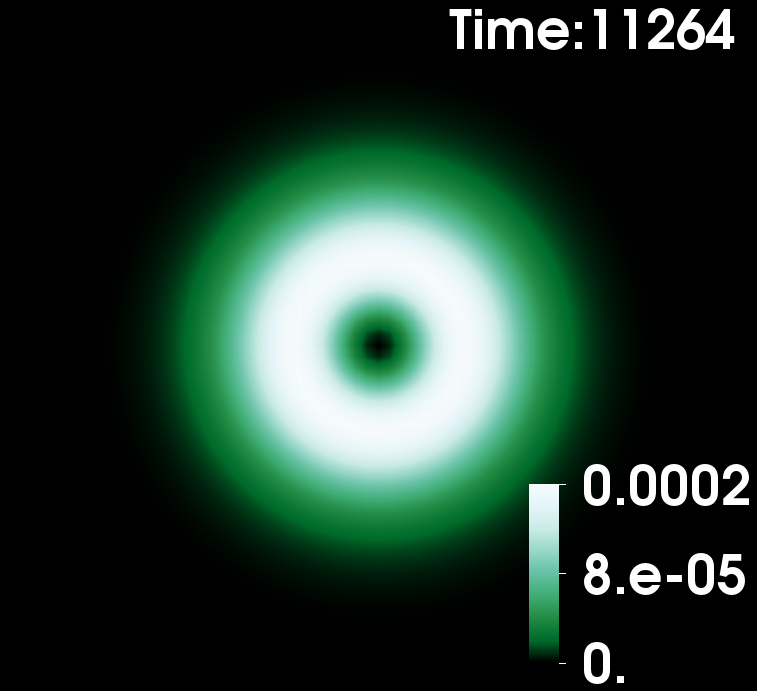}\hspace{-0.01\linewidth}
\includegraphics[width=0.195\linewidth]{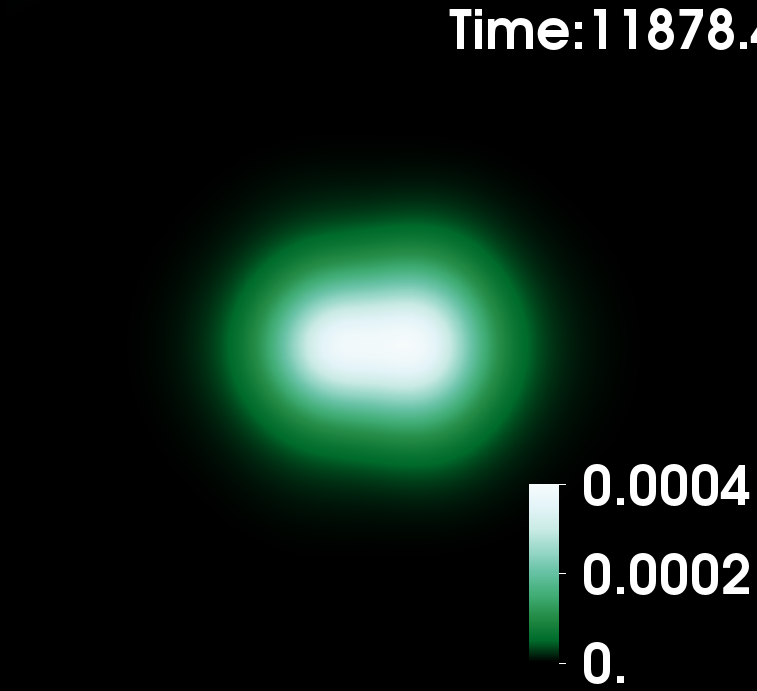}\hspace{-0.01\linewidth}
\includegraphics[width=0.195\linewidth]{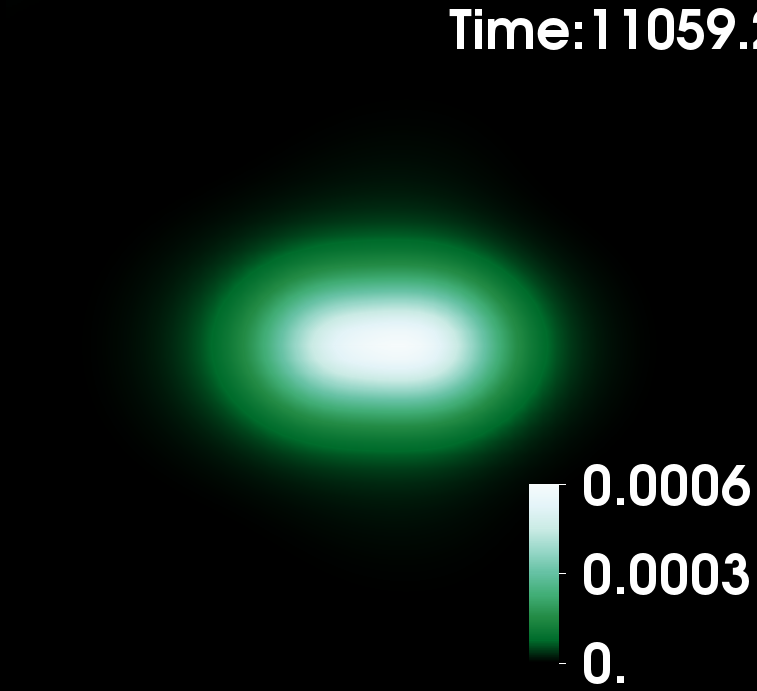}\hspace{-0.01\linewidth}\\
\caption{Time evolutions  of the energy density for five static BSs in  the $\ell=1$ family.   Each column shows six sequential snapshots of the $xz$ ($y$=0) plane (time running from top to bottom). 
From left to right the values of $w_1=w_3$ of the five models are: 0.955 (DBS$_0$), 0.945, 0.940 ($\ell-{\rm BS}$), 0.935, 0.933 (${\rm SBS}_{-1}+{\rm SBS}_{+1}$). }
\label{fig3}
\end{figure}

Fig.~\ref{fig3} exhibits the results for a sequence of static solutions ($i.e.$ along sequences {\bf 1} and {\bf 2} in~Fig.~\ref{fig1}, including the dipole, the $\ell$-BS and the toroidal static BS). We find that all solutions (except the $\ell-$BS) are dynamically unstable, decaying to a multi-field BS in which all fields have $\ell=m=0$. Including $J$ does not improve dynamical stability. The  SBS$_{\pm 1}$ are unstable against a non-axisymmetric instability~\cite{sanchis2019nonlinear} and all hybrid cases we have studied (such as SBS$_{\pm 1}+$DBS$_0$) also decay to the fundamental $\ell=0$ BSs.

{\bf {\em Hybrid $\ell$-BSs and a stabilization mechanism.}} 
Instead of focusing on a single $\ell$-BSs family we now allow superpositions of such families with different $\ell$'s. The most elementary example is to add $\ell=0$ BSs to the $\ell=1$ family. Thus we add a fourth complex scalar field $\Phi^{(0)}$,  obeying~\eqref{action} and~\eqref{ansatz-general} with  $j=0$ (hence, now $j=0,1,2,3$) and $m_j=0$. Its boundary conditions are $
\partial_r \phi_0\big|_{r=0}= \phi_0\big|_{r=\infty}= \partial_\theta \phi_0\big|_{\theta=0,\pi}=0$, besides being parity even.
 Keeping only $\Phi^{(0)}$, the basic solution is the single-field, fundamental, monopole BS (MBS$_0$). We now show  that adding MBS$_0$  can quench the instabilities observed in the $\ell=1$ family.

\begin{figure}[h!]
\centering
\includegraphics[width=0.24\linewidth]{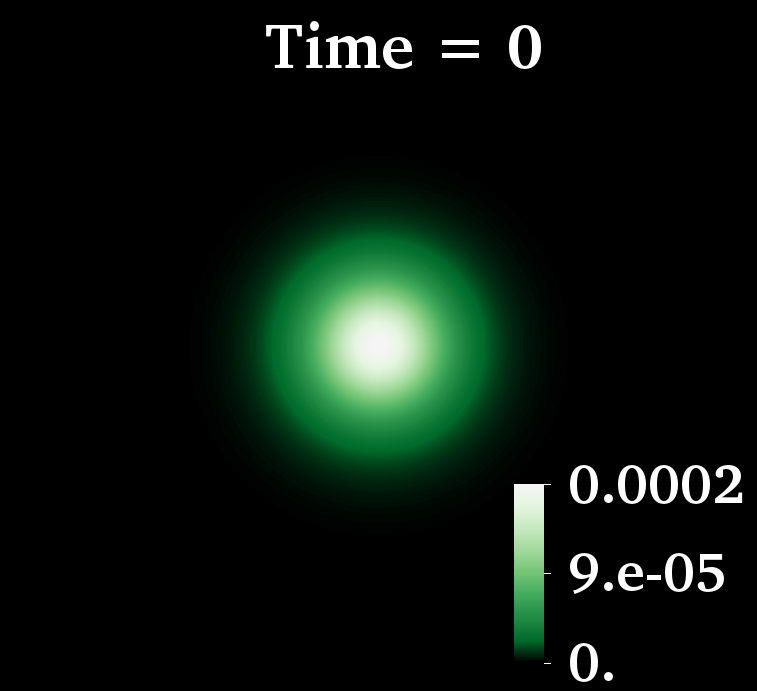}\hspace{-0.01\linewidth}
\includegraphics[width=0.24\linewidth]{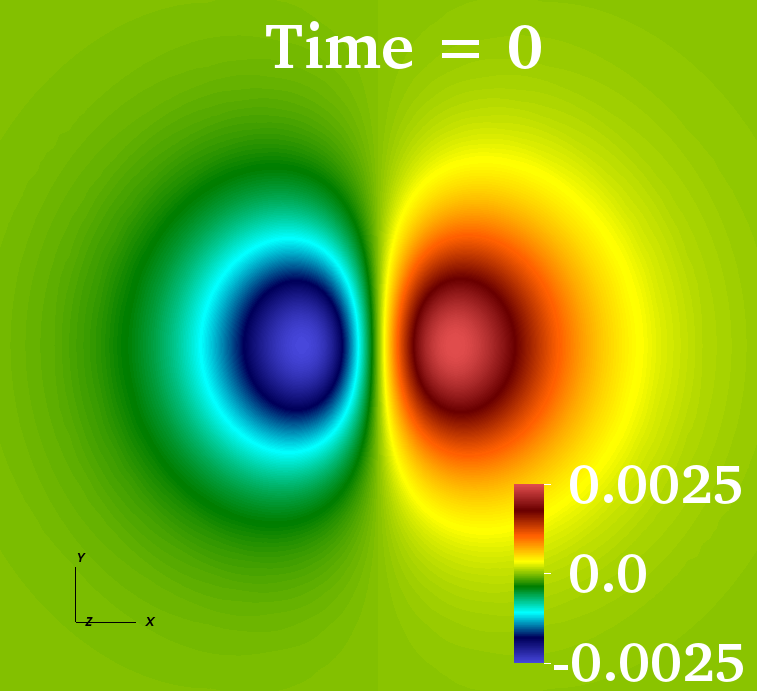}\hspace{-0.01\linewidth}
\includegraphics[width=0.24\linewidth]{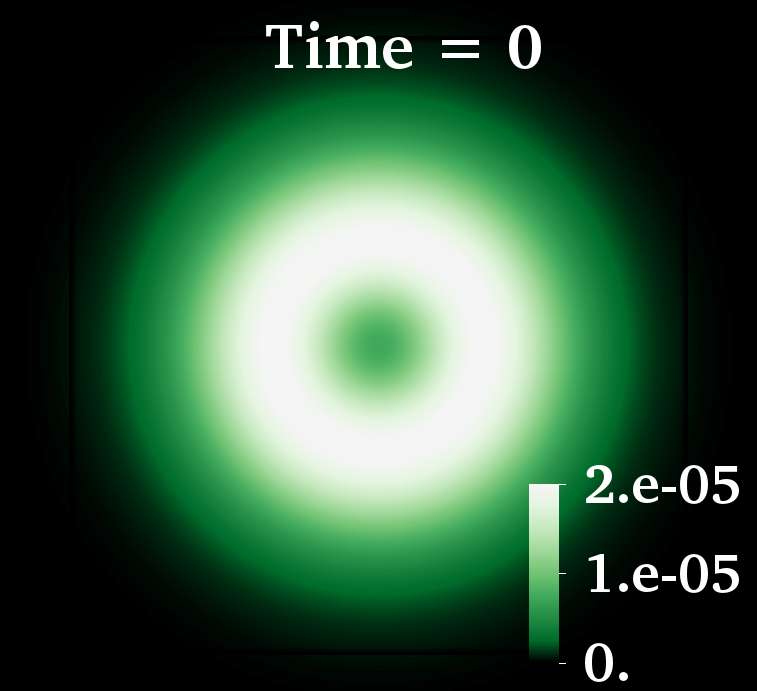}\hspace{-0.01\linewidth}
\includegraphics[width=0.24\linewidth]{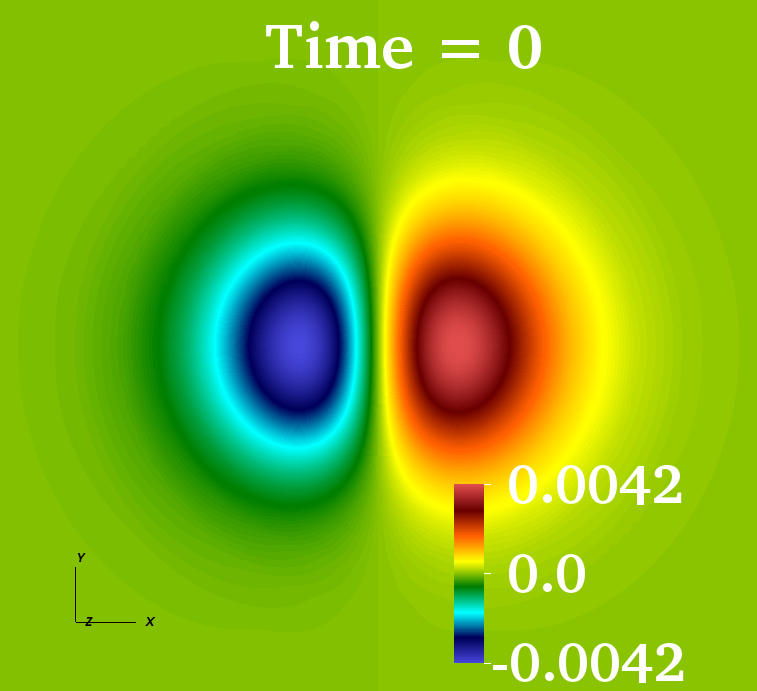}\hspace{-0.01\linewidth}\\
\includegraphics[width=0.24\linewidth]{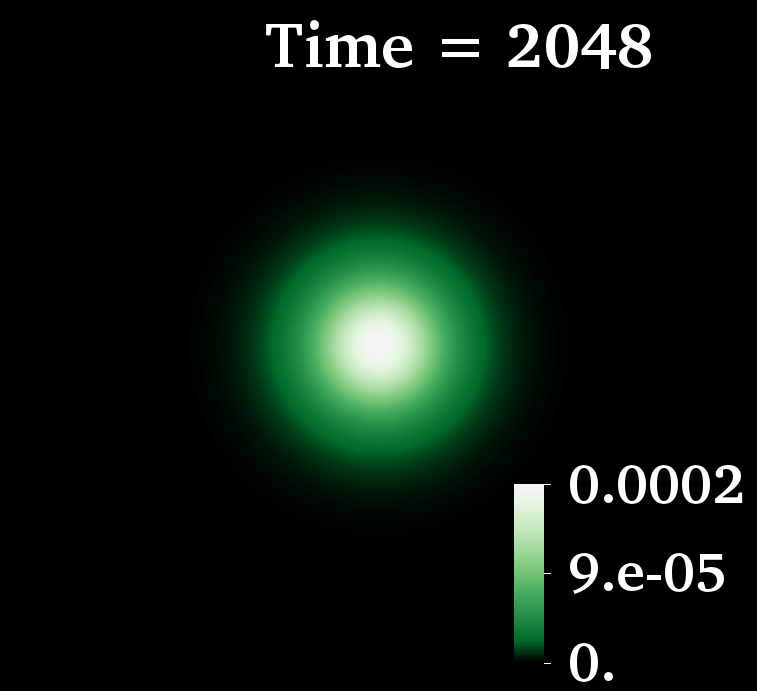}\hspace{-0.01\linewidth}
\includegraphics[width=0.24\linewidth]{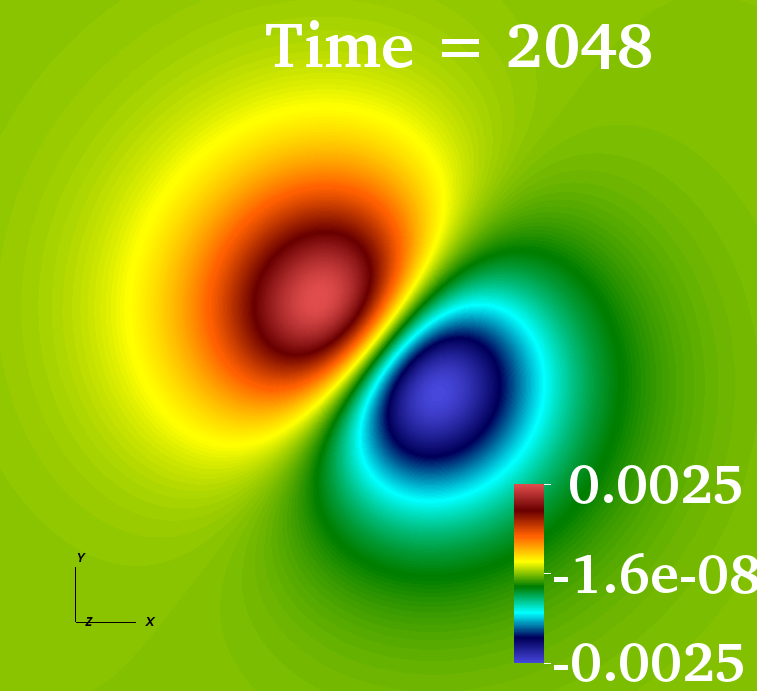}\hspace{-0.01\linewidth}
\includegraphics[width=0.24\linewidth]{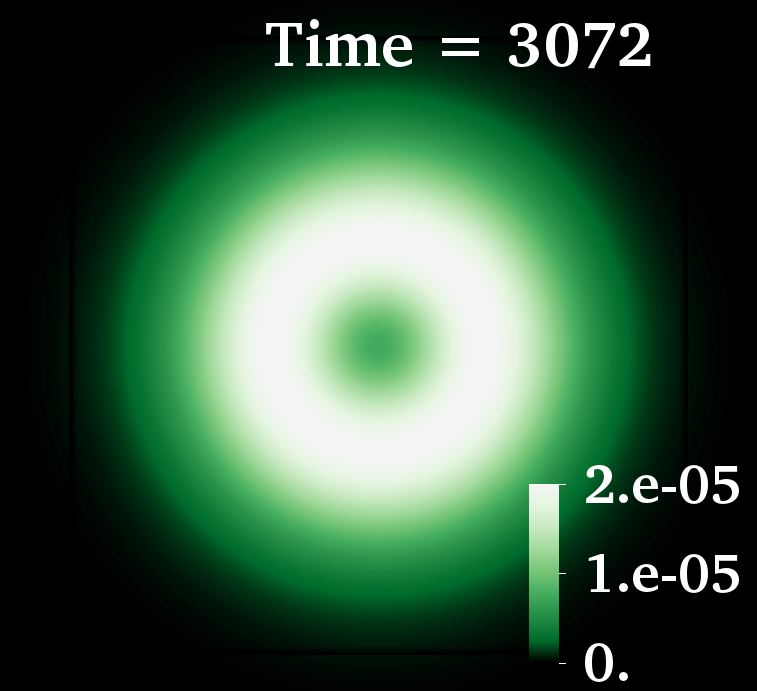}\hspace{-0.01\linewidth}
\includegraphics[width=0.24\linewidth]{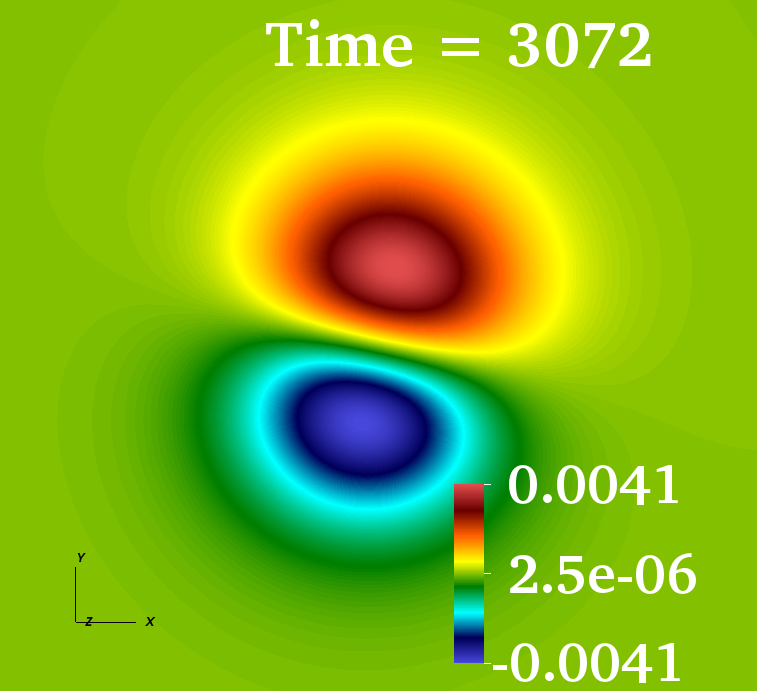}\hspace{-0.01\linewidth}\\
\includegraphics[width=0.24\linewidth]{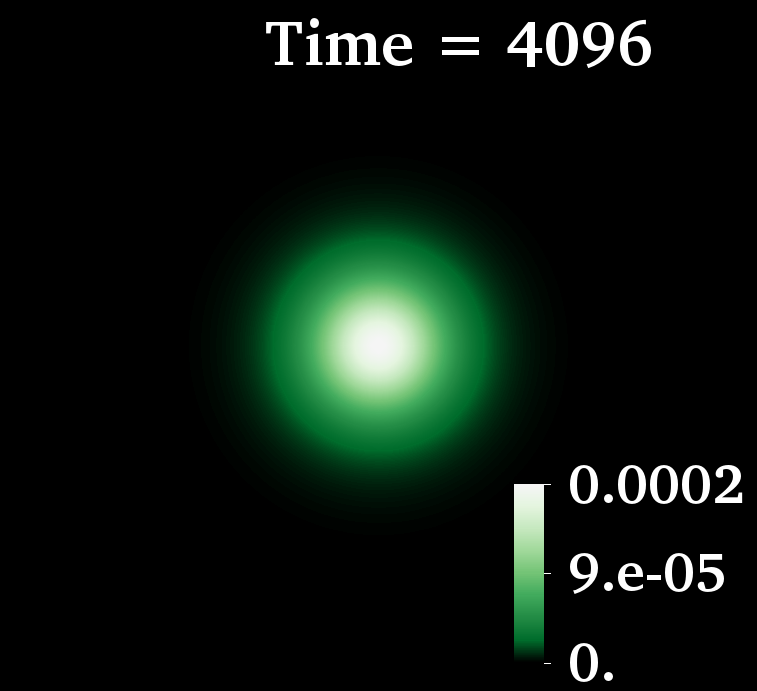}\hspace{-0.01\linewidth}
\includegraphics[width=0.24\linewidth]{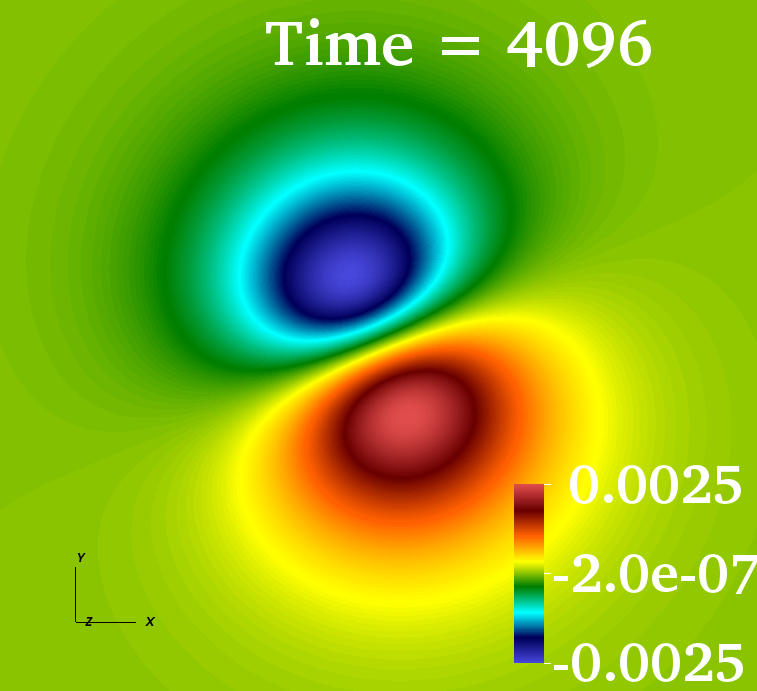}\hspace{-0.01\linewidth}
\includegraphics[width=0.24\linewidth]{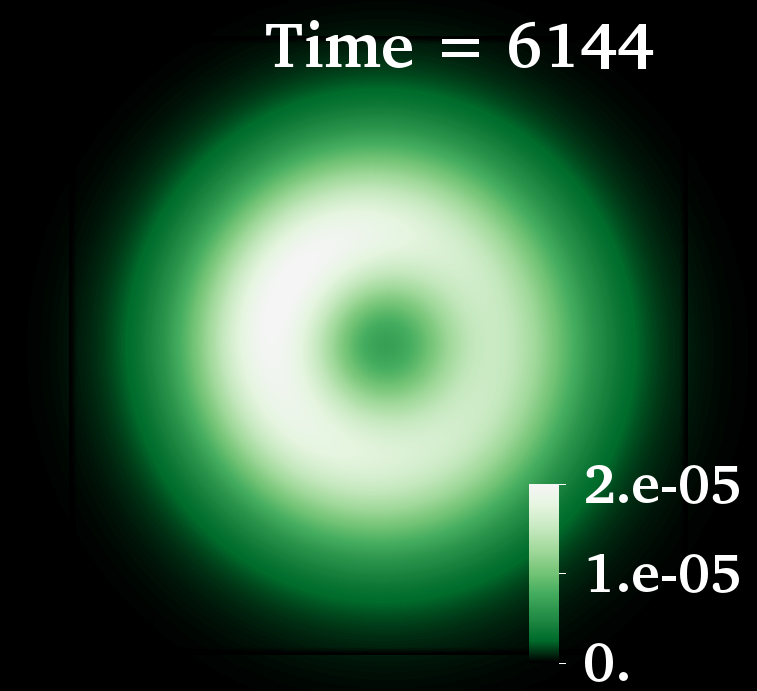}\hspace{-0.01\linewidth}
\includegraphics[width=0.24\linewidth]{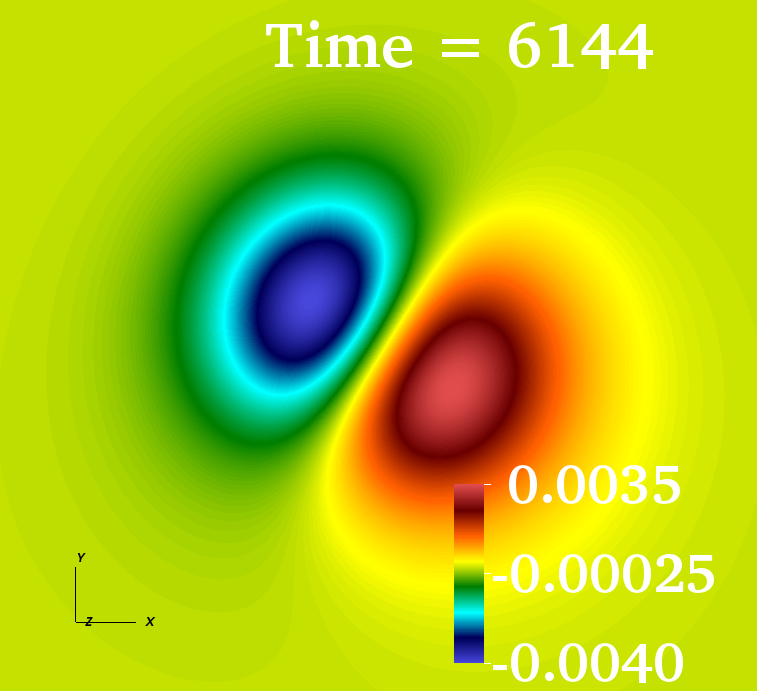}\hspace{-0.01\linewidth}\\
\includegraphics[width=0.24\linewidth]{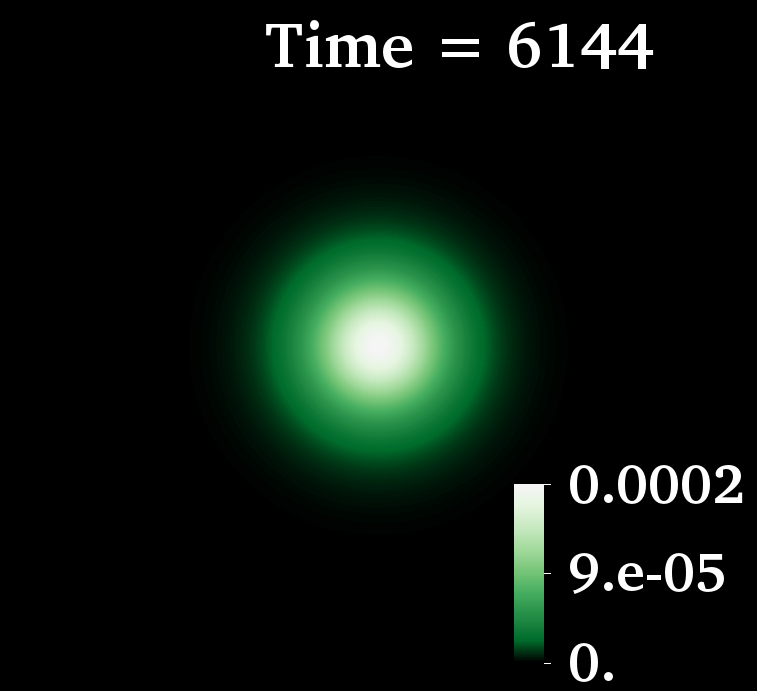}\hspace{-0.01\linewidth}
\includegraphics[width=0.24\linewidth]{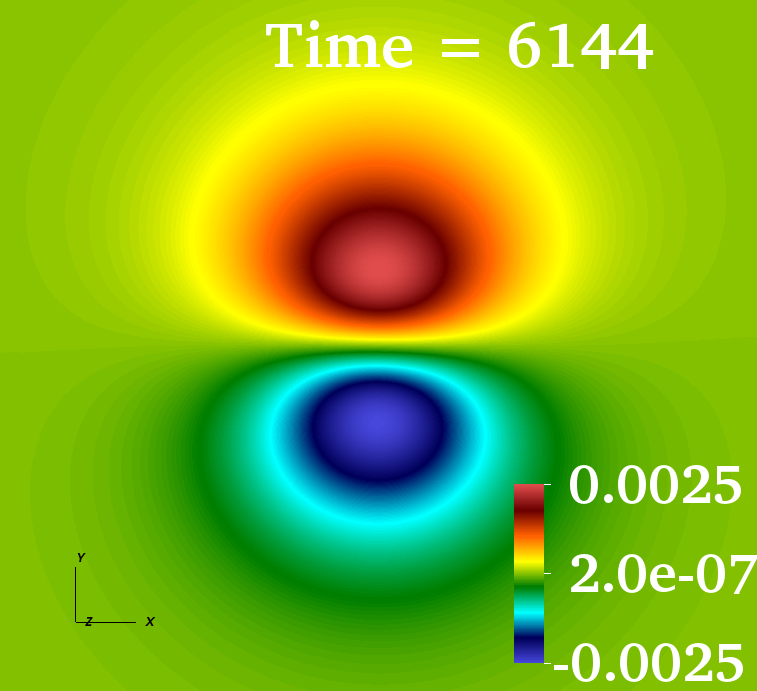}\hspace{-0.01\linewidth}
\includegraphics[width=0.24\linewidth]{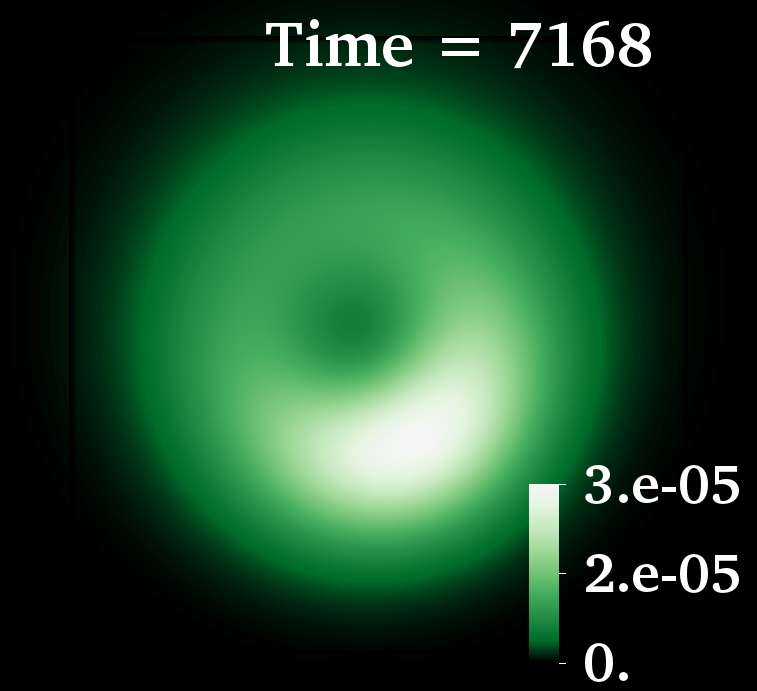}\hspace{-0.01\linewidth}
\includegraphics[width=0.24\linewidth]{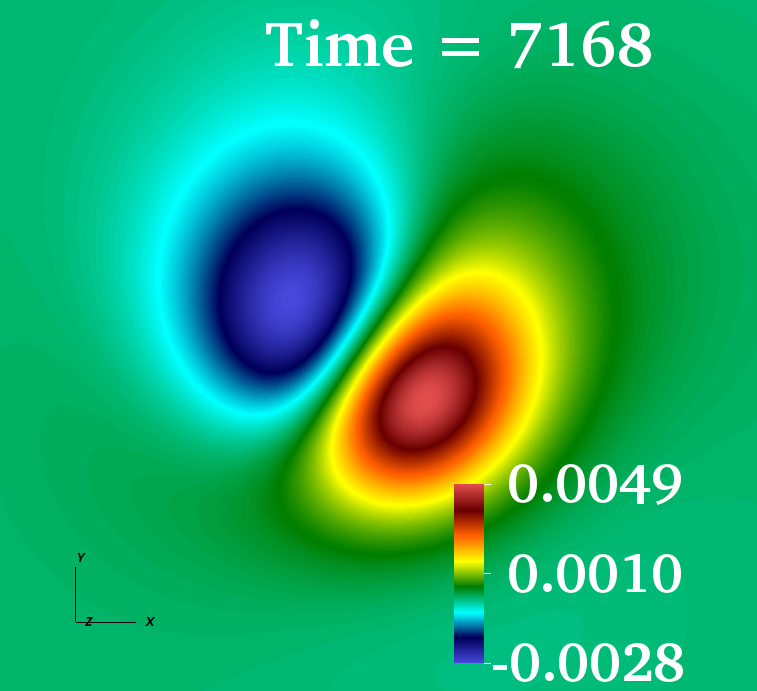}\hspace{-0.01\linewidth}\\
\includegraphics[width=0.24\linewidth]{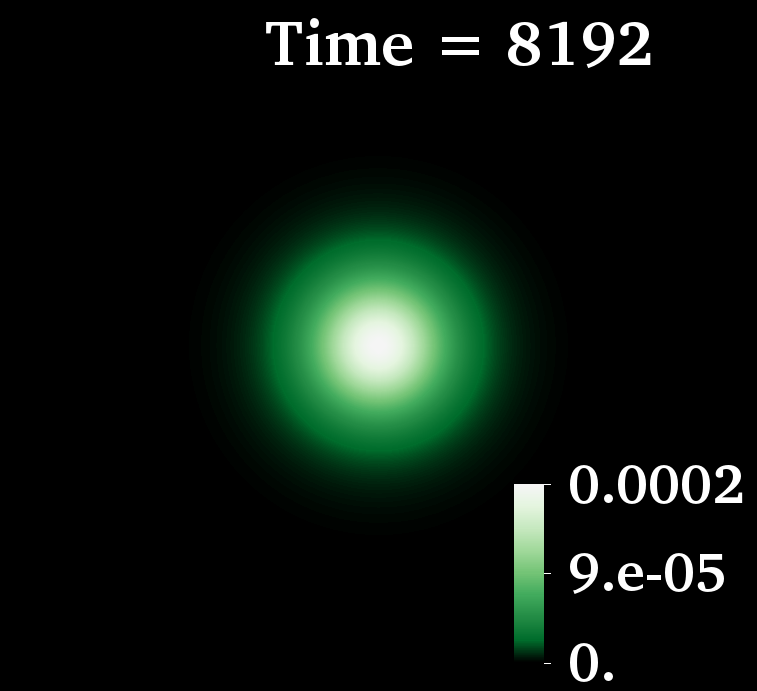}\hspace{-0.01\linewidth}
\includegraphics[width=0.24\linewidth]{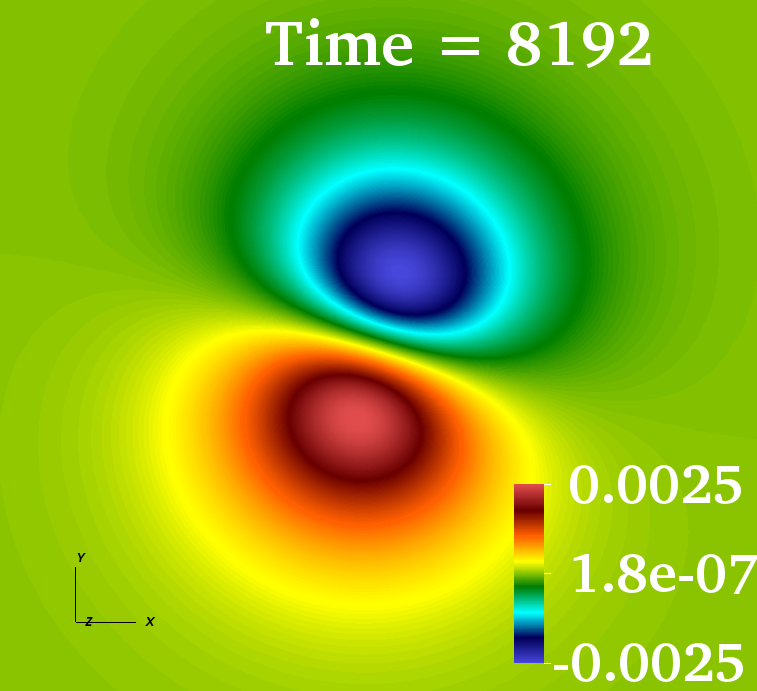}\hspace{-0.01\linewidth}
\includegraphics[width=0.24\linewidth]{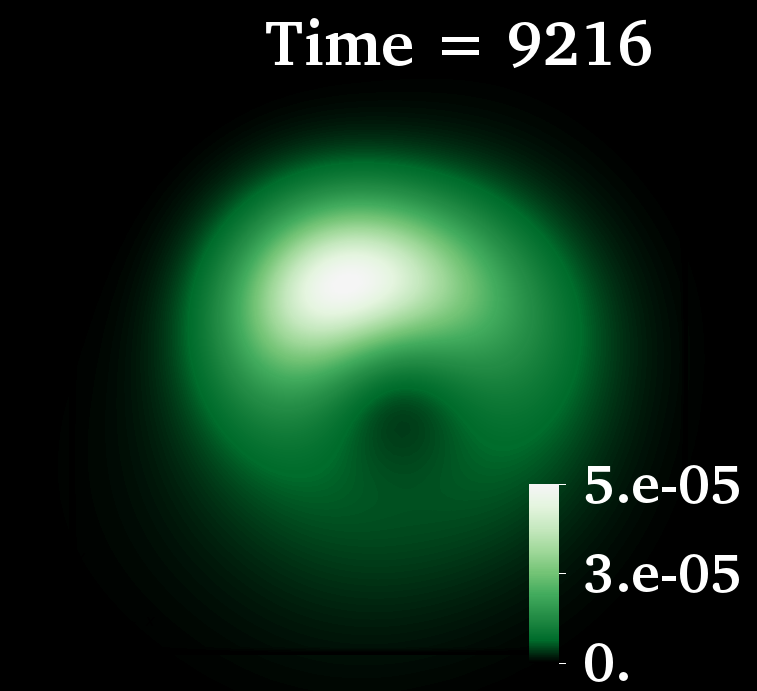}\hspace{-0.01\linewidth}
\includegraphics[width=0.24\linewidth]{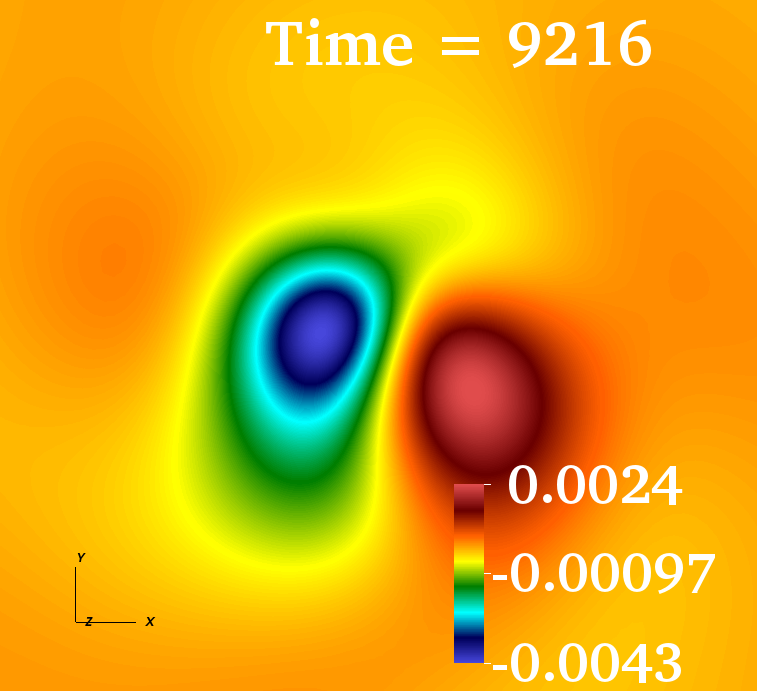}\hspace{-0.01\linewidth}\\
\includegraphics[width=0.24\linewidth]{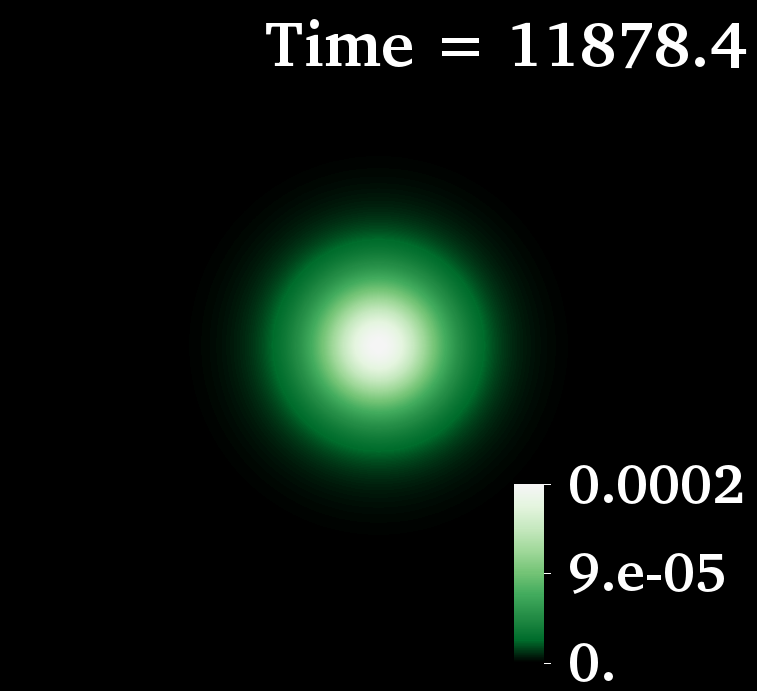}\hspace{-0.01\linewidth}
\includegraphics[width=0.24\linewidth]{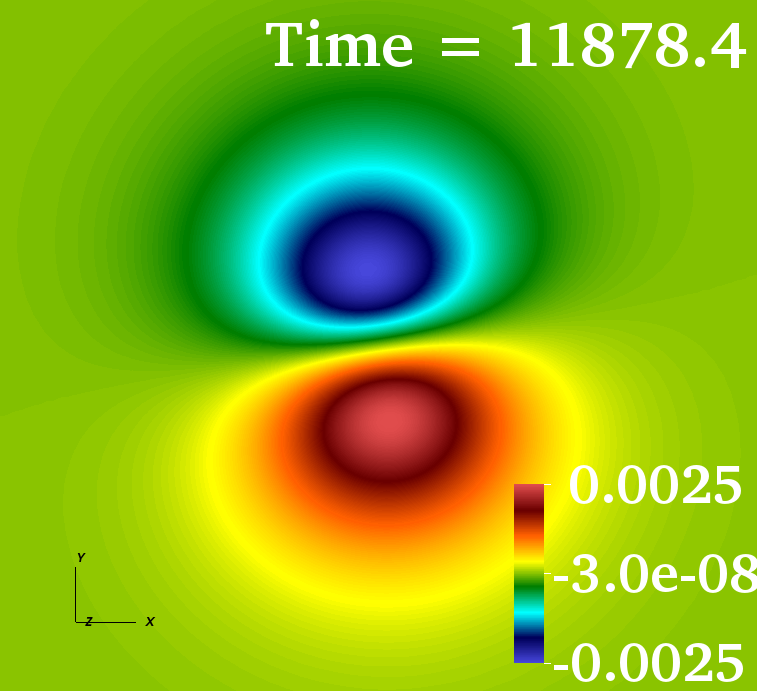}\hspace{-0.01\linewidth}
\includegraphics[width=0.24\linewidth]{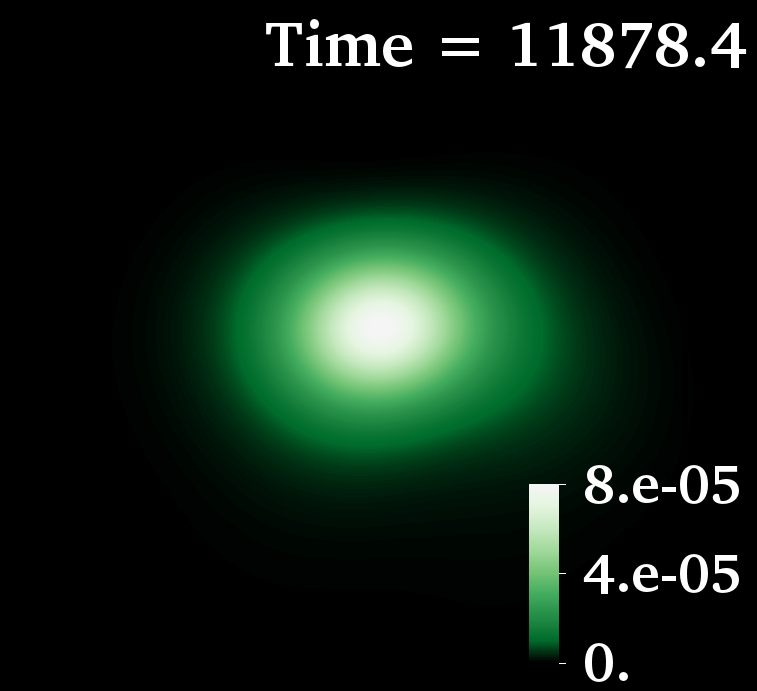}\hspace{-0.01\linewidth}
\includegraphics[width=0.24\linewidth]{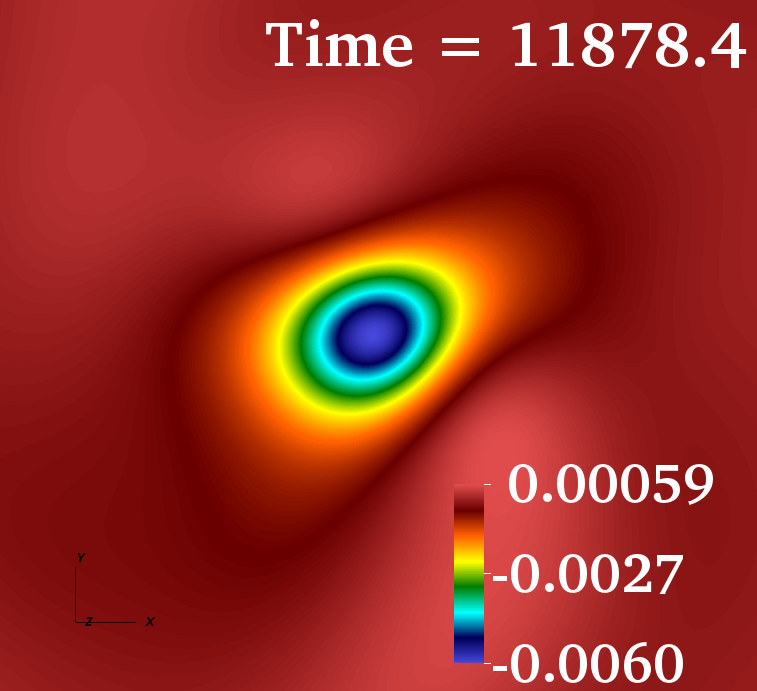}\hspace{-0.01\linewidth}\\
\caption{Time evolution (top to bottom) of two Saturns:  (left and middle left) $\omega_1/\mu=0.967$; (middle  right and right)  $\omega_1/\mu=0.974$. For  each case the energy density and the real part of $\Phi^{(3)}$ on the $xy$ plane are shown. The first (second) Saturn is close to  MBS$_0$ (SBS$_{+1}$)  and is stable (unstable).}
\label{fig4}
\end{figure}

\begin{figure}[h!]
\begin{tabular}{ p{0.5\linewidth}  p{0.5\linewidth} }
\end{tabular}
\\
\includegraphics[width=1.\linewidth]{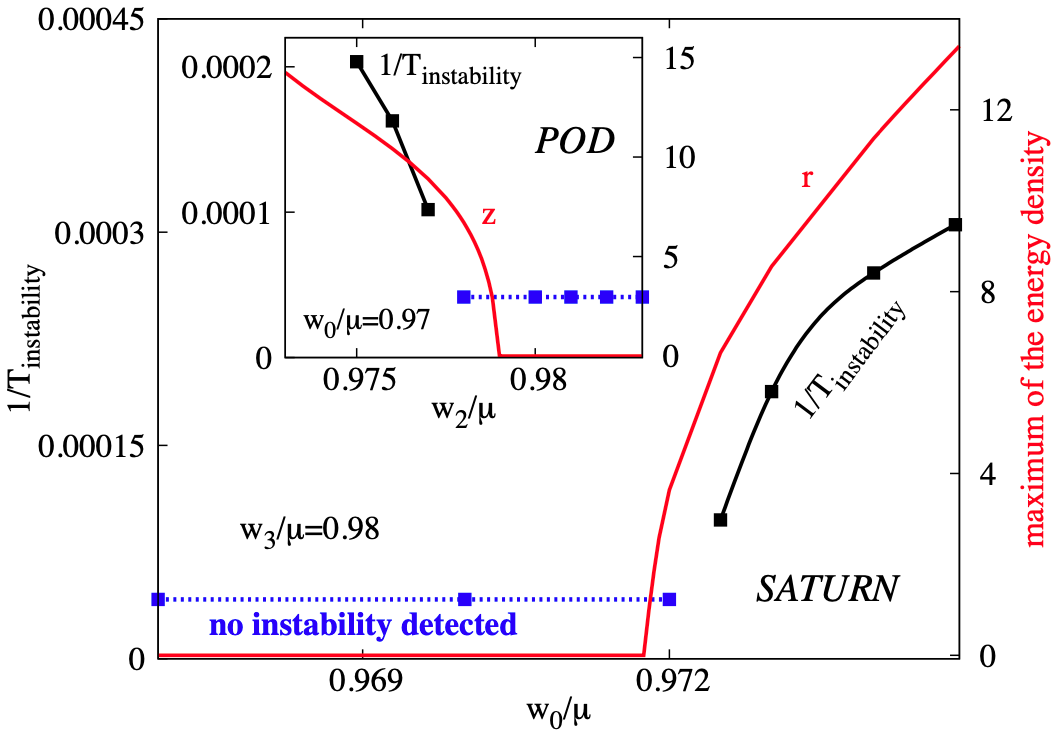}\hspace{-0.01\linewidth}
\caption{Instability timescale (black dots and lines) and  maximum of  the total energy density (red lines) for the sequence of Saturns (main panel) and pods (inset) described in the text.   The blue points and line are thresholds when no instability is seen. The $z$ coordinate in the inset is $r$ for $\theta=0,\pi$. $T_{\rm{instability}}$ refers to the time when we observe that the solution begins to clearly deviate from axisymmetry for Saturns and from equatorial symmetry for pods.
}
\label{figmax}
\end{figure}

To be concrete we consider the following superpositions: $(A)$ MBS$_0$+SBS$_{+ 1}$  and $(B)$ MBS$_0$+DBS$_0$. As an  illustration of $(A)$, fixing $w_3/\mu=0.98$, there is a continuous sequence of solutions reducing to  the MBS$_0$ (SBS$_{+ 1}$) for $w_0/\mu=0.964$ (0.975). We refer to the intermediate configurations as `\textit{Saturns}'. Their  dynamical evolutions - Fig.~\ref{fig4} -  exhibit a simple pattern: sufficiently close to the MBS$_0$ (SBS$_{+ 1}$) limit, Saturns are stable (unstable). Here, stability means no sign of instabilities for long evolutions ($t\simeq 24000$)~\footnote{The typical size of the BSs here is around $r\sim$ 15-30; then $t\simeq 24000$ means 400-800 light-crossing times.}.  Attempting to interpret the transition between the two regimes, we  observe a correlation between instability and the $r$ coordinate of the maximum  of  the energy density  -  Fig.~\ref{figmax}: when the latter is at the origin ($r=0$) no instability is observed.

As an  example of $(B)$, fixing $w_0/\mu=0.97$, there is a continuous sequence of solutions reducing to  the MBS$_0$ (DBS$_{0}$) for $w_2/\mu=0.983$ ($0.973$). We refer to the intermediate configurations as `\textit{pods}'. Evolving this sequence of pods reveals analogous patterns: $(a)$ sufficiently close to the MBS$_0$ (DBS$_{0}$) limit, pods are stable (unstable) - see Appendix  A for snapshots of the evolutions; $(b)$ when the energy density maximum, which, in  general, has two symmetric points located on the $z$-axis (i.e, $\theta=0,\pi$) is at the origin, no instability is observed.  - Fig.~\ref{figmax} (inset).

{\bf {\em Generality  and remarks.}} 
An analogous family of \textit{vector} $\ell=1$ BSs should exist.  Preliminary results show an important difference: the whole sequence {\bf 3}   (see Fig.~\ref{fig1}) is stable in the vector case,  including the SBS$_{-1}$+SBS$_{+ 1}$ static configuration, which is now spheroidal rather than toroidal (see Appendix B for details). This is a consequence of the stability of vector SBS$_{\pm 1}$~\cite{sanchis2019nonlinear}. On the other hand, we have evidence that the vector DBS$_{0}$ is unstable, as in the scalar case (see Appendix C).

A byproduct  of our construction is the realization that all single-frequency BSs arising  in (combinations of)  models of type~\eqref{action} are continuously connected within a multidimensional solutions manifold, interpolated by multi-frequency solutions. For instance, spherical (MBS$_0$) and spinning (SBS$_{\pm 1}$) BSs,  typically described as disconnected, are connected (via Saturns).

Adding the fundamental MBS$_0$, which is the ground state of the whole family, stabilizes different types of unstable BSs, such as excited monopole BSs~\cite{Bernal:2009zy}, spinning  and dipole BSs. Such stable configurations can actually form from the (incomplete) gravitational  collapse of dilute distributions of the corresponding fields and multipoles - see Appendix C. It would be interesting to probe the generality of this cooperative stabilization mechanism, and if other (higher $\ell$, say) multipolar BSs can be stabilized similarly~\footnote{See also~\cite{1852629} for another multi-state construction.}.

Another mechanism for mitigating instabilities is adding self-interactions~\cite{DiGiovanni:2020ror}, which was suggested to quench the instability of spinning BSs, without requiring the energy density to  be maximized at the origin~\cite{Siemonsen:2020hcg}. It would be interesting to  construct the corresponding  $\ell=1$ BSs family in models with  self-interactions,  and investigate whether other members of the family can be stabilized by self-interactions.


\bigskip

{\bf {\em Acknowledgements.}} 
We thank  Dar\'io N\'u\~nez, V\'ictor Jaramillo, Argelia Bernal, Juan Carlos Degollado, Juan Barranco, Francisco Guzm\'an and Luis Ure\~na-L\'opez, for useful discussions and valuable
comments. This work was supported by the Spanish Agencia Estatal de Investigaci\'on (grant PGC2018-095984-B-I00),
by the Generalitat Valenciana (PROMETEO/2019/071 and GRISOLIAP/2019/029), by the Center for Research and Development in Mathematics and Applications (CIDMA) through the Portuguese Foundation for Science and Technology (FCT - Funda\c c\~ao para a Ci\^encia e a Tecnologia), references UIDB/04106/2020 and UIDP/04106/2020, by national funds (OE), through FCT, I.P., in the scope of the framework contract foreseen in the numbers 4, 5 and 6
of the article 23, of the Decree-Law 57/2016, of August 29,
changed by Law 57/2017, of July 19 and by the projects PTDC/FIS-OUT/28407/2017,  CERN/FIS-PAR/0027/2019 and PTDC/FIS-AST/3041/2020. This work has further been supported by  the  European  Union's  Horizon  2020  research  and  innovation  (RISE) programme H2020-MSCA-RISE-2017 Grant No.~FunFiCO-777740 and by FCT through
Project~No.~UIDB/00099/2020. We would like to acknowledge networking support by the COST Action GWverse CA16104.
Computations have been performed at the Servei d'Inform\`atica de la Universitat
de Val\`encia, the Argus and Blafis cluster at the U. Aveiro and on the ``Baltasar Sete-Sois'' cluster at IST.

\bibliography{num-rel2}

\bigskip

\appendix

\begin{figure}[h!]
\centering
\includegraphics[width=0.24\linewidth]{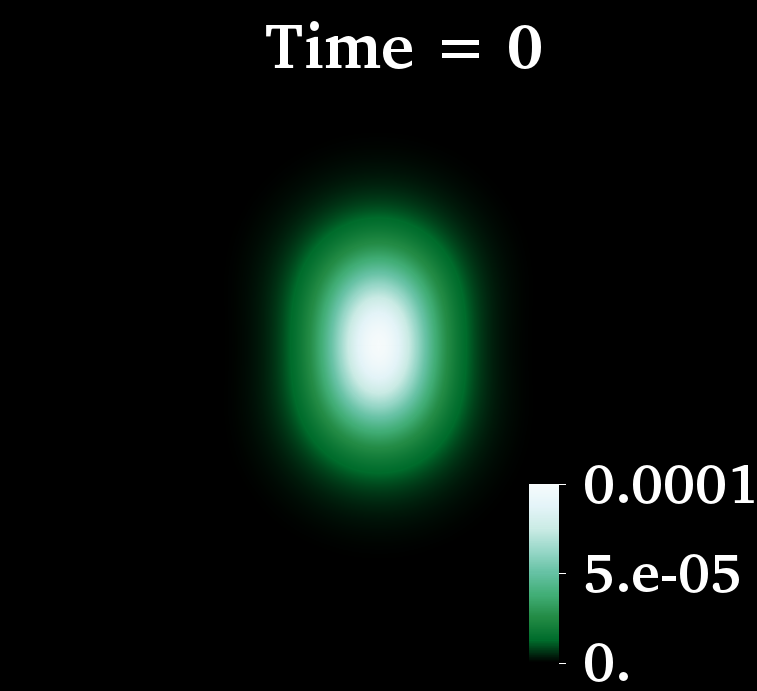}\hspace{-0.01\linewidth}
\includegraphics[width=0.24\linewidth]{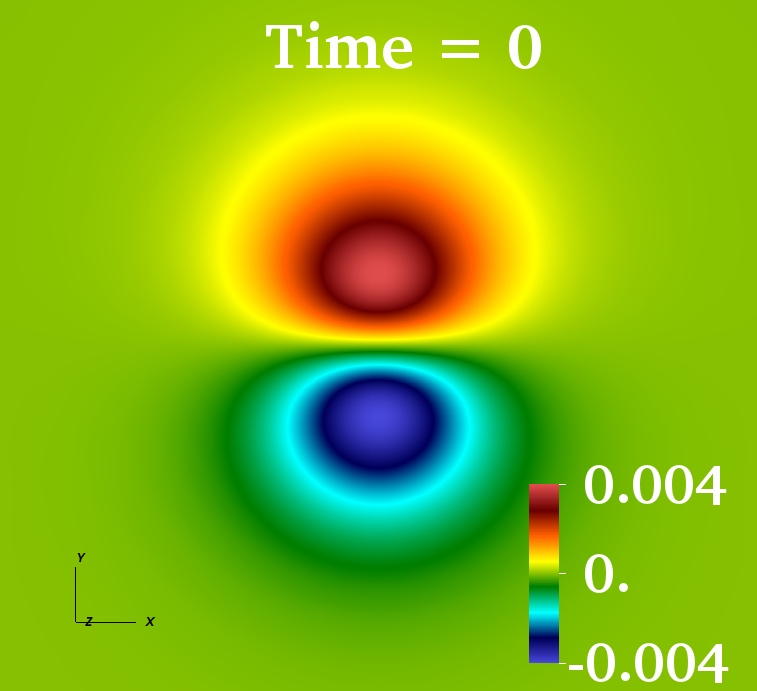}\hspace{-0.01\linewidth}
\includegraphics[width=0.24\linewidth]{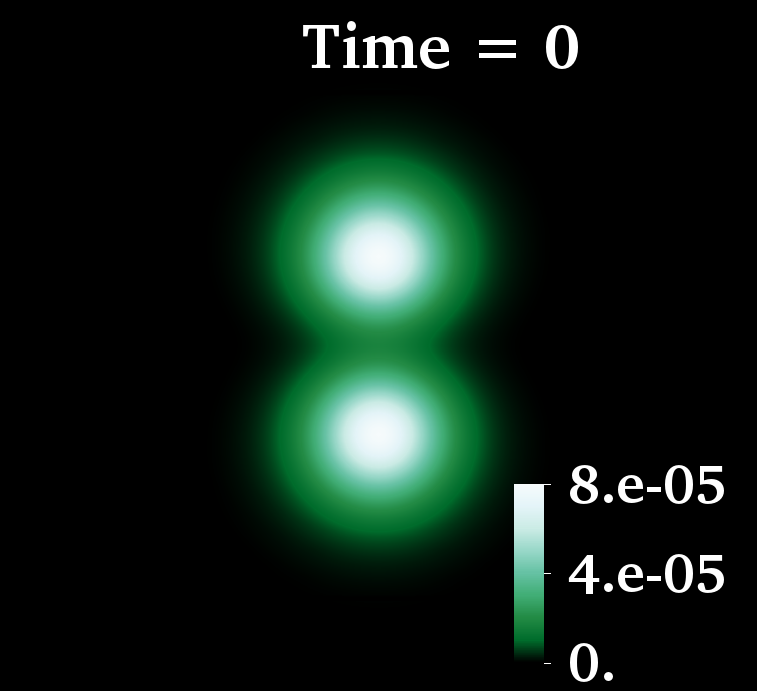}\hspace{-0.01\linewidth}
\includegraphics[width=0.24\linewidth]{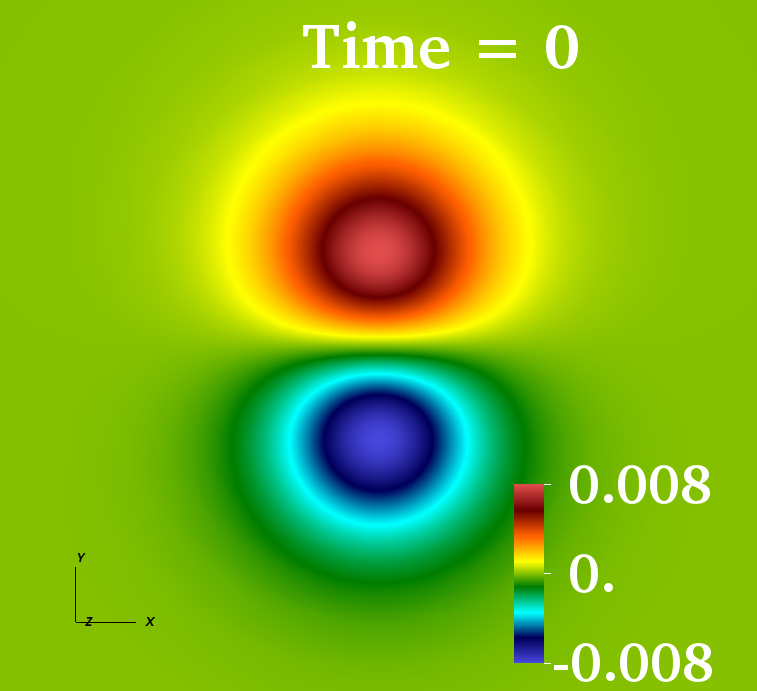}\hspace{-0.01\linewidth}\\
\includegraphics[width=0.24\linewidth]{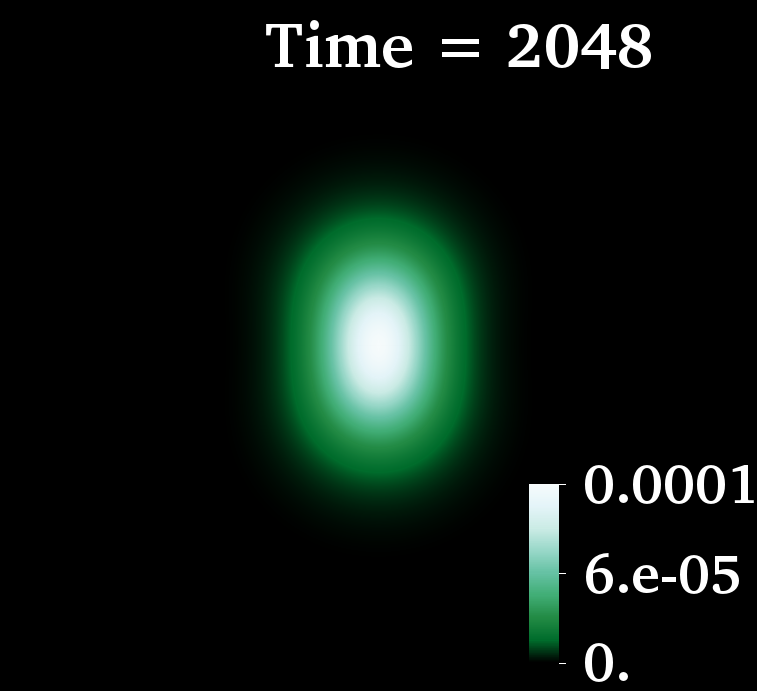}\hspace{-0.01\linewidth}
\includegraphics[width=0.24\linewidth]{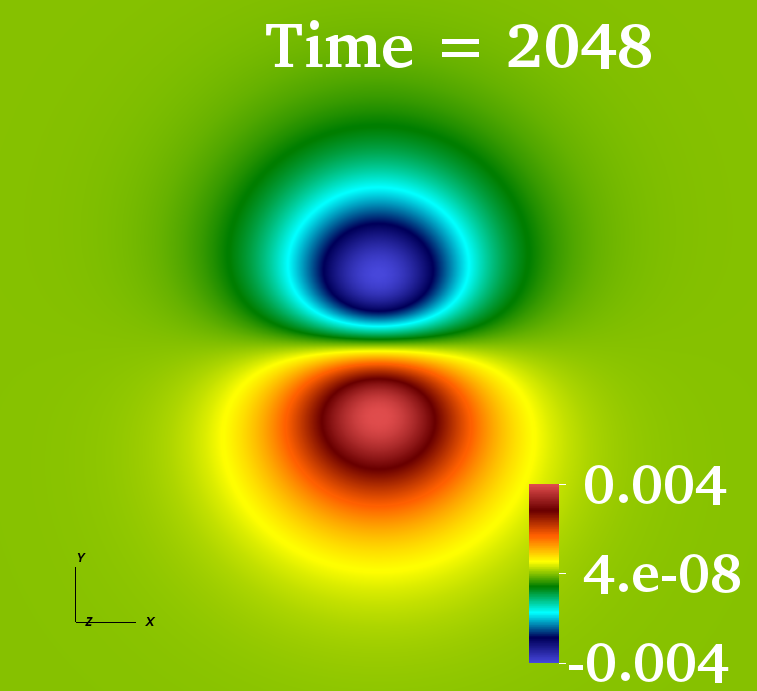}\hspace{-0.01\linewidth}
\includegraphics[width=0.24\linewidth]{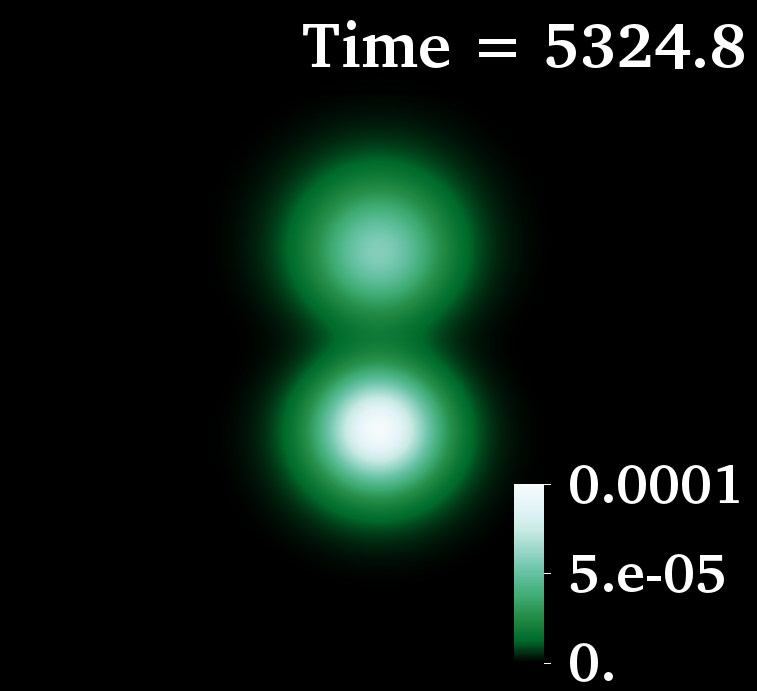}\hspace{-0.01\linewidth}
\includegraphics[width=0.24\linewidth]{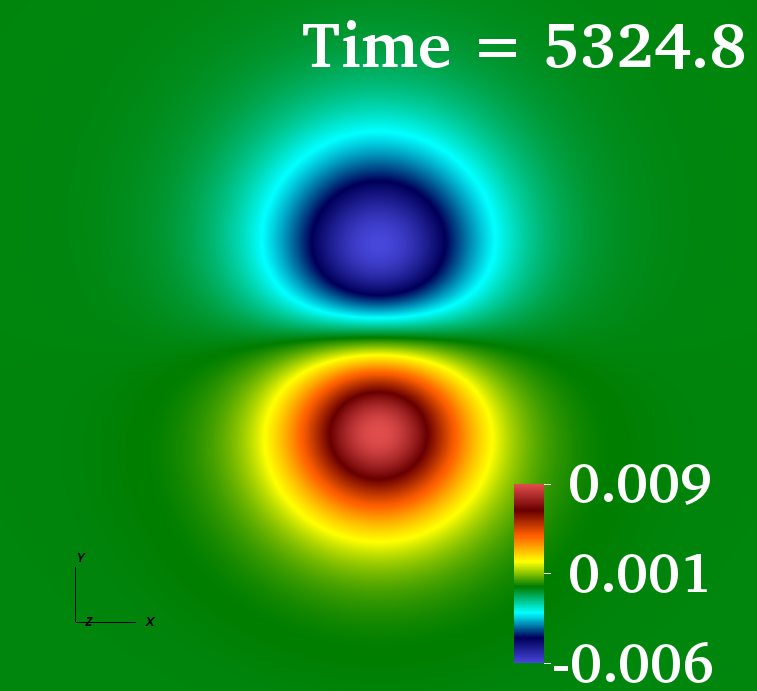}\hspace{-0.01\linewidth}\\
\includegraphics[width=0.24\linewidth]{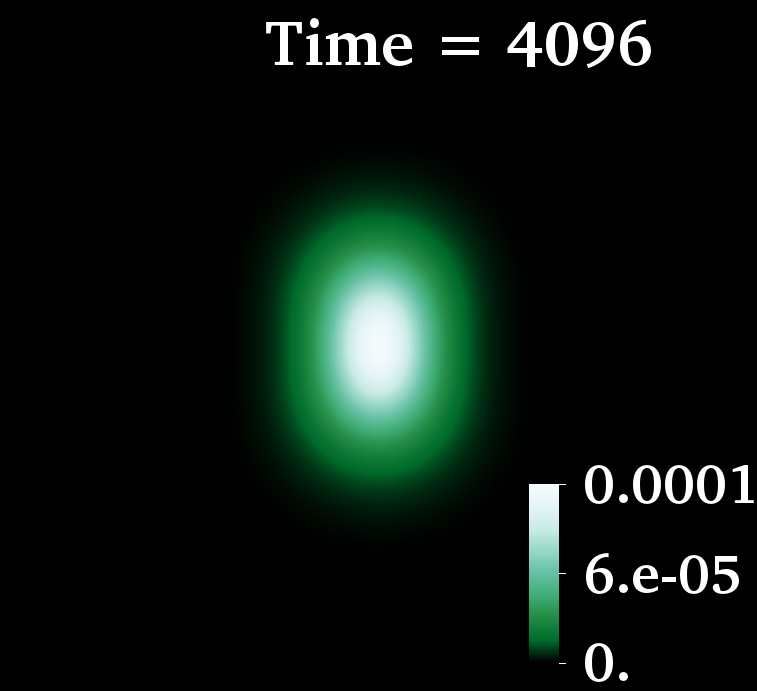}\hspace{-0.01\linewidth}
\includegraphics[width=0.24\linewidth]{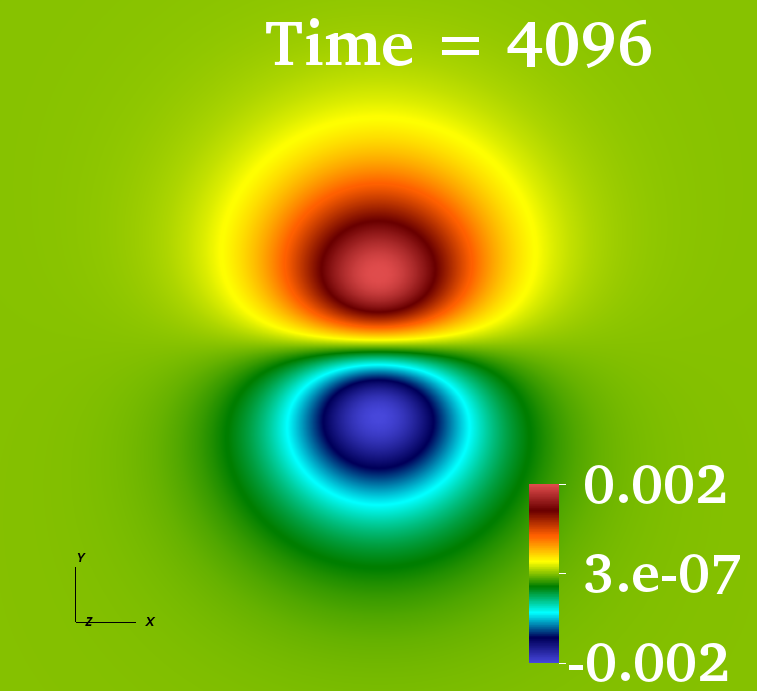}\hspace{-0.01\linewidth}
\includegraphics[width=0.24\linewidth]{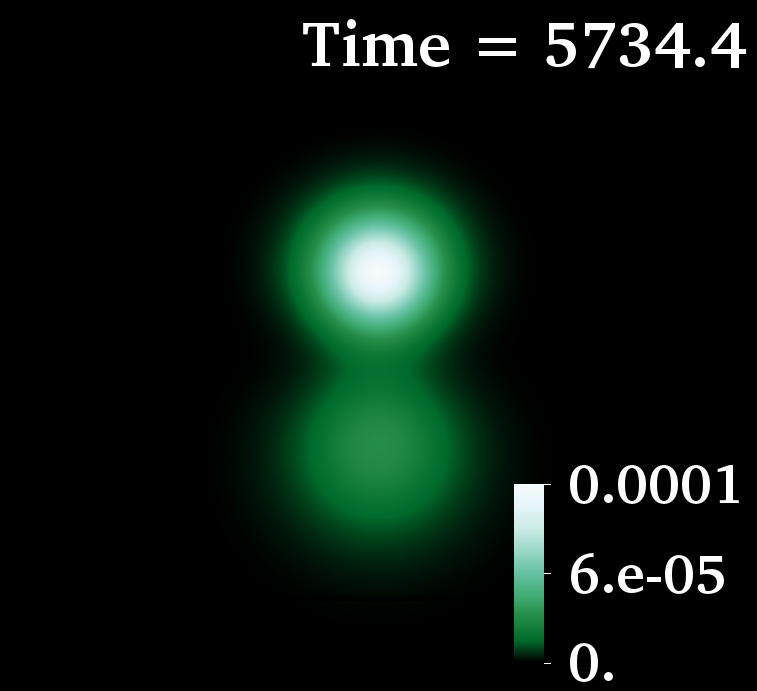}\hspace{-0.01\linewidth}
\includegraphics[width=0.24\linewidth]{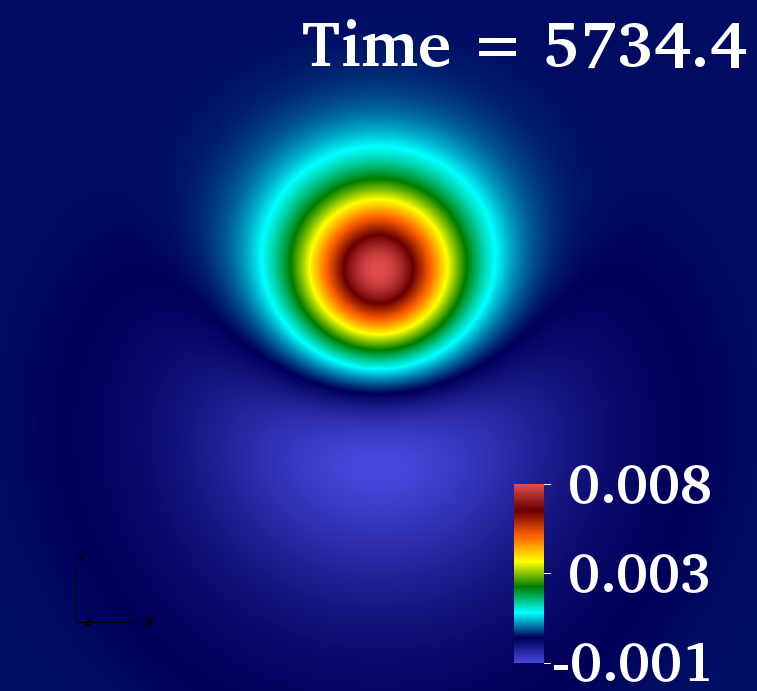}\hspace{-0.01\linewidth}\\
\includegraphics[width=0.24\linewidth]{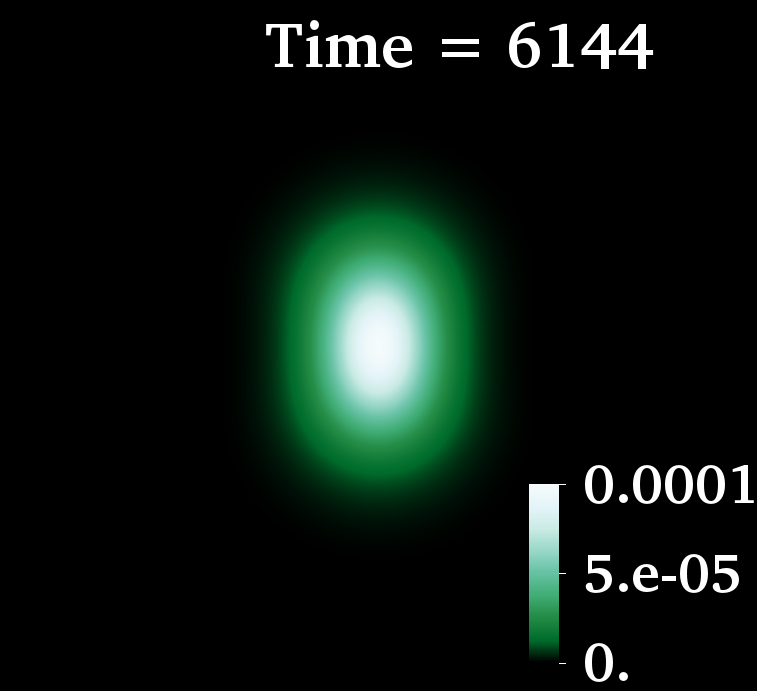}\hspace{-0.01\linewidth}
\includegraphics[width=0.24\linewidth]{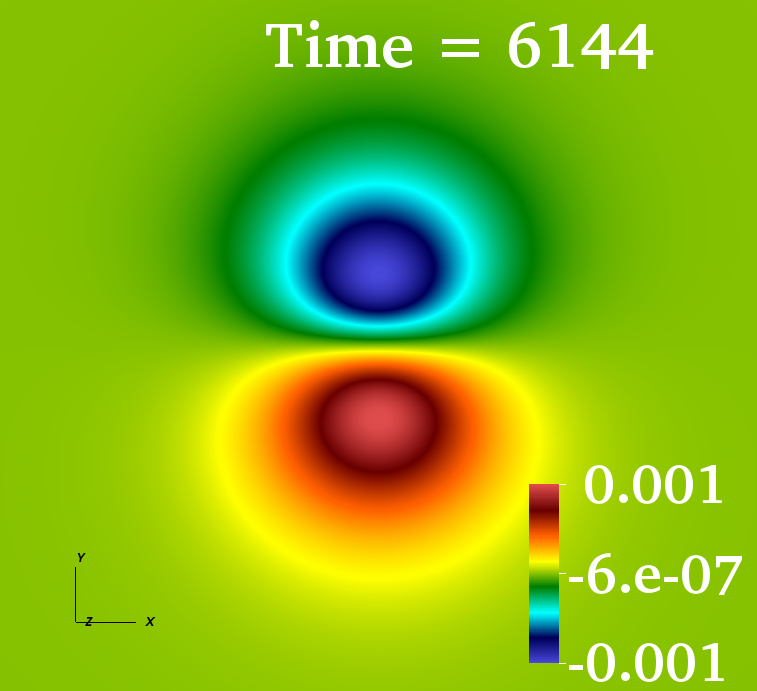}\hspace{-0.01\linewidth}
\includegraphics[width=0.24\linewidth]{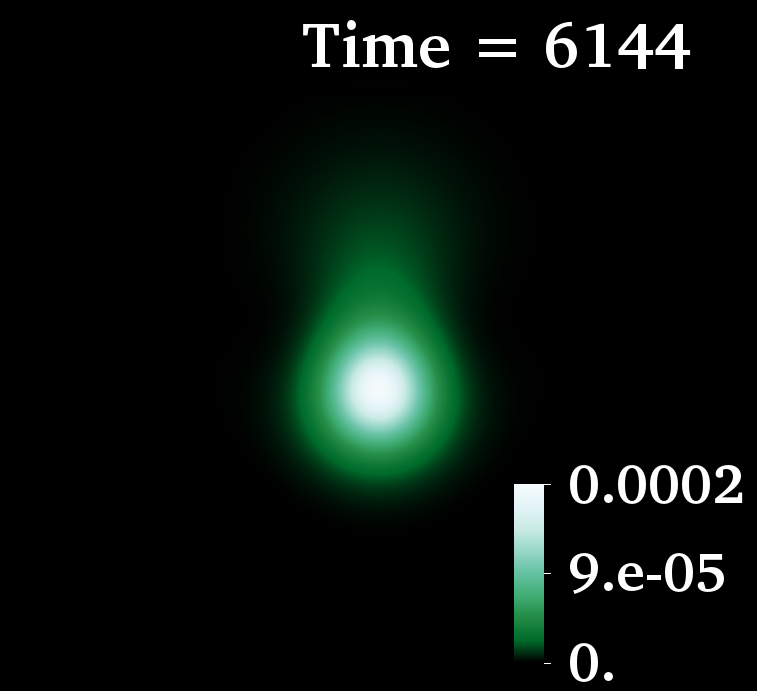}\hspace{-0.01\linewidth}
\includegraphics[width=0.24\linewidth]{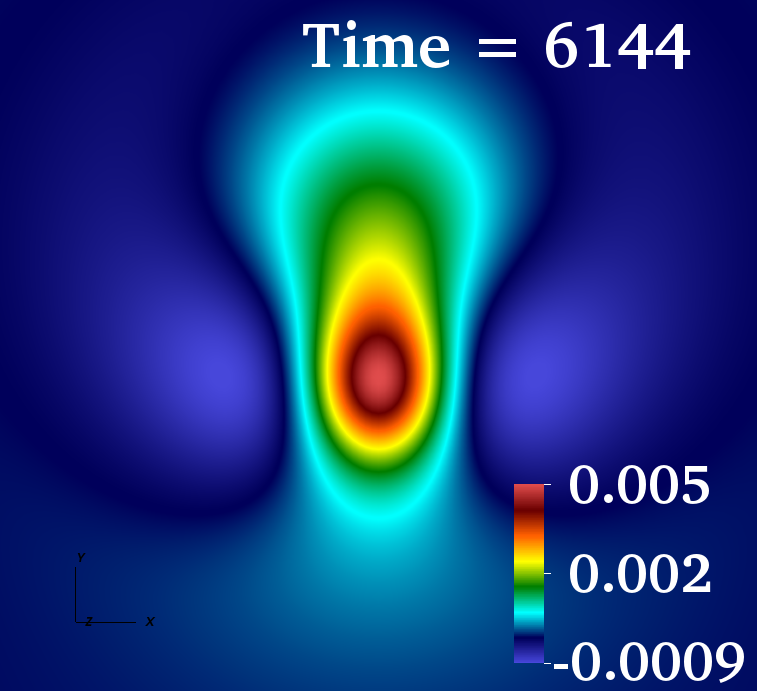}\hspace{-0.01\linewidth}\\
\includegraphics[width=0.24\linewidth]{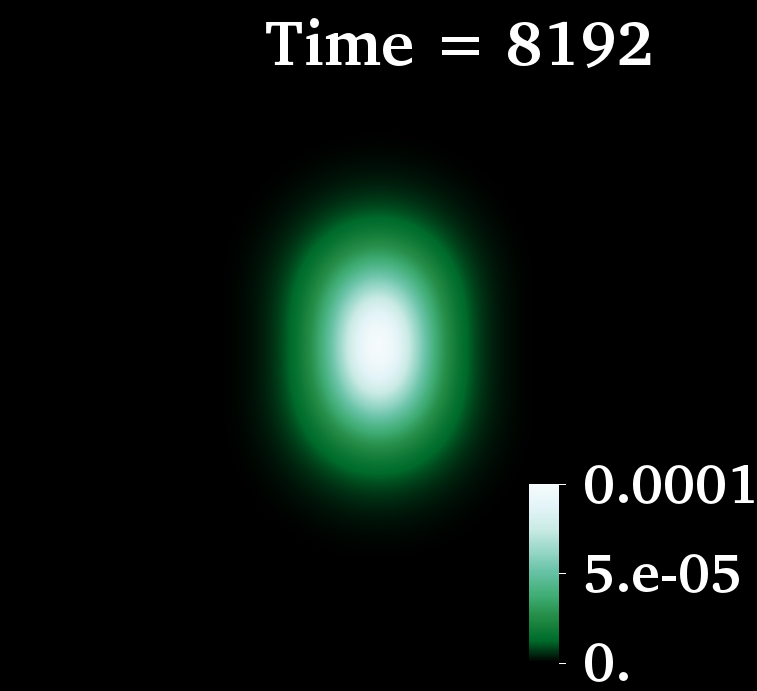}\hspace{-0.01\linewidth}
\includegraphics[width=0.24\linewidth]{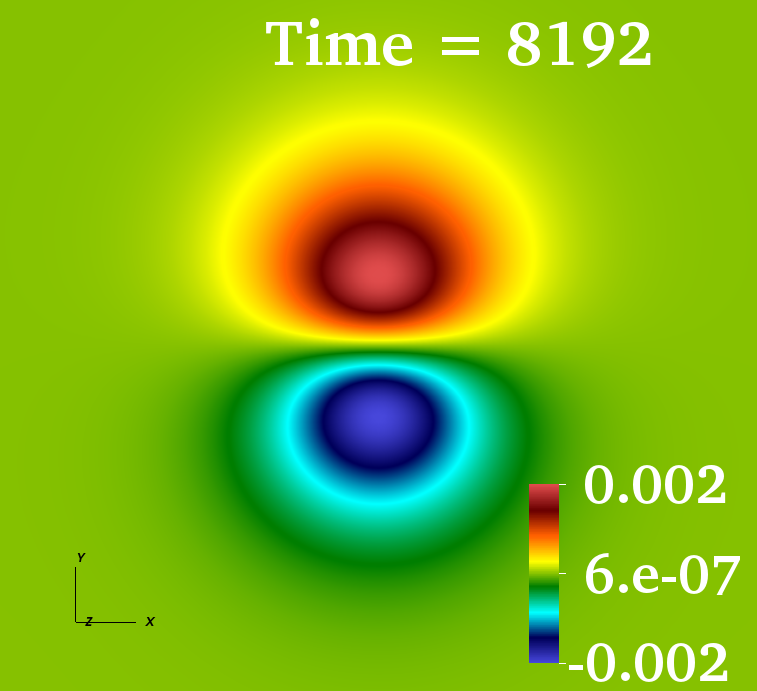}\hspace{-0.01\linewidth}
\includegraphics[width=0.24\linewidth]{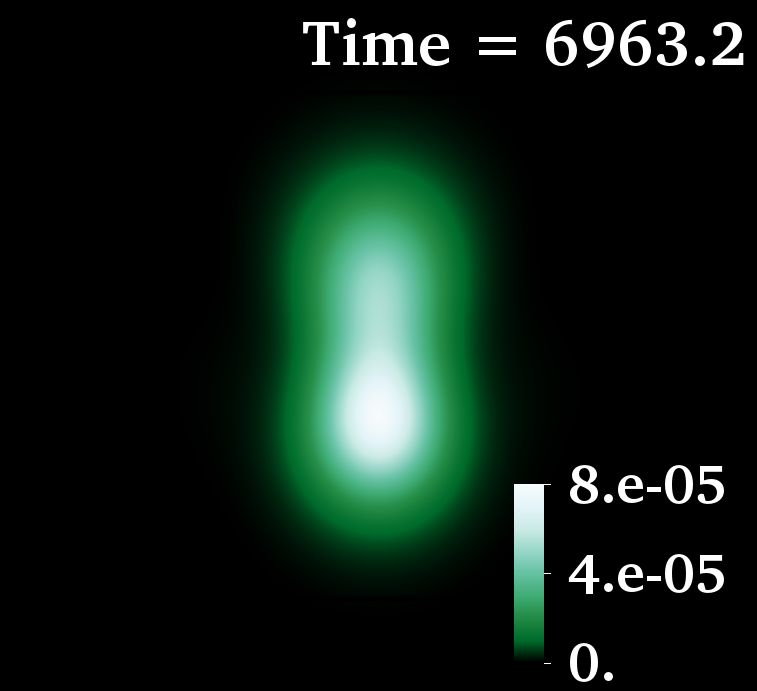}\hspace{-0.01\linewidth}
\includegraphics[width=0.24\linewidth]{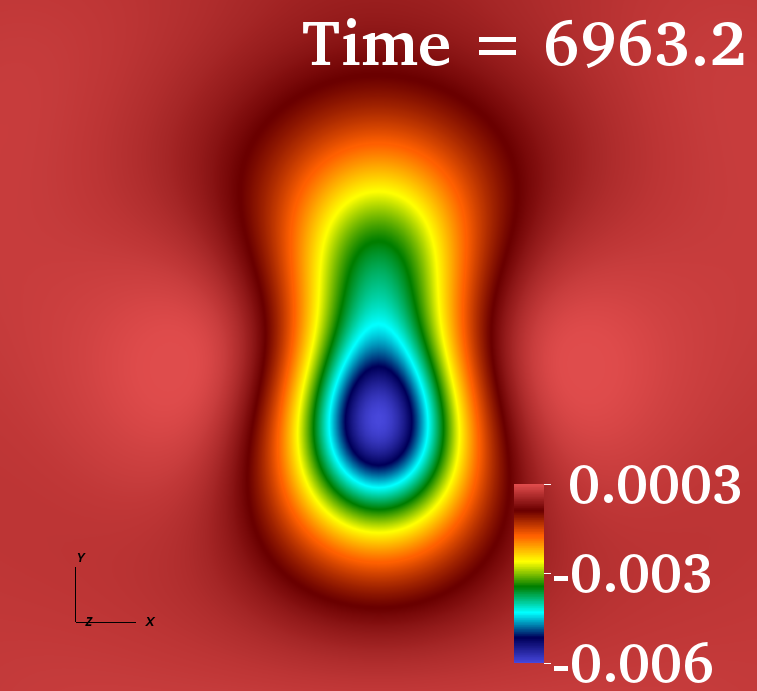}\hspace{-0.01\linewidth}\\
\includegraphics[width=0.24\linewidth]{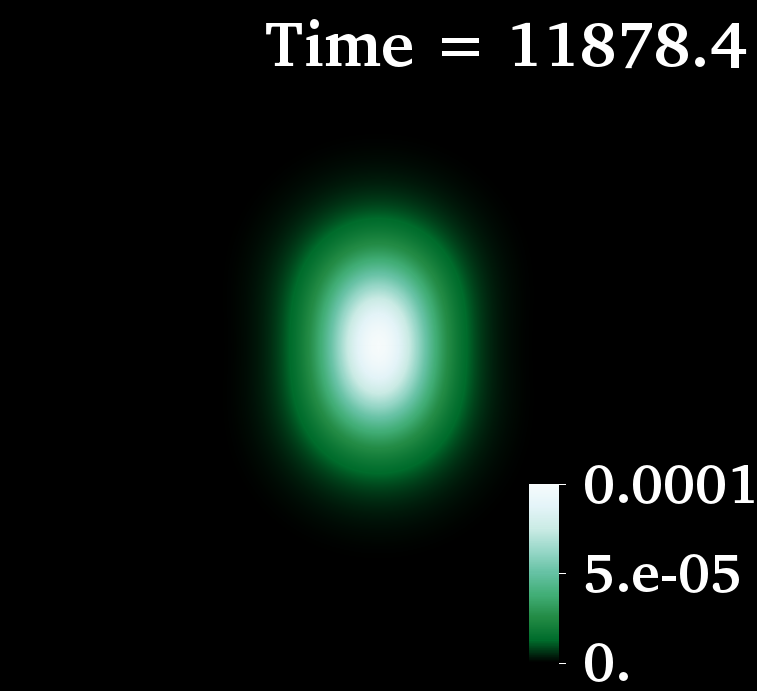}\hspace{-0.01\linewidth}
\includegraphics[width=0.24\linewidth]{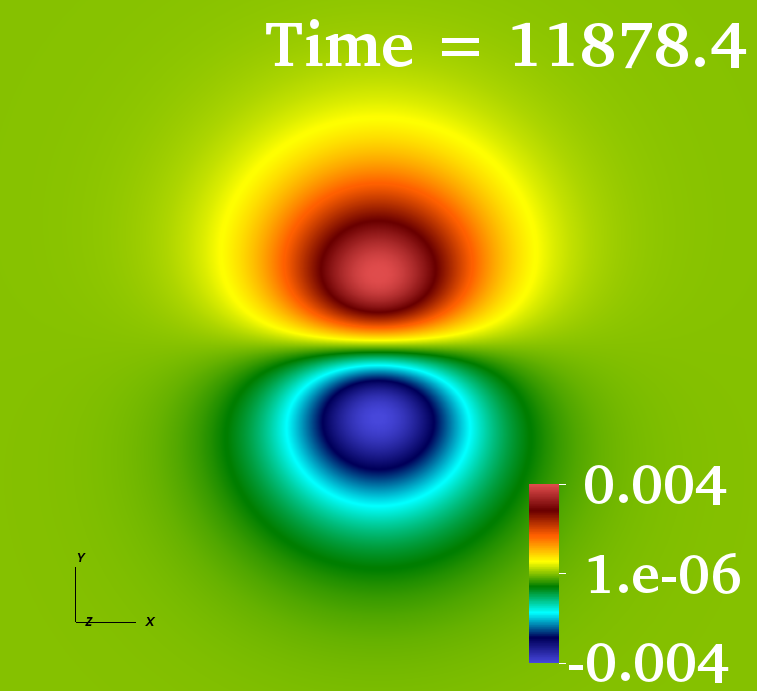}\hspace{-0.01\linewidth}
\includegraphics[width=0.24\linewidth]{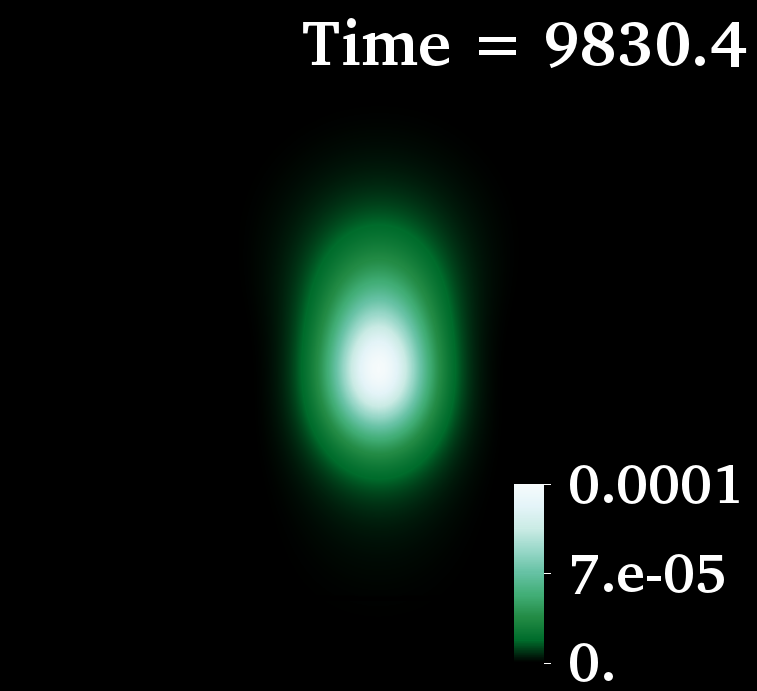}\hspace{-0.01\linewidth}
\includegraphics[width=0.24\linewidth]{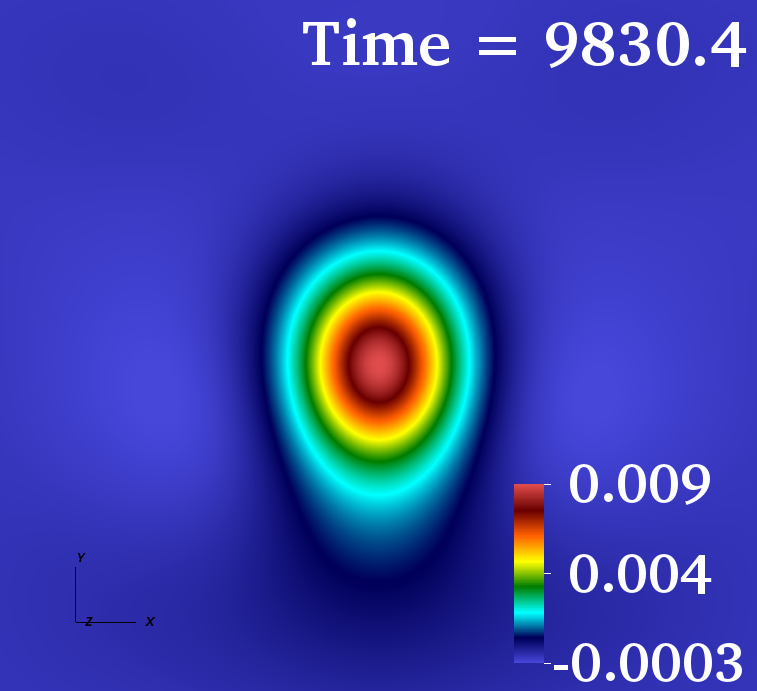}\hspace{-0.01\linewidth}\\
\caption{Time evolution (top to bottom) of two pods:  (left and middle left) $\omega_2/\mu=0.98$; (middle  right and right)  $\omega_2/\mu=0.975$. For  each case the energy density and the real part of $\Phi^{(2)}$ on the $xz$ plane are shown. The first (second) pod is close to  MBS$_0$ (DBS$_{0}$)  and is stable (unstable).}
\label{fig6}
\end{figure}

{\bf {\em Appendix A. Evolution of  pods.}} 
In Fig.~\ref{fig6} we exhibit the time evolution of two illustrative pods, one stable  and one unstable,  along the sequence described in the  main text, with $w_0/\mu=0.97$. 

The decay of the  unstable one is similar to the one of the pure dipoles, exhibited in Fig.~\ref{fig3} (left column). Observe that the decay  of the unstable Saturns shown in Fig.~\ref{fig4}, is triggered by  a  non-axisymmetric perturbation, inherited from  the instability of  the pure SBS$_{\pm 1}$; this is not the case for the decay of the unstable pods, which would occur even in axi-symmetry.

{\bf {\em Appendix B. Vector BSs evolutions.}} 
%
Dynamical evolutions of representative members of the $\ell=1$ family of vector BSs are performed using the same computational infrastructure of ~\cite{sanchis2019nonlinear}.
The vector BSs (\textit{aka} Proca) equations are solved using a modification of the \textsc{Proca} thorn in the \textsc{Einstein Toolkit} \cite{Canuda_2020_3565475, Zilhao:2015tya} to include a complex field. Here we illustrate these evolutions with the  vector SBS$_{-1}$+SBS$_{+1}$ configuration, which we succeded in constructing, and it is static  as in the scalar case  but now spheroidal, a property inherited from  the spheroidal nature of the spinning vector BSs (SBS$_{\pm1}$).
\begin{figure}[t!]
\centering
\includegraphics[width=0.24\linewidth]{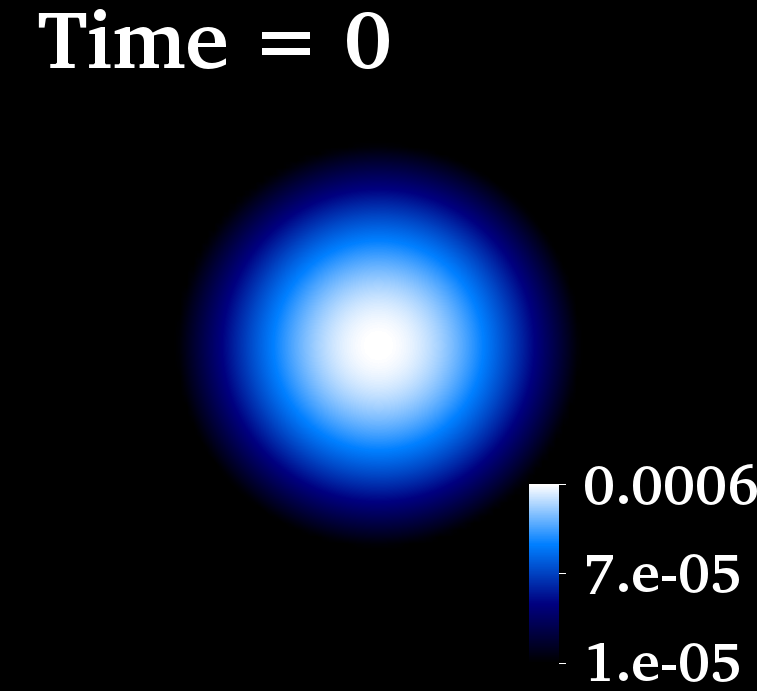}
\includegraphics[width=0.24\linewidth]{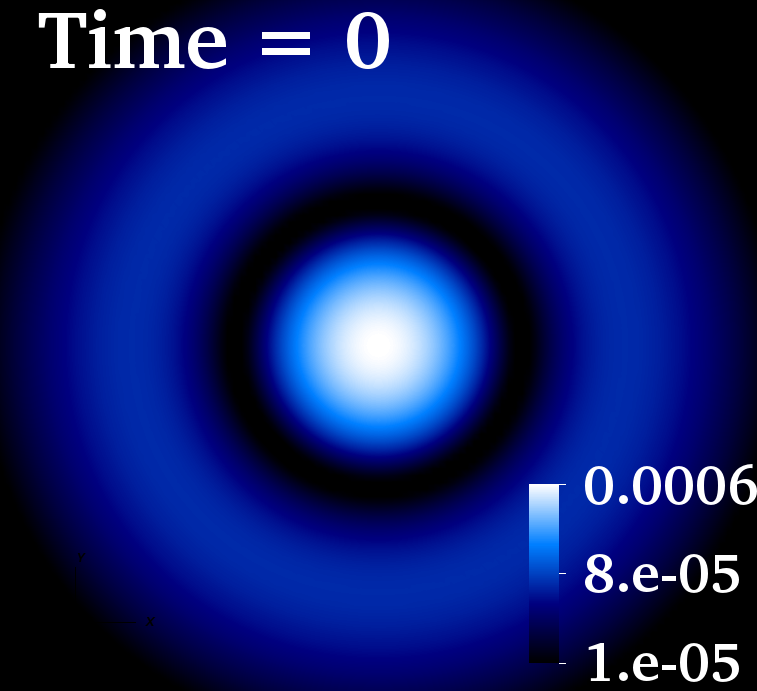}\\
\includegraphics[width=0.24\linewidth]{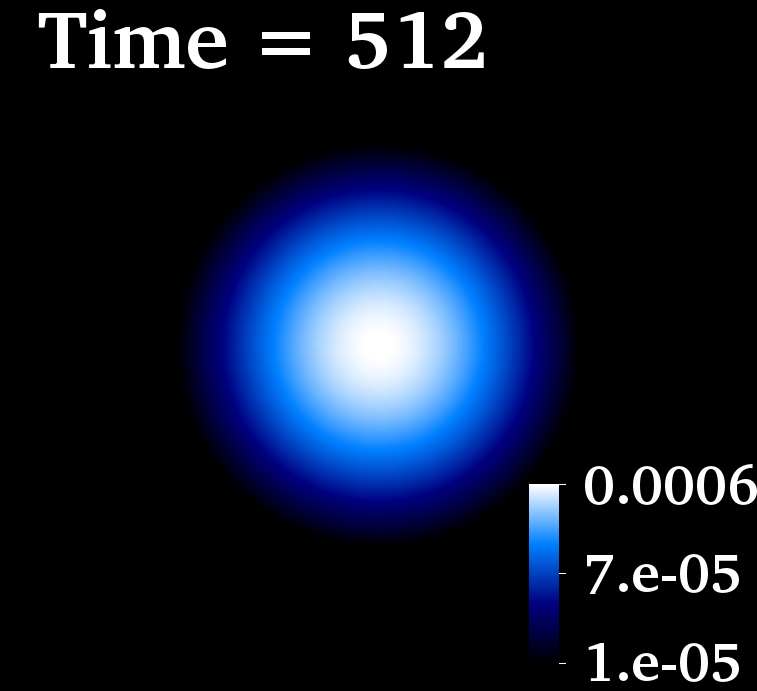}
\includegraphics[width=0.24\linewidth]{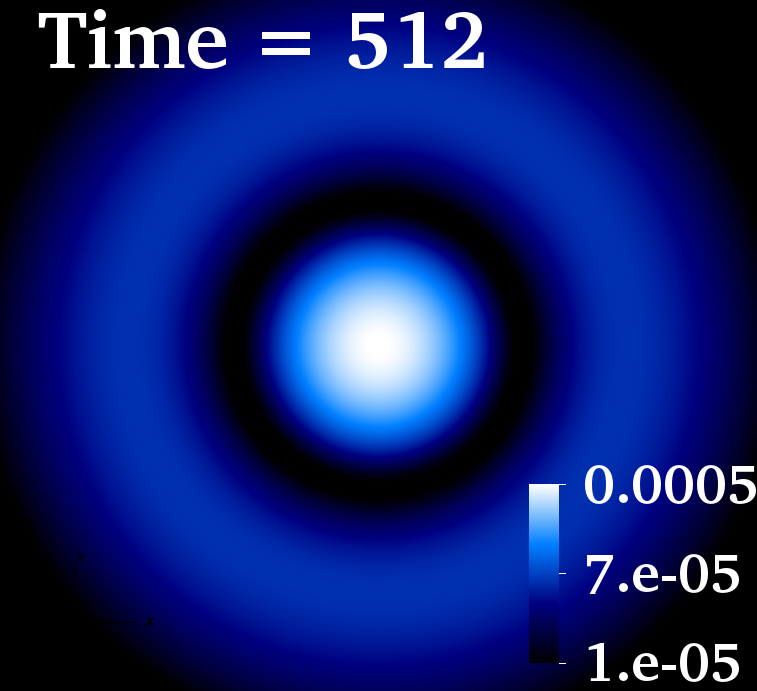}\\
\includegraphics[width=0.24\linewidth]{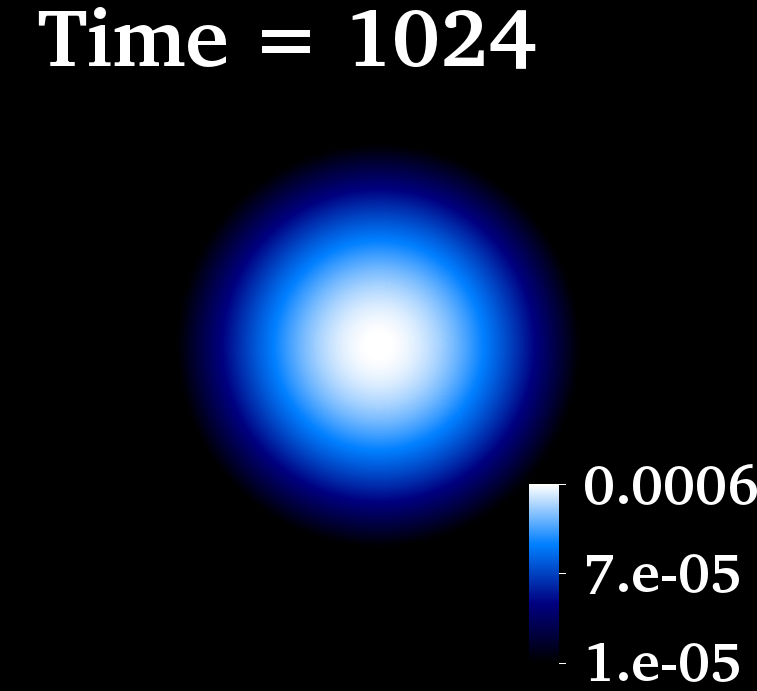}
\includegraphics[width=0.24\linewidth]{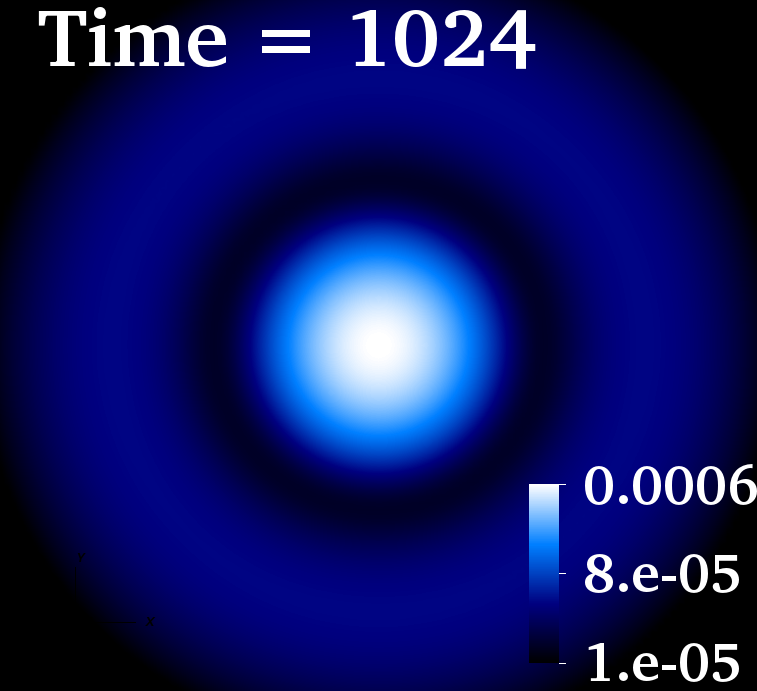}\\
\includegraphics[width=0.24\linewidth]{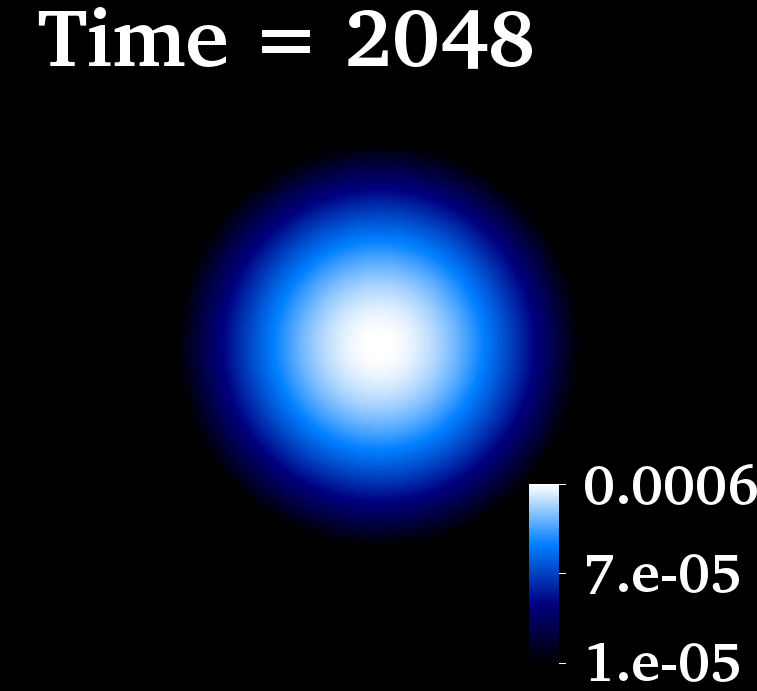}
\includegraphics[width=0.24\linewidth]{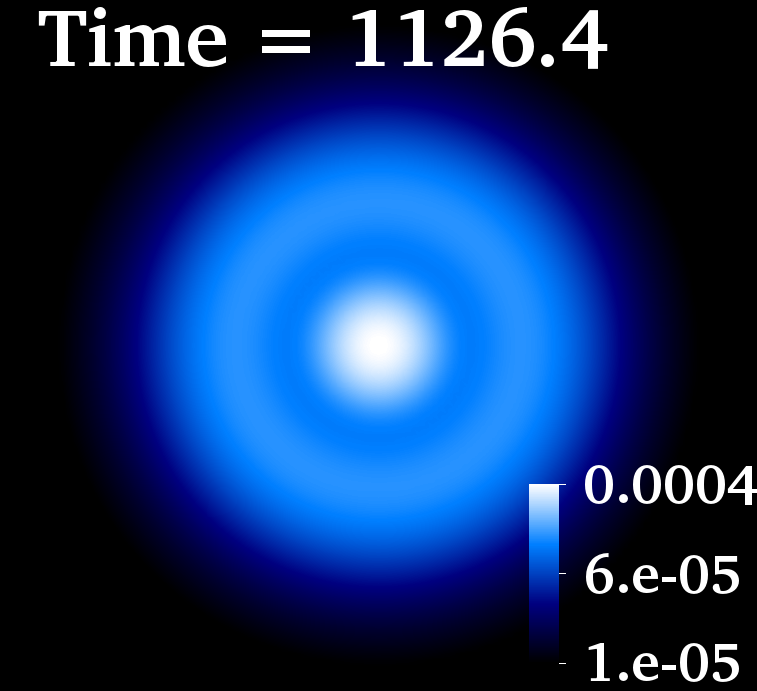}\\
\includegraphics[width=0.24\linewidth]{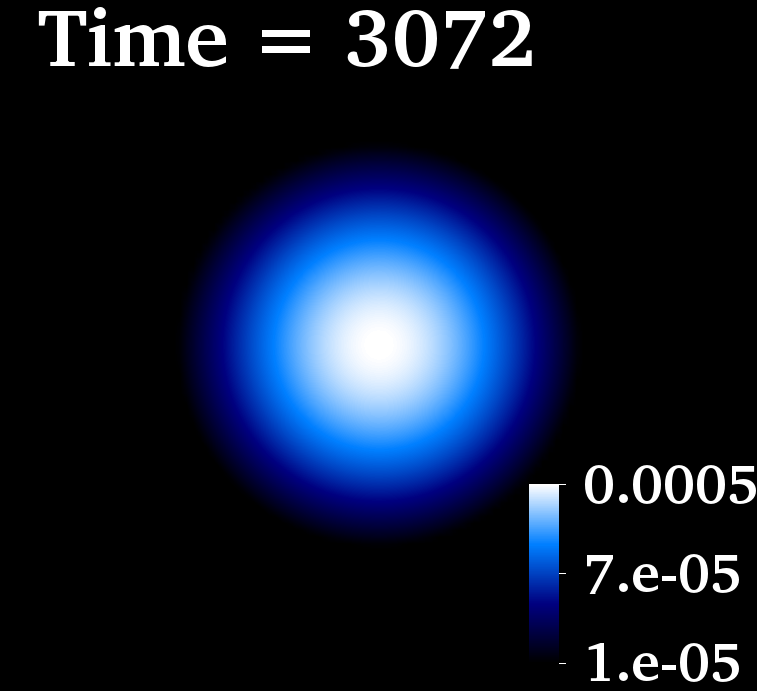}
\includegraphics[width=0.24\linewidth]{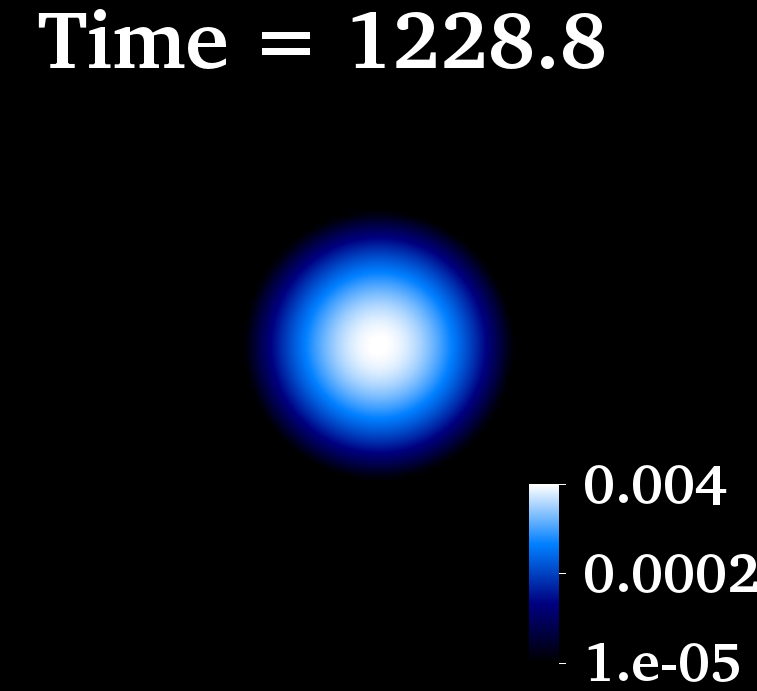}\\
\includegraphics[width=0.24\linewidth]{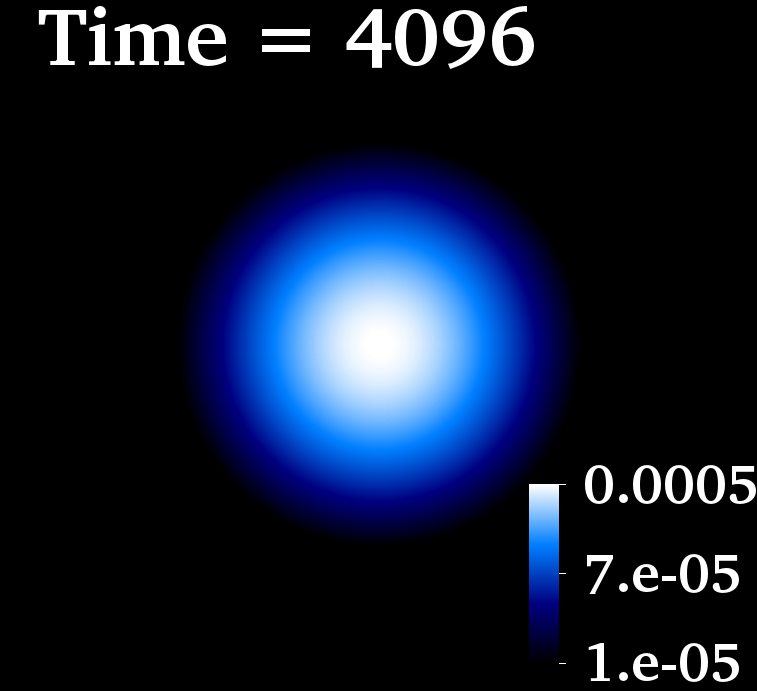}
\includegraphics[width=0.24\linewidth]{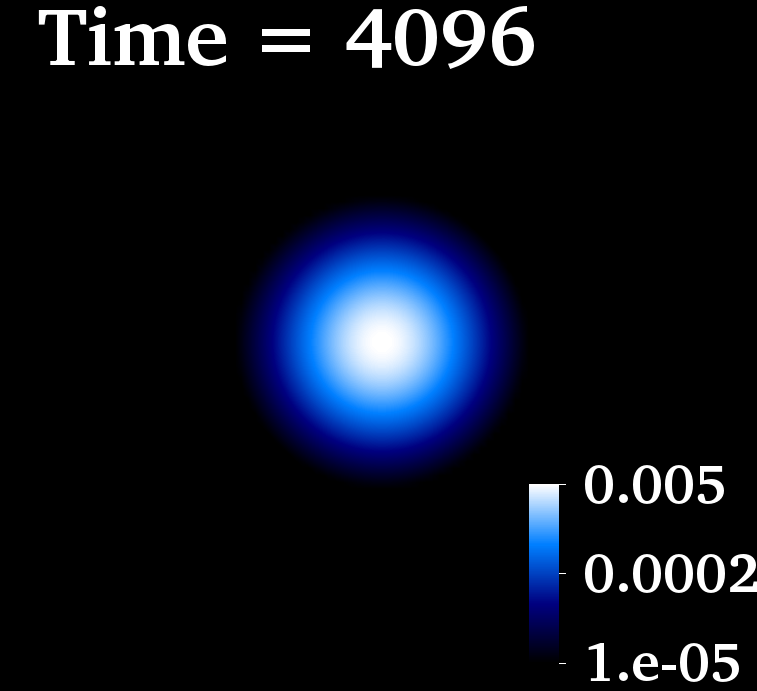}\\
\caption{Time evolution (top to bottom) of two vector SBS$_{-1}$+SBS$_{+1}$ with  $\omega_2/\mu=0.95$. The  fundamental (excited state) is shown in the left (right) column and is found to be stable  (unstable).}
\label{fig7}
\end{figure}

As in the scalar case, the individual (complex, vector) fields carry angular momentum density, but the total angular momentum is zero. We observe that the vector SBS$_{-1}$+SBS$_{+1}$ is dynamically  stable - Fig.~\ref{fig7} (left column), as their individual components (SBS$_{\pm1}$)~\cite{sanchis2019nonlinear}.
 In Fig.~\ref{fig7} (right column) we also plot the time evolution of the energy density for one excited state (with one node) of the vector SBS$_{-1}$+SBS$_{+1}$. This excited star decays to the nodeless solution. Interestingly, the maximum of the energy density of  the fundamental solution is at the centre. These results indicate that the non-axisymmetric instability observed in spinning scalar BSs (as well as for the vector stars with $m>1$~\cite{sanchis2019nonlinear}) is also present in these multi-field BSs. Moreover, the same correlation observed in the main text for the scalar case applies to the vector case:  the maximum of the energy density is at  the centre for stable stars.

\begin{figure}[h!]
\begin{tabular}{ p{0.24\linewidth} p{0.24\linewidth} p{0.24\linewidth}  p{0.24\linewidth} }
\centering Scalar \\ DBS$_0$&\centering Vector \\ DBS$_0$ &\centering Scalar \\ DBS$_0$+SBS$_{+ 1}$& \centering Vector SBS$_{+ 1}$+SBS$_{-1}$
\end{tabular}
\begin{tabular}{ p{0.48\linewidth} p{0.0\linewidth}  p{0.48\linewidth} }
\centering ($y=0$, $xz$ plane)&\centering $|$& \centering ($z=0$, $xy$ plane)
\end{tabular}
\\
\includegraphics[width=0.24\linewidth]{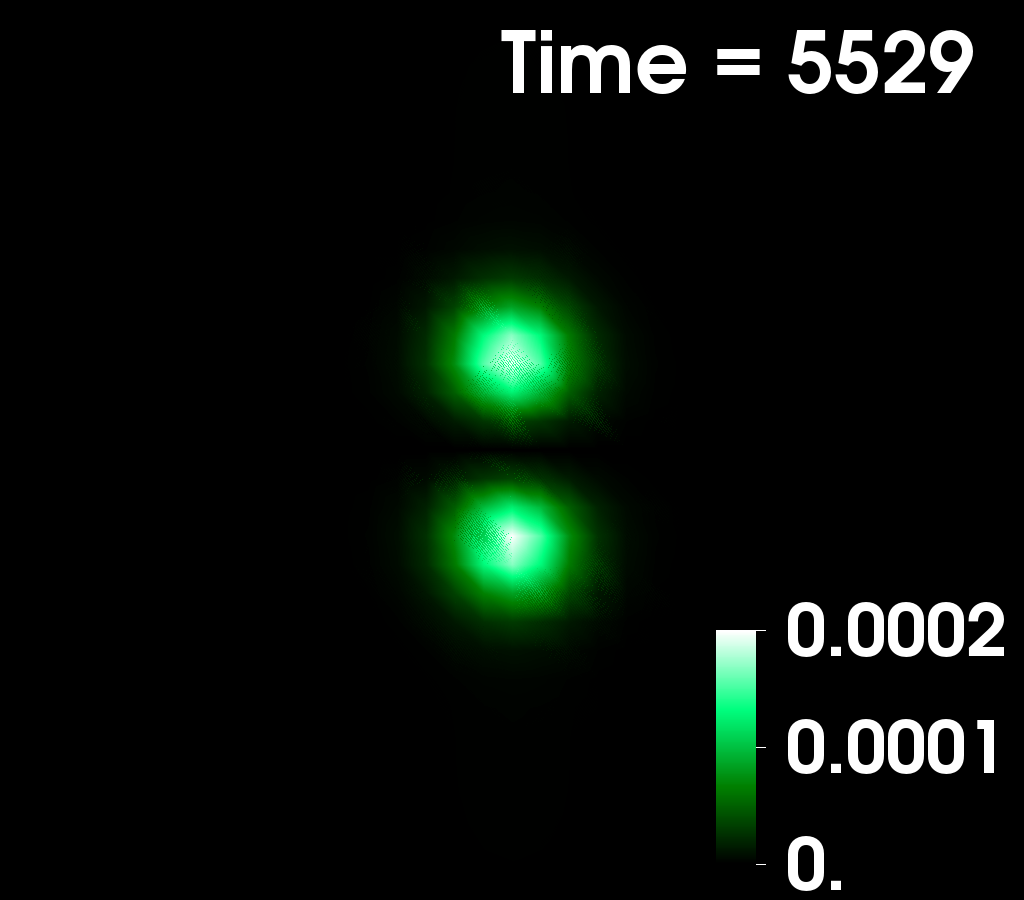}\hspace{-0.01\linewidth}
\includegraphics[width=0.236\linewidth]{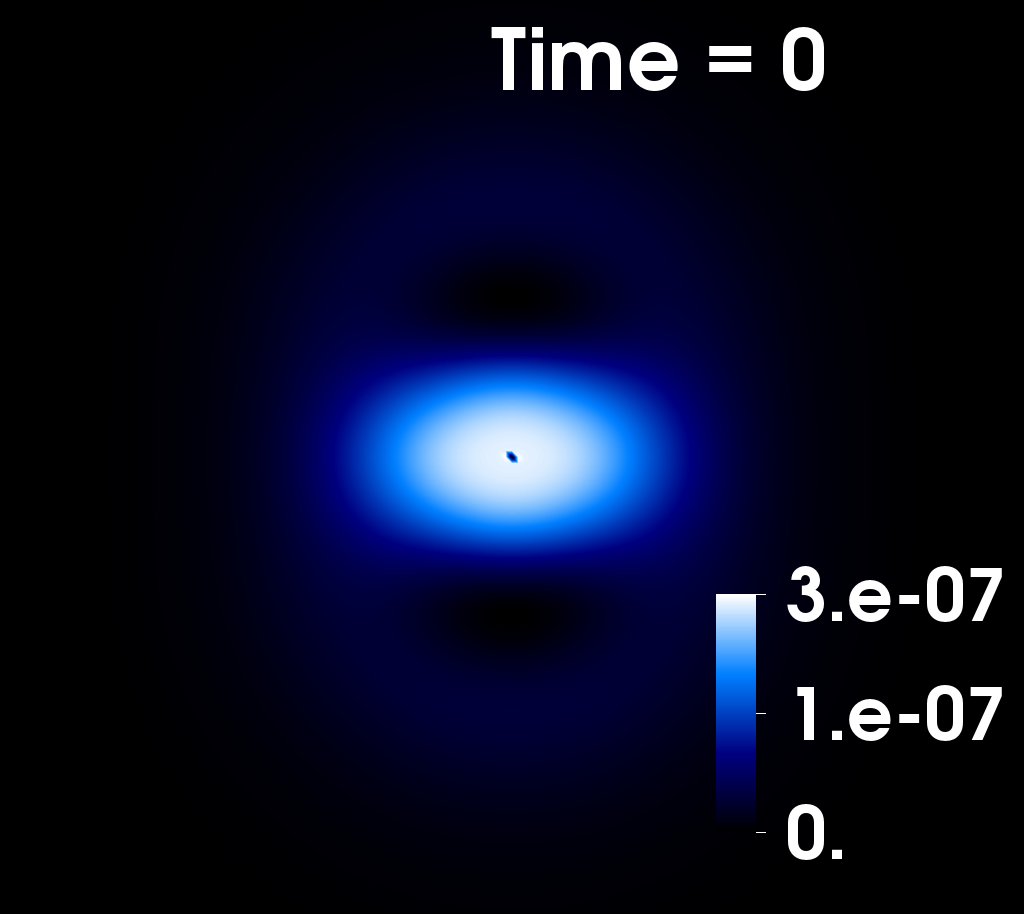}\hspace{-0.01\linewidth}
\includegraphics[width=0.24\linewidth]{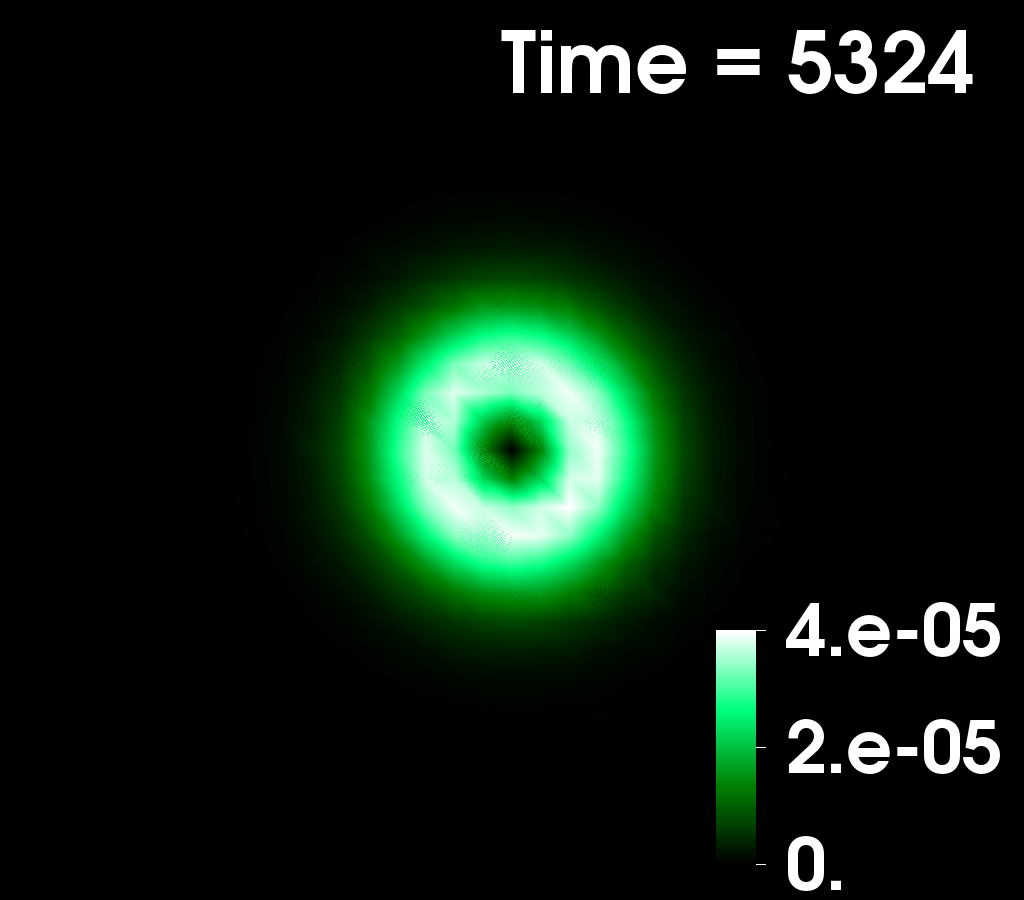}\hspace{-0.01\linewidth}
\includegraphics[width=0.24\linewidth]{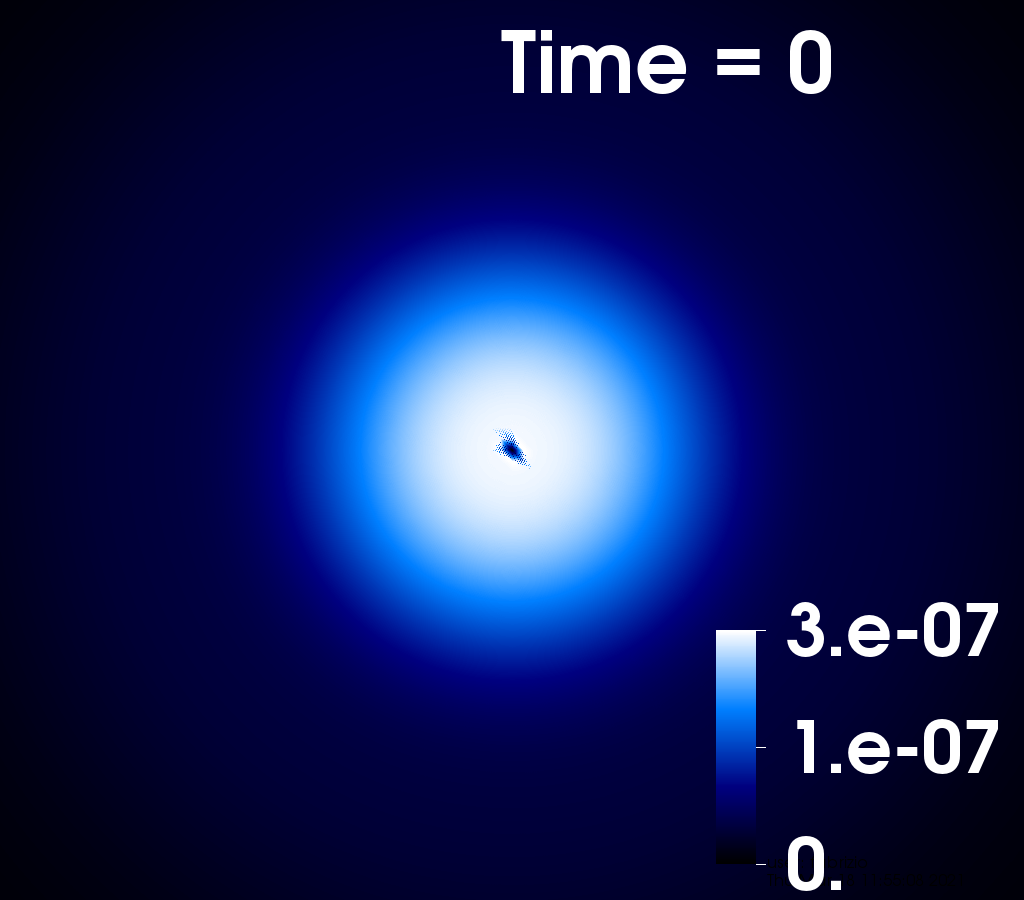}\hspace{-0.01\linewidth}

\includegraphics[width=0.24\linewidth]{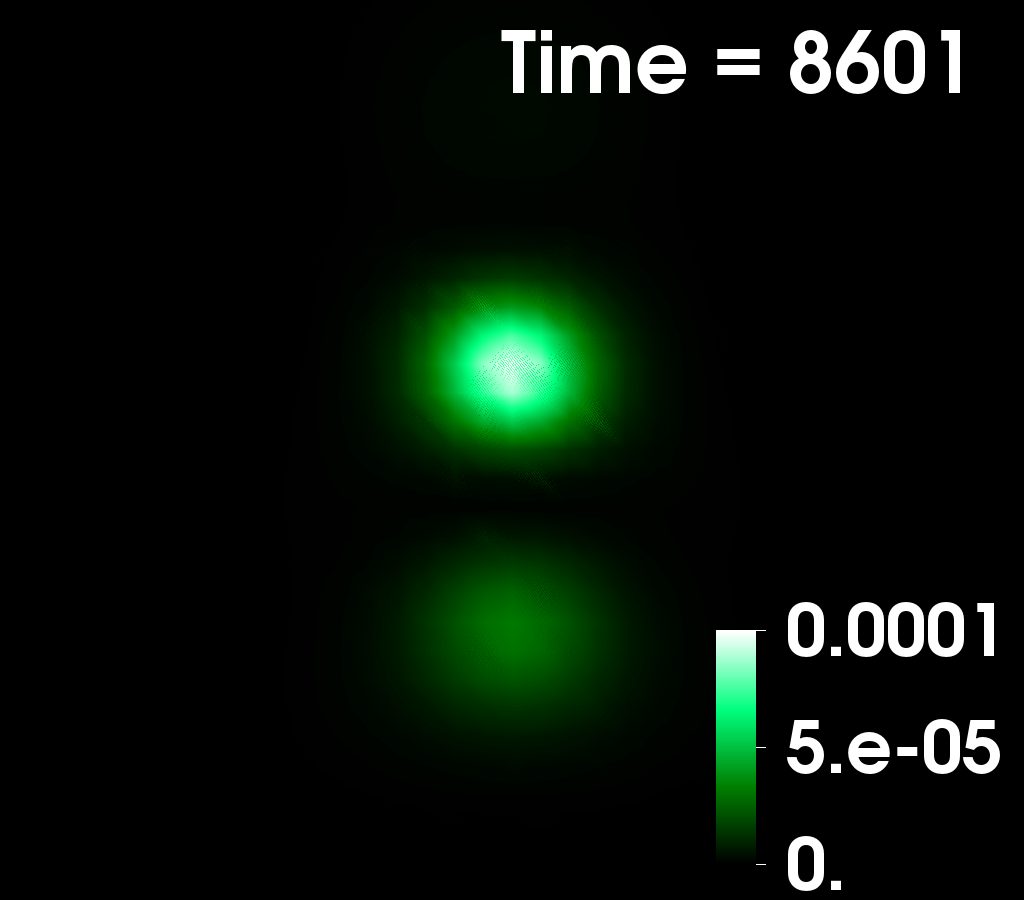}\hspace{-0.01\linewidth}
\includegraphics[width=0.236\linewidth]{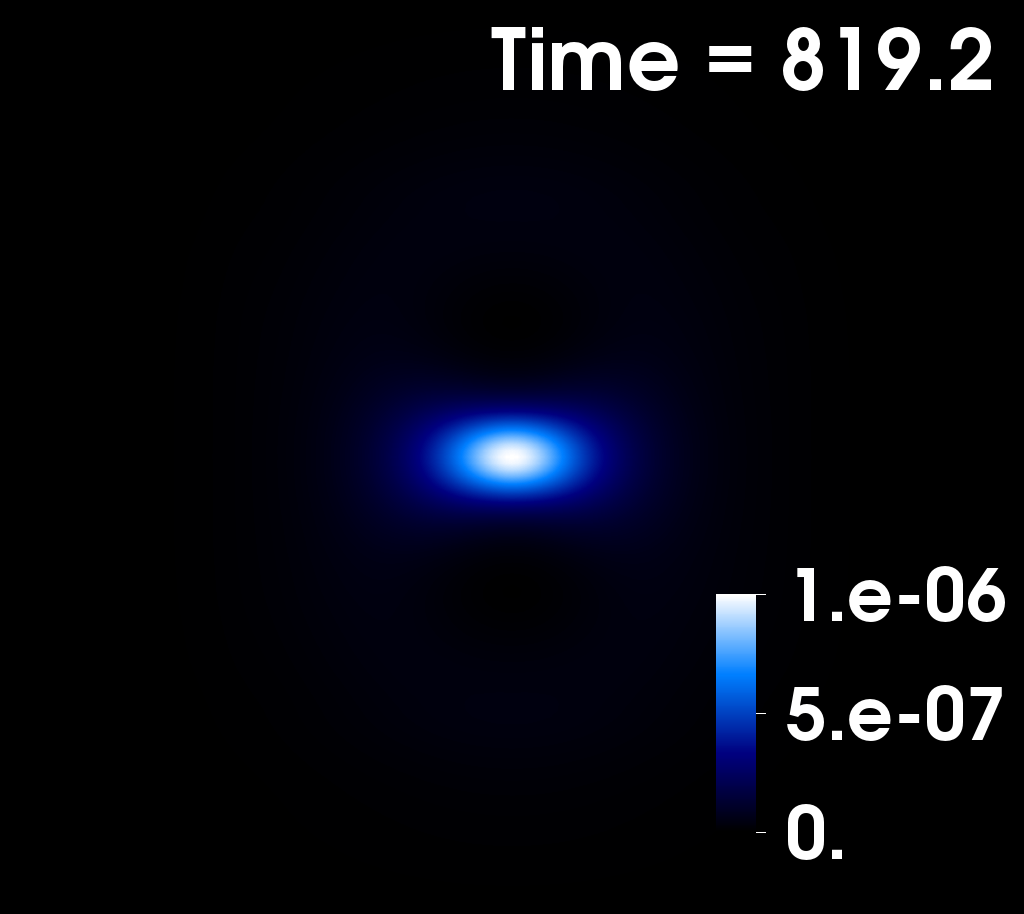}\hspace{-0.01\linewidth}
\includegraphics[width=0.24\linewidth]{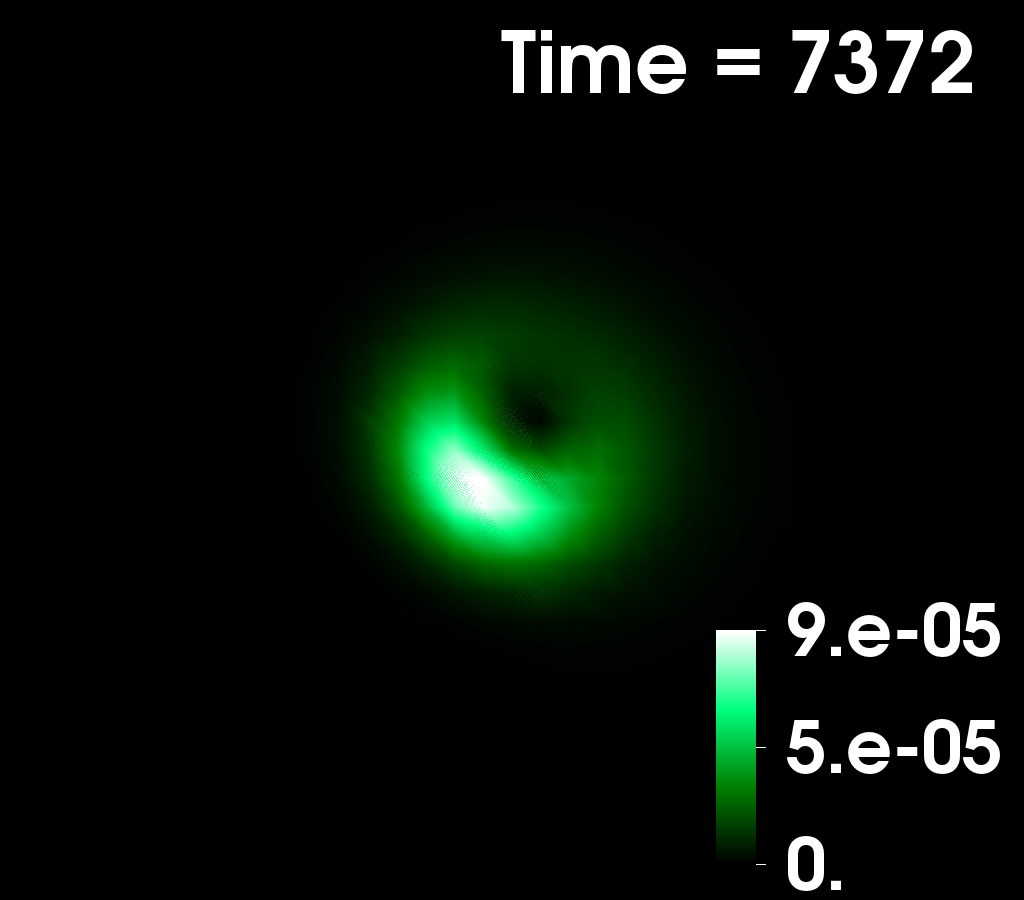}\hspace{-0.01\linewidth}
\includegraphics[width=0.24\linewidth]{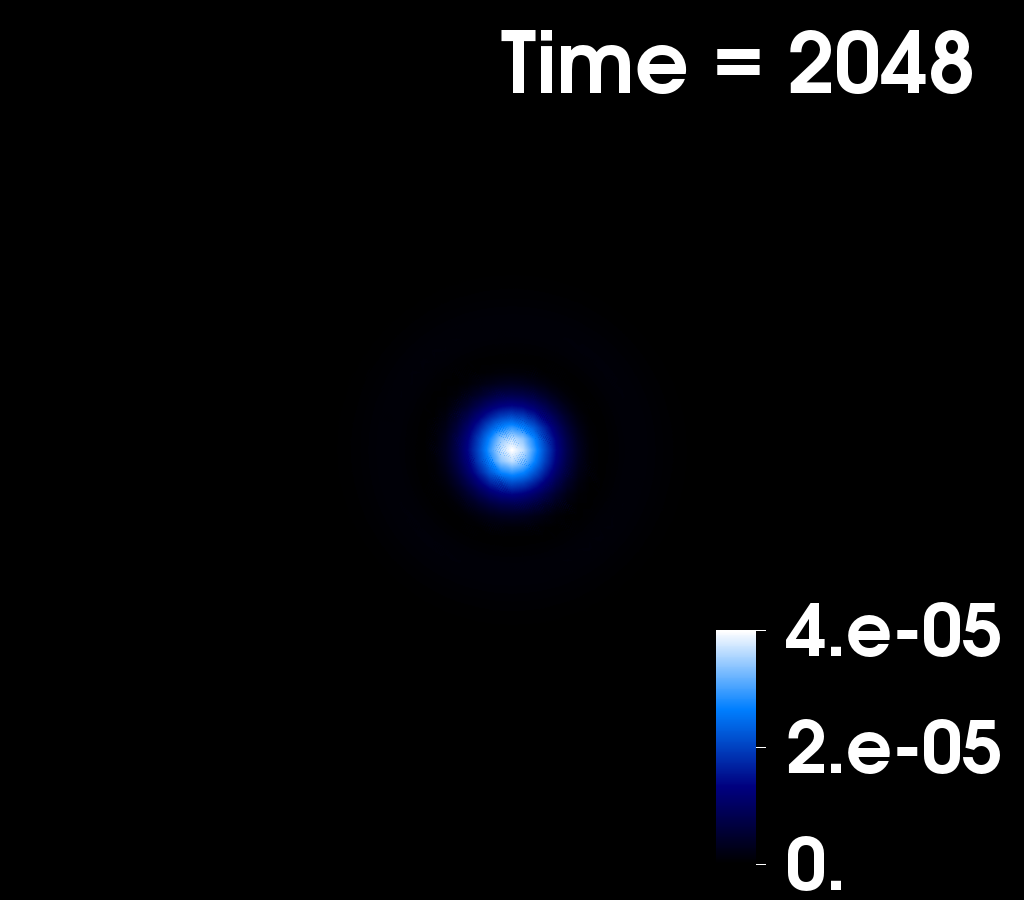}\hspace{-0.01\linewidth}

\includegraphics[width=0.24\linewidth]{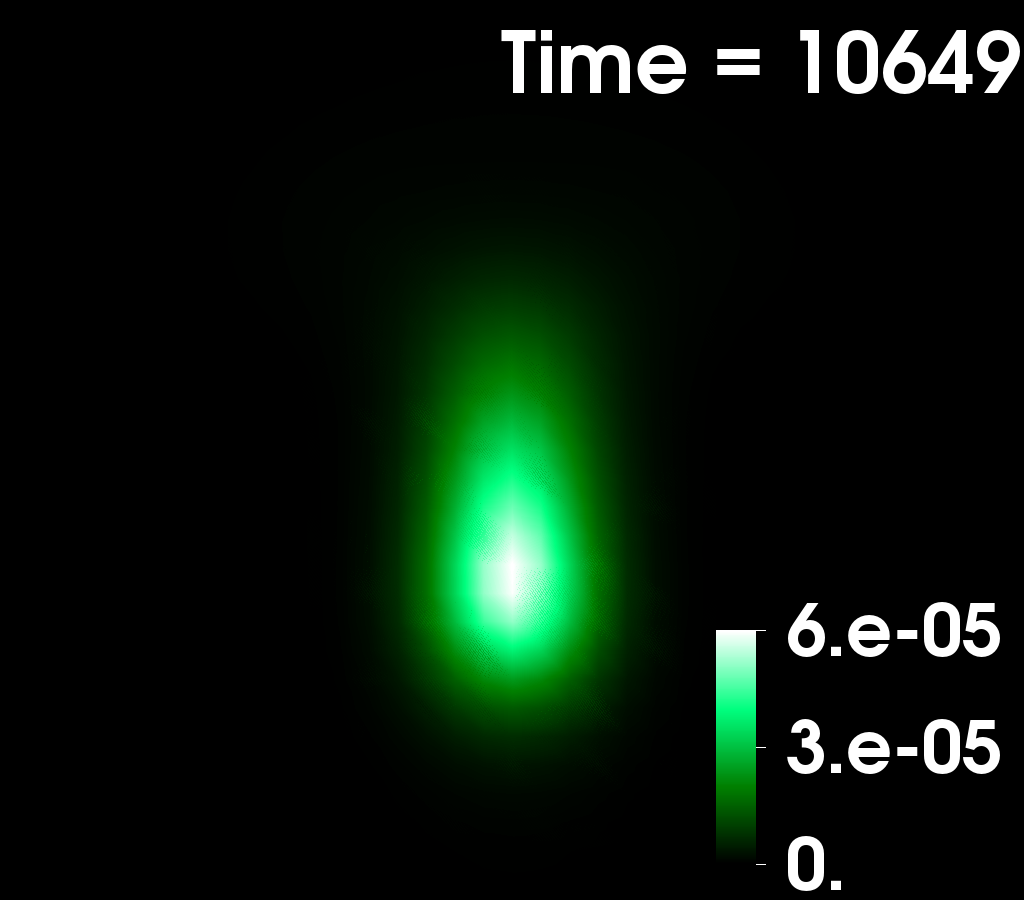}\hspace{-0.01\linewidth}
\includegraphics[width=0.236\linewidth]{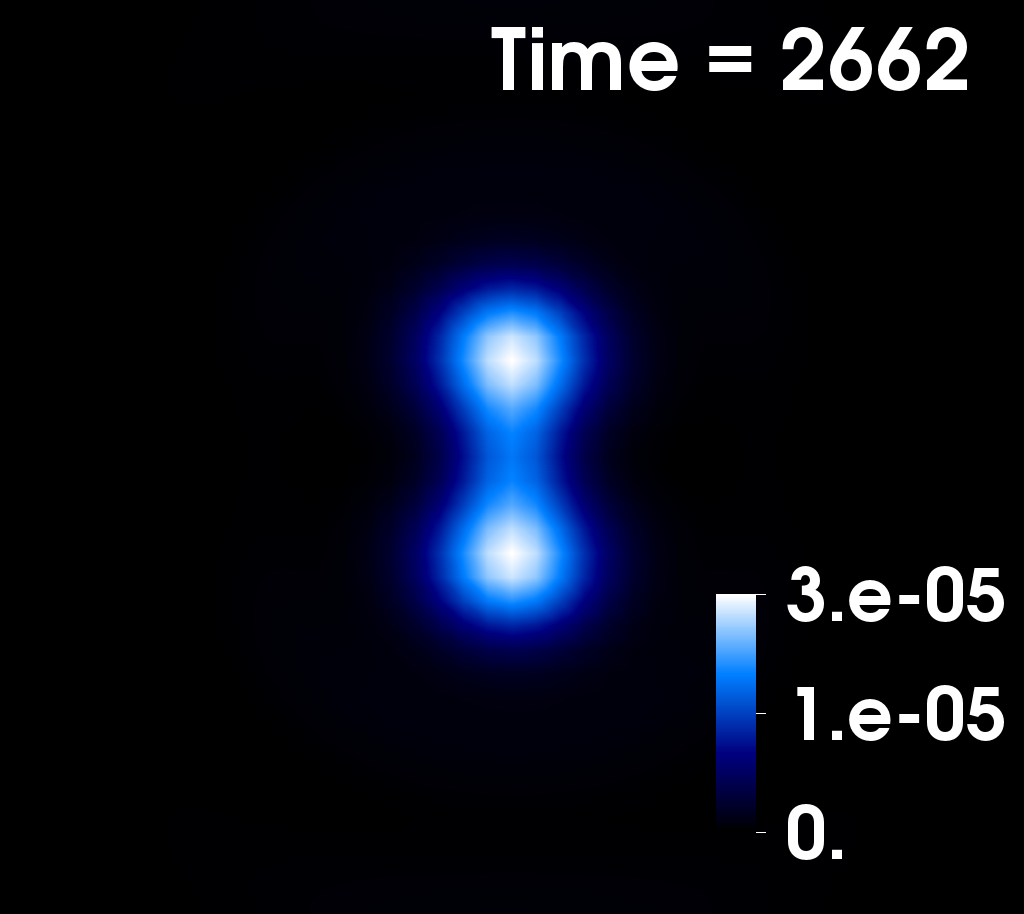}\hspace{-0.01\linewidth}
\includegraphics[width=0.24\linewidth]{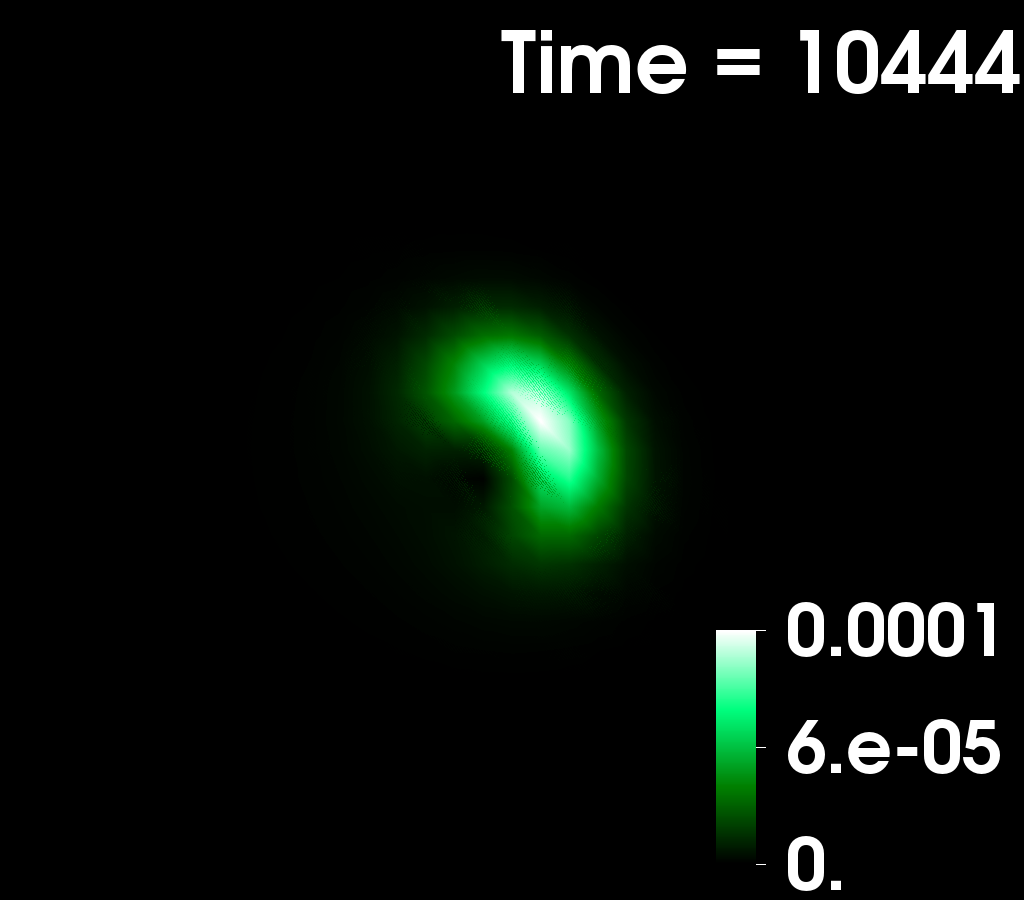}\hspace{-0.01\linewidth}
\includegraphics[width=0.24\linewidth]{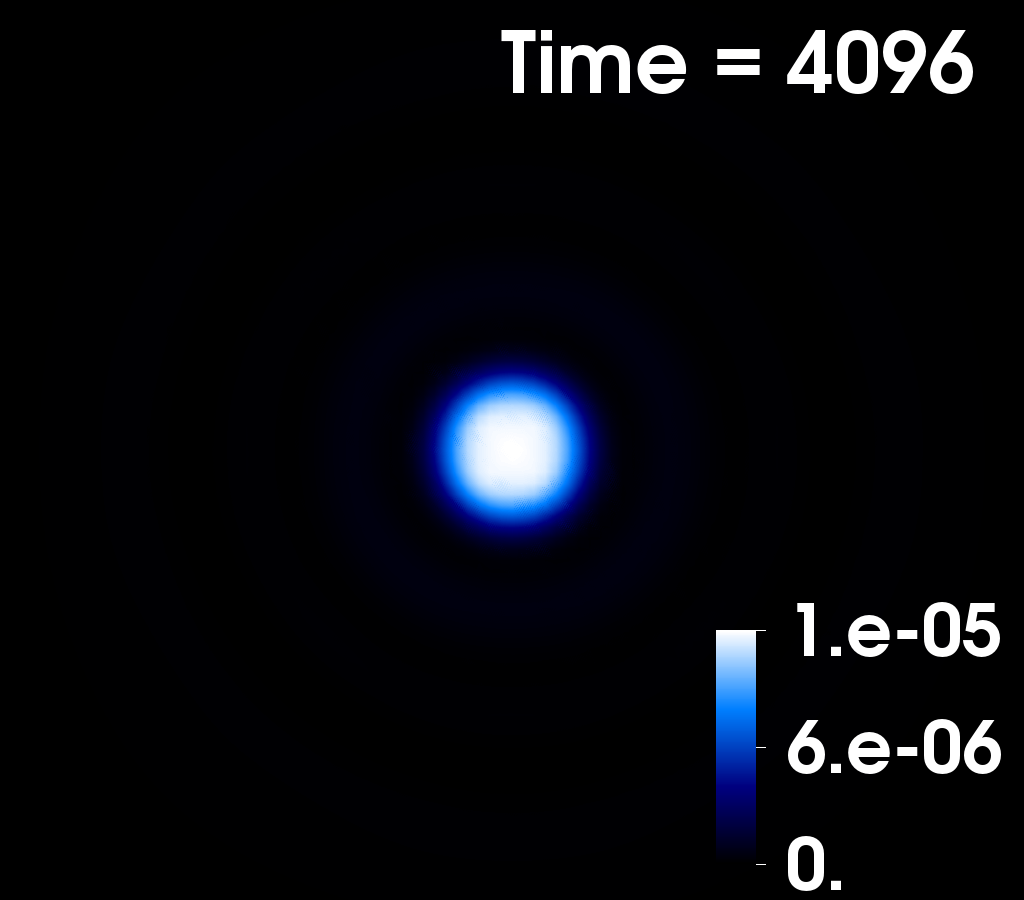}\hspace{-0.01\linewidth}

\includegraphics[width=0.24\linewidth]{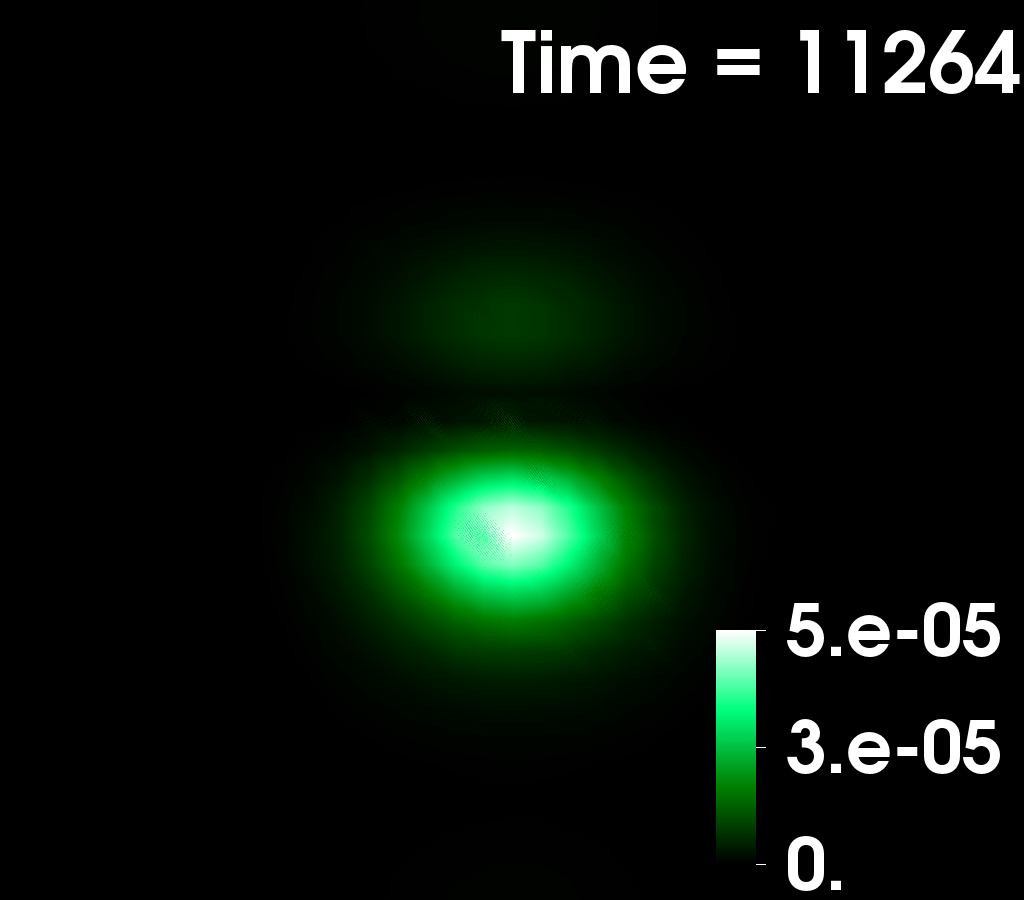}\hspace{-0.01\linewidth}
\includegraphics[width=0.236\linewidth]{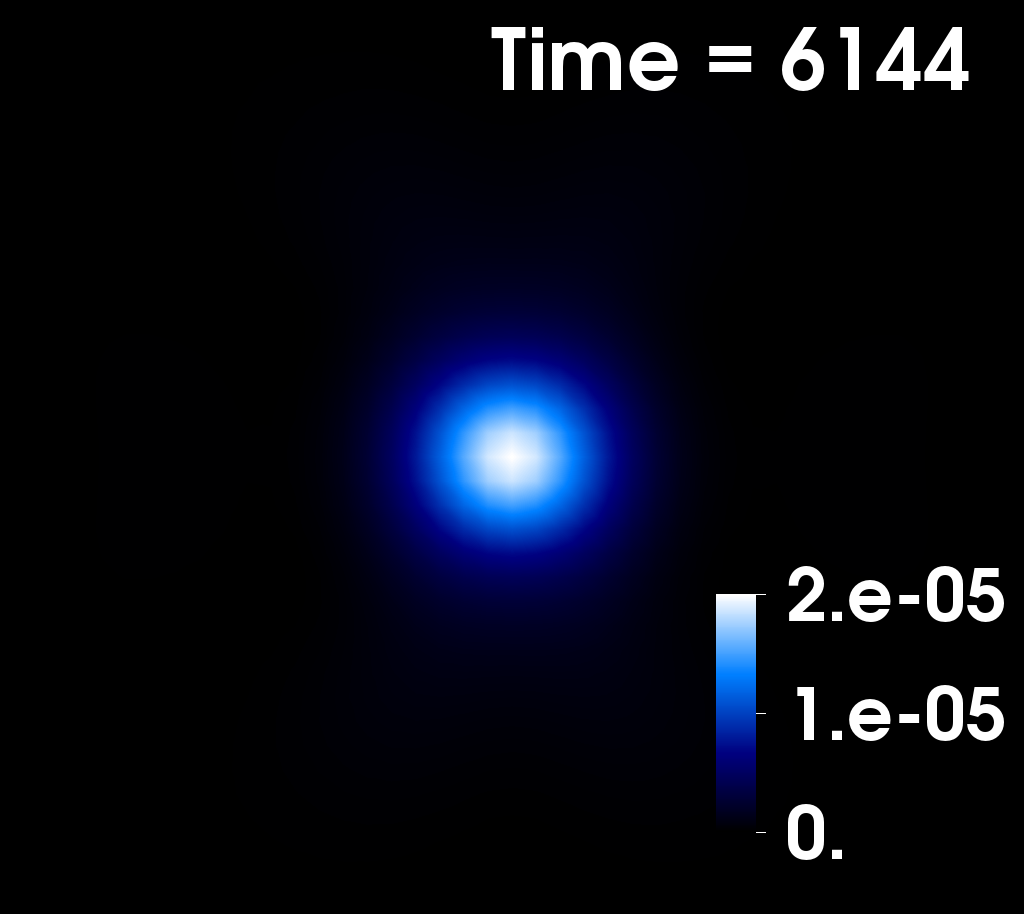}\hspace{-0.01\linewidth}
\includegraphics[width=0.24\linewidth]{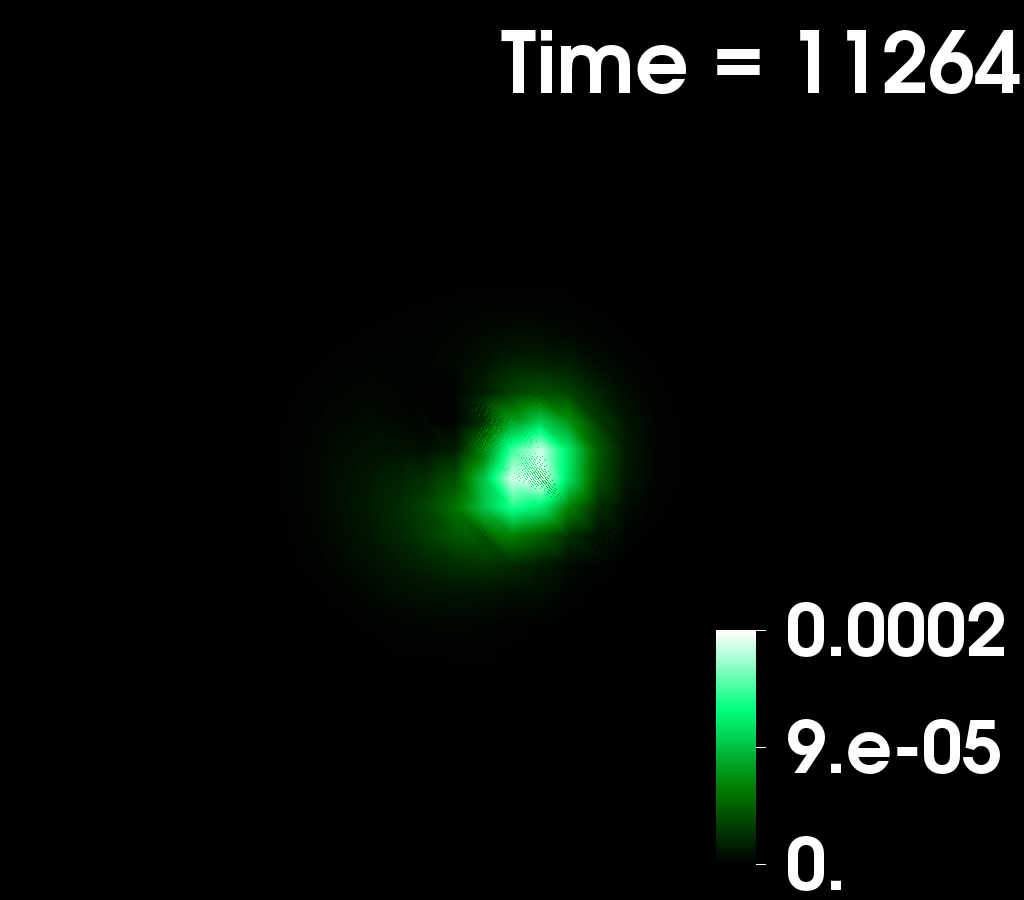}\hspace{-0.01\linewidth}
\includegraphics[width=0.24\linewidth]{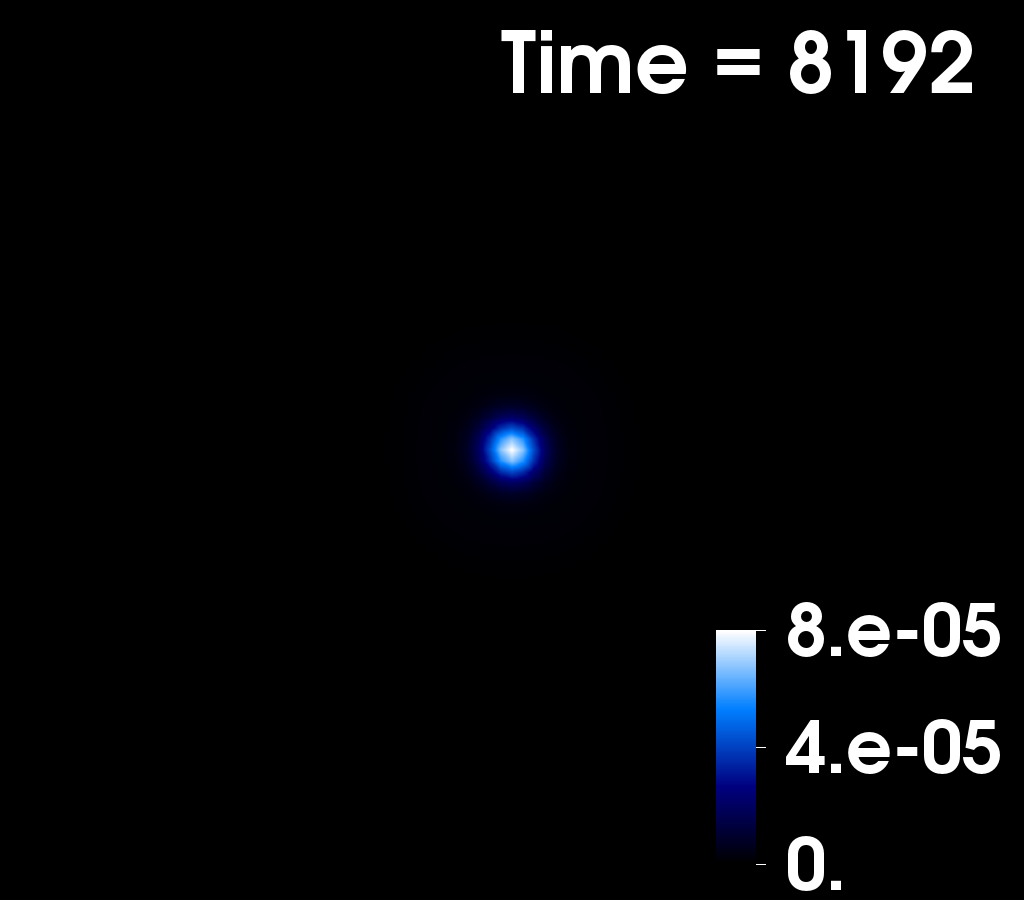}\hspace{-0.01\linewidth}

\includegraphics[width=0.24\linewidth]{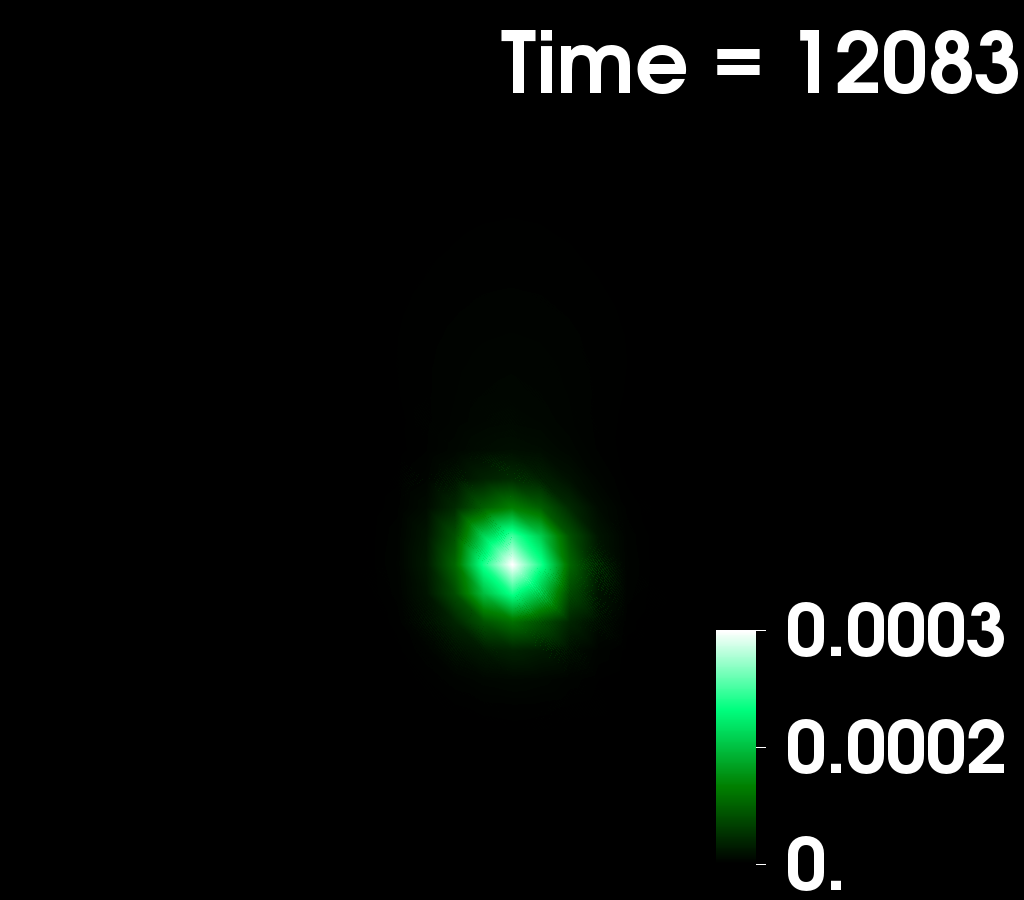}\hspace{-0.01\linewidth}
\includegraphics[width=0.236\linewidth]{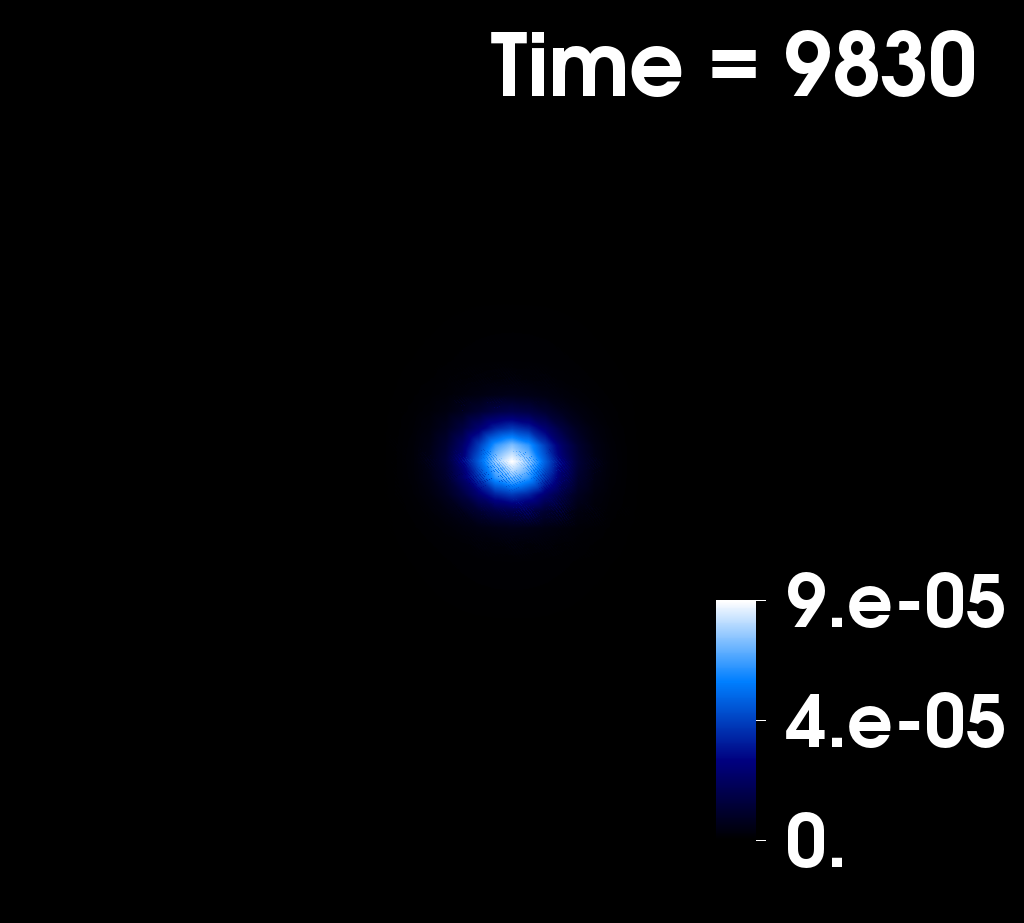}\hspace{-0.01\linewidth}
\includegraphics[width=0.24\linewidth]{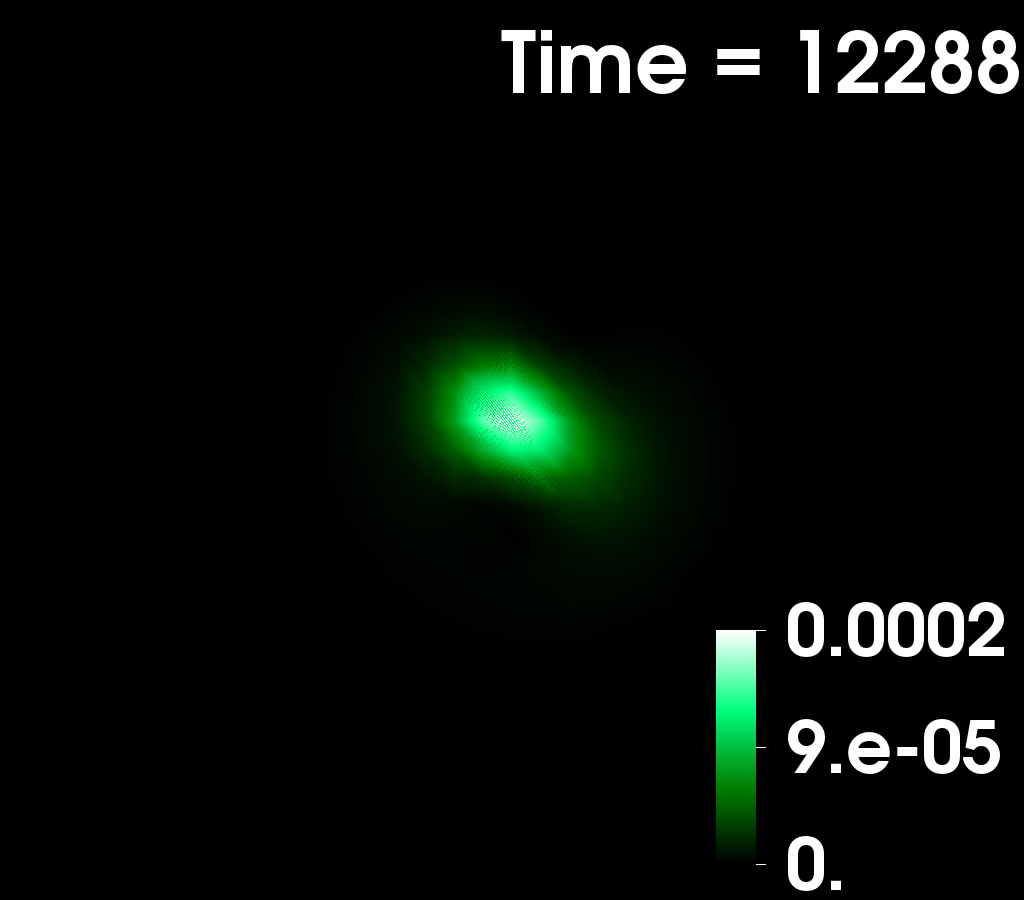}\hspace{-0.01\linewidth}
\includegraphics[width=0.24\linewidth]{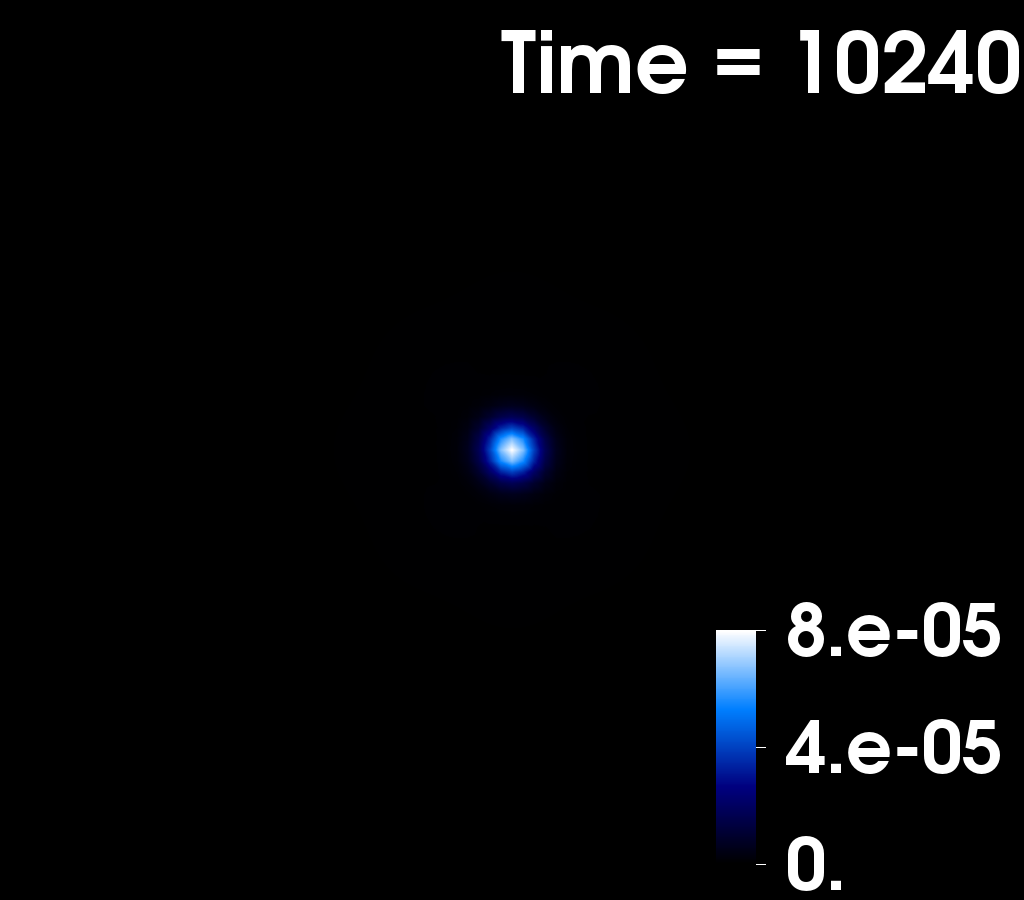}\hspace{-0.01\linewidth}
\caption{Time evolution (top to bottom) 
of the energy density (green for scalar/blue for vector)  in the formation scenario of different BSs. The models in the first three columns become disrupted during the collapse. The model in the fourth column keeps its structure (it only becomes more compact) which shows that vector SBS$_{+ 1}$+SBS$_{-1}$ are stable.}
\label{figformation}
\end{figure}

{\bf {\em Appendix C. Dynamical formation.}} 
%
A further, complementary, confirmation of dynamical robustness is  the existence of a formation mechanism. For the fundamental (monopole) BSs such mechanism exists: gravitational cooling~\cite{Seidel:1993zk,di2018dynamical}. We thus investigate the dynamical formation of some multi-field BSs  from the collapse of an initial dilute cloud of bosonic fields.  Our results show that the stable models can be formed through the gravitational cooling mechanism, while unstable models decay to the spherical non-spinning solutions, see Fig.~\ref{figformation}.
 For instance, starting with dilute distributions of  scalar field with the multipolar components of DBS$_0$, SBS$_{\pm 1}$, and DBS$_{0}$+SBS$_{+1}$, or of a vector  field with the distribution of DBS$_0$, the collapse ends up disrupting the star; on  the other hand,  starting  with dilute distributions of  vector field with the multipolar components of SBS$_{+ 1}$+SBS$_{-1}$, a compact object with these modes survives. The interested reader is addressed to~\cite{sanchis2019nonlinear} for further details on the construction of constraint-solving initial data describing the cloud of bosonic matter that leads to the formation of SBS$_{\pm 1}$. 

The procedure we follow for the new configurations DBS$_{0}$ and for the multi-field BSs is the same
, but with a different ansatz for the fields. In the case of scalar DBS$_{0}$ we consider the following 
\begin{equation}
\Phi^{(0)} (t, r, \theta, \varphi) = A e^{-\frac{r^2}{\sigma^2}} Y_{10}(\theta,\varphi) e^{-i\omega t},
\end{equation} 
where $Y_{10}(\theta,\varphi) = \cos{\theta}$ is the $\ell = 1$, $m = 0$ spherical harmonic, and $A$ and $\sigma$ are free parameters.

The formulation we adopt for the evolution of the Proca field is the one described in~\cite{Zilhao:2015tya} where the Proca field is split up into its scalar potential $\Aphi$, its 3-vector potential $\A_{i}$ and the ``electric" and ``magnetic" fields $E^{i}$ and $B^{i}$. The vector case is more involved than the scalar one because one must solve also the Gauss constraint. Nevertheless, the procedure to solve the Gauss constraint for the vector DBS$_{0}$ is the same as the one we already used for vector SBS$_{1}$, which is described in Appendix A of~\cite{sanchis2019nonlinear}; one must make an ansatz for the scalar potential, which for DBS$_{0}$ is chosen as
\begin{eqnarray} 
\Aphi(t, r, \theta, \varphi)  = R(r) Y_{10}(\theta,\varphi)\,e^{-i\omega t} \,,
\label{scalar_potential}
\end{eqnarray}
where $R(r) = r^2 e^{-\frac{r^2}{\sigma^2}}$.
Once the Gauss constraint is solved for the chosen scalar potential, we obtain the shape of all the other fields:
\begin{align}
E^r(r,\theta) =&\ \frac{A\, \sigma^2}{12 r^3} \biggl( 8 \sigma^4 - 2 e^{-\frac{r^2}{\sigma^2}} [3 r^4 +  \\
     &\ 4r^2\sigma^2 + 4\sigma^4] -\sqrt{\pi}r^3\sigma \,\rm{Erf} \left(\frac{r}{\sigma}\right) \biggr)\cos{\theta}\ , \notag \\
E^{\theta}(r,\theta) =&\ \frac{A\, \sigma e^{-\frac{r^2}{\sigma^2}}}{24 r^3} \biggl( -e^{\frac{r^2}{\sigma^2}}\biggl[2 \sqrt{\pi} \sigma^2 r^3 + 8 \sigma^5  \\
		 & -2\sqrt{\pi}\sigma^3r^2 \,\rm{Erf} \left(\frac{r}{\sigma}\right)\biggr] + 8 r^2 \sigma^3 + 8 \sigma^5 \biggr) \sin{\theta}\ , \notag \\
E^{\varphi}(r,\theta) =&\ \frac{A\, \sigma e^{-\frac{r^2}{\sigma^2}}}{24 r^3} \biggl( -e^{\frac{r^2}{\sigma^2}}\biggl[2 \sqrt{\pi} \sigma^2 r^3 + 8 \sigma^5  \\
		 & -2\sqrt{\pi}\sigma^3r^2 \,\rm{Erf} \left(\frac{r}{\sigma}\right)\biggr] + 8 r^2 \sigma^3 + 8 \sigma^5 \biggr) \frac{\cos{\theta}}{\sin^2{\theta}} \ , \notag \\
\A_i(r,\theta) =&\ \frac{1}{\alpha} (\omega + \beta^{\varphi}) \gamma_{ij}E^j(r,\theta) \ .
\end{align}

As we are considering the case $m=0$ there is no dependence on the azimuthal coordinate $\varphi$ in the fields. Notice also that the radial dependence of the field is the same as for SBS$_{+1}$, but its angular dependence is different.

\end{document}